\documentclass[seceqn]{elsart}
\usepackage{natbib}
\usepackage{color}

\usepackage{graphicx}
\journal{\url{http://hal.inria.fr/inria-00178189}}

\usepackage{xypic}
\usepackage{latexsym,amsmath,amssymb}
\usepackage{stmaryrd}
\usepackage{theorem,url}
\newenvironment{proof}{\textsc{proof:} }{\ensuremath{\Box}}
\newenvironment{compy}{\textsc{Computational Note:} }{\ensuremath{\bullet}}
\newcommand{\further}[1]{}
\newcommand{\sil}[1]{}

\theoremheaderfont{\bfseries\slshape}
\theorembodyfont{\slshape}
\newtheorem{exy}{Example}[section]
\newtheorem{propy}[exy]{Proposition}
\newtheorem{coly}[exy]{Corollary}
\newtheorem{theoy}[exy]{Theorem}
\newtheorem{lemy}[exy]{Lemma}
\newtheorem{defy}[exy]{Definition}

\newtheorem{noty}[exy]{Notation}

\theorembodyfont{\slshape}
\theorembodyfont{\slshape} 

\newcommand{\algf}{\sffamily  }
\newcommand{\italgf}{\slshape  }

\newcommand{\hb}[1]{\hbox{ #1 }}
\newcommand{\Ref}[1]{(\ref{#1})}
\newcommand{\secr}[1]{Section~\ref{#1}}

\newcommand{\propr}[1]{Proposition~\ref{#1}}
\newcommand{\colr}[1]{Corollary~\ref{#1}}
\newcommand{\theor}[1]{Theorem~\ref{#1}}
\newcommand{\lemr}[1]{Lemma~\ref{#1}}
\newcommand{\exr}[1]{Example~\ref{#1}}
\newcommand{\defr}[1]{Definition~\ref{#1}}
\newcommand{\notr}[1]{Notation~\ref{#1}}

\newcommand{\seq}[3][1]{\ensuremath{{#2}_{#1},\ldots,{#2}_{#3}}}

\newcommand{\ds}{\displaystyle}

\newcommand{\pd}[1]{\ensuremath{\frac{\partial \hbox{ }}{\partial #1}}}
\newcommand{\pdif}[2]{\frac{\partial #1}{\partial #2}}
\newcommand{\equi}{\; \Leftrightarrow \; }

\newcommand{\dt}[1]{\frac{\hbox{d} #1}{\hbox{d}t  }}

\newcommand{\Ni}{\ensuremath{\mathbb{N}}}

\newcommand{\Ri}{\ensuremath{\mathbb{R}}}

\newcommand{\gva}{\ensuremath{\mathcal{G}} }
\newcommand{\dimg}{\ensuremath{r} }

\newcommand{\liea}[1][]{\ensuremath{{{\mathrm{v}}}_{#1}}}

\newcommand{\lliea}[1][]{\ensuremath{{\hat{{\mathrm{v}}}}_{#1}}}

\newcommand{\ac}[1][]{\ensuremath{g_{#1}} }
\newcommand{\vf}[1][]{\ensuremath{{\mathrm{V}}^{#1}}}

\newcommand{\zva}{\ensuremath{\mathcal{M}} }

\newcommand{\ova}{\ensuremath{\mathcal{O}} } 
\newcommand{\dimo}{\ensuremath{d} }

\newcommand{\sid}{\ensuremath{P} } 
\newcommand{\sva}[1][]{\ensuremath{\mathcal{P}^{#1}} }

\newcommand{\ncs}[1][]{\ensuremath{\mathcal{U}^{#1}}}

\newcommand{\inv}{\ensuremath{\bar\iota}}

\newcommand{\fiu}[2][\alpha]{\ensuremath{\mathfrak{u}_{#1}^{#2}}}
\newcommand{\fix}[2][]{\ensuremath{\mathfrak{x}^{#2}_{#1}}}
\newcommand{\fiy}[1]{\ensuremath{\mathfrak{z}^{#1}}}

\newcommand{\icv}{\ensuremath{\bar\iota}}

\newcommand{\dd}{\hbox{d}}

\newcommand{\mfm}[1][ ]{moving frame#1} 
\newcommand{\cnis}[1][ ]{normalized invariants#1} 
\newcommand{\ngen}{\ensuremath{N}}
\newcommand{\A}[1][]{\ensuremath{\mathfrak{A}^{#1}}}

\newcommand{\xva}{\ensuremath{\mathcal{X}}}
\newcommand{\xx}[1][]{\ensuremath{x_{#1}}}

\newcommand{\dimx}{\ensuremath{\ensuremath{m}}}

\newcommand{\uva}{\ensuremath{\mathcal{U}}}
\newcommand{\uu}[1][]{\ensuremath{u^{#1}}}

\newcommand{\dimu}{\ensuremath{n}}

\newcommand{\jva}[1][]{\ensuremath{{\mathrm{J}}^{#1}}}
\newcommand{\ninv}[1][]{\ensuremath{\mathcal{I}^{#1}}}

\newcommand{\moinv}[1][]{\ensuremath{\mathcal{E}^{#1}}}
\newcommand{\ndinv}[1][]{\ensuremath{\mathcal{N}^{#1}}}
\newcommand{\cminv}[1][]{\ensuremath{\Lambda_{#1}}}
\newcommand{\mcm}[1][]{\ensuremath{K}_{#1}} 

\newcommand{\F}[1][]{\ensuremath{\FR(\jva[{#1}])}}
\newcommand{\FG}[1][]{\ensuremath{\FRG(\jva[{#1}])}}
\newcommand{\FA}[1][]{\ensuremath{\FR(\A[{#1}])}}
\newcommand{\FR}{\ensuremath{\mathcal{F}}}
\newcommand{\FRG}{\ensuremath{\mathcal{F}^\gva}}

\newcommand{\D}[1][]{{\mathrm{D}}^{#1}}
\newcommand{\pD}[1][]{\tilde{\mathrm{D}}^{#1}}
\newcommand{\der}[1][]{\ensuremath{\mathrm{D}_{#1}}}

\newcommand{\ider}[1][]{\ensuremath{\mathcal{D}_{#1}}}
\newcommand{\iD}[1][]{\ensuremath{\mathcal{D}^{#1}}}
\newcommand{\idx}[1][]{\ensuremath{\mathfrak{x}_{#1}}}
\newcommand{\fder}[1][]{\ensuremath{\mathfrak{D}_{#1}}}
\newcommand{\fD}[1][]{\ensuremath{\mathfrak{D}^{#1}}}
\newcommand{\Syzr}{\ensuremath{\mathfrak{R}}}
\newcommand{\Syzs}{\ensuremath{\mathfrak{S}}}
\newcommand{\Syzt}{\ensuremath{\mathfrak{T}}}
\newcommand{\syzs}[2]{\ensuremath{{S}^{#1}_{#2}}}
\newcommand{\syzt}[2]{\ensuremath{{T}^{#1}_{#2}}}

\newcommand{\bas}[1][\beta]{\ensuremath{\bar{#1}}}
\newcommand{\cre}[1][\beta]{\ensuremath{\hat{#1}}}
\newcommand{\fs}[1][\beta]{\ensuremath{f(#1)}}
\newcommand{\ls}[1][\beta]{\ensuremath{l(#1)}}

\newcommand{\sep}{\hbox{\algf sep}}

\begin{document}

\begin{frontmatter}
\title{Differential invariants of a Lie group action: 
             syzygies on a generating set} 
\author[eh]{Evelyne Hubert}
\ead[url]{www.inria.fr/cafe/Evelyne.Hubert}
\ead{evelyne.hubert@sophia.inria.fr}
\address[eh]{INRIA Sophia Antipolis, France}

\begin{abstract}
Given a group action, known by its infinitesimal generators,
we exhibit a complete set of syzygies on a generating 
set of differential invariants.
For that we elaborate on  the reinterpretation of 
Cartan's moving frame by \cite{fels99}.
This provides  constructive tools for 
exploring algebras of differential invariants. 
\end{abstract}


\end{frontmatter}

\section*{Introduction} 
  \label{intro} 

A great variety of group actions arise in mathematics, 
physics, science  and engineering and their invariants, 
 whether algebraic or differential, are commonly used for 
symmetry reduction or to solve equivalence problems and  
determining canonical forms. 
Classifying  invariants is consequently an essential task. 
One needs to determine a
generating set of invariants and their syzygies, 
i.e. the relations they satisfy.

With minimal amount of data on the group action,
we shall characterize two generating sets of differential invariants.
Though not computing them explicitly, we describe inductive
processes to rewrite any differential invariants in terms of them 
and their invariant derivatives. 
For one of those generating 
set we determine a complete set of
 differential relationships, which we call syzygies. 
The other generating set  is  of bounded cardinality and a 
 complete set of syzygies can be computed from the previous one
by the generalized  differential elimination scheme 
provided by  \citet{hubert05}.

The results in this paper are constructive and our 
presentation describes very closely  their symbolic
implementation in \textsc{aida}   \citep{aida}. 
They are indeed part of a bigger project the aim of which is to
develop the foundations for  symmetry reduction of differential systems
with a view towards differential elimination.
This is  outlined in the motivational example of \cite{hubert05}.
The computational requirements  include four main components: 
the explicit computation of a generating set of invariants (1), 
and the relations among them (2);
procedures for rewriting the problem in terms of the invariants (3);
 and finally  procedures for computing in the algebra of invariants (4). 
In this paper we focus on (2) and (3)
while (1) and (4) were consistently addressed by 
\citet{hubert07,hubert07b} and \citet{hubert05} respectively.
This paper thus completes an algorithmic suite. 
While component (4) has been implemented as a generalization 
of the  \textsc{maple}  library  \emph{diffalg} \citep{diffalg,ncdiffalg},
components (1-3) is implemented in  our Maple package 
\textsc{aida}   \citep{aida} that works on top of 
the \textsc{maple}  library
\emph{DifferentialGeometry} \citep{vessiot}, 
as well as \emph{diffalg} and \emph{Groebner}.
In this paper we also use component (4) to reduce the number
of generators, while still providing the complete syzygies.

On one hand, the question of the finite generation 
of differential invariants was addressed by 
\citet{tresse94,kumpera74,kumpera75a,kumpera75b,munoz03}, in the more
general case of pseudo-groups - see also
 \cite{ovsiannikov78,olver:purple} for Lie groups.
On the other hand, \citeauthor{griffiths74}'s (\citeyear{griffiths74})  
interpretation of Cartan's (\citeyear{cartan35,cartan37,cartan53})
moving frame  method solved equivalence problem in many geometries
\citep{green78, jensen77,gardner89,ivey03}.
Alternatively, the approach of \cite{gardner89} and its
recent symbolic implementation \citep{neut03} 
lead to computational solutions for the 
classification of differential equations \citep{neut02,dridi06,dridi06a}.
Besides \cite{fels99} offered another  interpretation of 
Cartan's moving frame method, the application of which goes 
beyond geometry \citep{olver05s}.  
In particular it includes an explicit approach to the generation properties.

The main original contribution in this paper
is to formalize the notion of differential syzygies 
for a generating set of  differential invariants
and prove the completeness of a finite set of those. 
To this end we redevelop the construction of normalized invariants
and invariant derivations of \citet{fels99} in a spirit we believe 
closer to the audience of this journal. 
We offer alternative proofs, and sometimes more general results.
In particular we shall put the emphasis on derivations, 
rather than differential forms.

One is interested in  the action (effective on subsets)  
of a group $\gva$ on a manifold $\xva\times \uva$ and 
its prolongation to the higher order jets $\jva[k](\xva,\uva)$.
In other words, $\xva$ is the space of
independent variables while $\uva$ is the set of dependent variables.
The jet space is  parameterized by the derivatives of the
dependent variables with respect to the independent variables.
At each order $k$, a local cross-section to the orbits
defines a  finite set of \cnis[.] Those latter form a 
generating set for differential invariants of order $k$, in a functional sense.
Rewriting those latter in terms of the normalized invariants 
is furthermore a trivial substitution. We review this material in
 \secr{diffinv}, following the presentation of \citet{hubert07b}. 

As the orbit dimension stabilizes at order $s$  
the action becomes locally free and, 
to any local cross-section, we can associate a 
\emph{moving frame}, i.e. an equivariant map 
 $\rho: \jva[s](\xva,\uva) \rightarrow \gva$ \citep{fels99}. 
The moving frame defines in turn a basis of 
invariant derivations. The great value of this particular set of
invariant derivations is the fact that we can write explicitly their
action on invariantized functions. This is captured in the so called
\emph{recurrence formulae}. 
They are the key to proving generation, rewriting and syzygies.
\cite{fels99} gave the recurrence formulae for the normalized 
invariants in the case of a coordinate cross-section.
We propose generalized recurrence formulae in the case of any
cross-section and offer an alternate proof, close in spirit to
the one of \cite{mansfield:bk}. 

We can then show that \cnis of order $s+1$ form a generating set 
with respect to those invariant derivations. Rewriting any differential
invariant in terms of those and their derivative is  a simple
application of the recurrence formulae (\secr{geni}).
By exhibiting a canonical rewriting, we can prove the completeness of
a set of differential syzygies for those differential invariants,
 after giving this concept a definition (\secr{syzygies}). 

We formalize the notion of syzygies through the introduction
of the  algebra of \emph{monotone derivatives}. 
Along the lines of   \citet{hubert05}, 
this algebra is equipped with derivations that are 
defined inductively so as to encode the nontrivial commutation rules of the invariant derivations. 
The syzygies are the elements of the kernel of the differential
morphism between the algebra of monotone derivatives and 
the algebra of differential invariants, equipped with the invariant derivations.
The type of differential algebra introduced at this stage 
was shown to be a natural generalization
of classical differential algebra \citep{ritt,kolchin}. 
In the polynomial case, 
it is indeed endowed with an effective differential elimination 
theory that has been implemented \citep{ncdiffalg,hubert05}.

For  cross-sections of minimal order we can also prove that
the set of \emph{edge invariants} is generating.
This latter set has a cardinality bounded  by 
$\dimx \,\dimg + d_0$, where $\dimx$, $\dimg$ are the dimensions of
$\xva$ and $\gva$ while $d_0$ is the codimension of the orbits on
$\xva\times \uva$.
This is a generalization of the result of \cite{olver07}
that bears on coordinate cross-sections. The edge invariants
then form a subset of the normalized invariants of order $s+1$.
\cite{fels99} first conjectured  syzygies 
on this set of generating invariants. 
We feel that constructing directly a complete and 
finite set  of syzygies for the set of edge invariants
is challenging, the problem bearing a high combinatorial difficulty.
To obtain those, we  suggest to apply 
generalized differential elimination \citep{ncdiffalg,hubert05}
on the set of syzygies for the normalized invariants.
This is illustrated in the examples of \secr{syzygies}, 
\ref{examples} and \ref{d3}.

Similarly, to reduce further the  number of generators 
for the differential invariants we can apply  the same
generalized differential elimination techniques 
  to the syzygies. 
This substantially reduces the work of computing explicitly a
generating set  for a given action. This is an approach that was
applied for surfaces in Euclidean, affine, conformal and projective
geometry \citep{olver07a,hubert07c}.

Let us stress here the minimal amount of data indeed needed 
for the determination of a generating set, the rewriting in terms of those 
and  the differential syzygies. All is based on the recurrence
formulae that can be written with only the knowledge of 
 the infinitesimal generators of the action 
and the equations of the cross-section.
Furthermore the operations needed consist of derivations, 
arithmetic operations and test to zero. 
Provided the coefficients of the infinitesimal generators
are rational functions, which provide a general enough class,
we are thus in the realm of symbolic computation since
we can indeed always choose linear equations for the cross-section. 
On the other hand, the explicit expression of the invariant derivations, 
or the  differential invariants,
requires the knowledge of the moving frame. 
This latter is obtained by application of the implicit function
theorem on the group action.
This is therefore not  constructive in general,
but there are algorithms in the algebraic case 
 \citep{hubert07,hubert07b}.

In \secr{action} we extract from the books of 
\citet{olver:yellow,olver:purple} the essential material 
we need for describing actions and their prolongations. 
In \secr{amf} we define invariantization and \cnis 
 for the action of a group on a manifold
along the lines of \cite{hubert07b}. 
We then extend those notions to differential invariants.
In \secr{invder} we define invariant derivations as the derivations
 that commute with the infinitesimal generators of the action. 
We introduce the construction of invariant derivations
 of \cite{fels99} based on the \mfm[] together with the
 \emph{recurrence formulae}.
We write those latter in a more general form (\theor{recurrences}):
the derivations of the invariantization of a function
are given explicitly in terms of invariantizations. 
\secr{geni} discusses then the generation property of the \cnis[] and 
effective rewriting. We furthermore show the generalization of
\cite{olver07a}, the generation property of the \emph{edge invariants}
in the case of minimal order cross-section.
In \secr{syzygies} we emphasize the non uniqueness of the rewriting in
terms of the normalized invariants.
We then introduce the algebra of monotone derivatives,
and the inductive derivations acting on it, in order to 
formalize the concept of syzygies.
We can then write a finite set of syzygies and prove its completeness. 

In the penultimate section we present geometric examples that many readers are
familiar with in order to illustrate our general approach: the action
of the Euclidean group on space curves and surfaces. 
In the last section we undertake the challenging analysis for 
the action of the indefinite orthogonal groups on three independent
variables, and their affine extensions.
To the best of our knowledge,
the structure of their differential algebra had not been explored so far.
Additional  non trivial applications of the results in this paper,
 and the related software, were developed by \cite{hubert07c}.

\section{Group action and their prolongations}
   \label{action} \label{prolongation}  

This is a preliminary section introducing the definition
 and notations for Lie group actions and their prolongation to derivatives.
We essentially follow the books of \citet{olver:yellow, olver:purple}.

\subsection{Local action of a Lie group on a manifold}

\subsubsection*{Pullbacks and push-forwards of maps}

Consider a smooth manifold  $\zva$.
$\FR(\zva)$ denotes the ring of smooth functions
on $\zva$ while $\hbox{Der}(\zva)$ denotes the 
$\FR(\zva)$-module of derivations on $\FR(\zva)$.

If $ \mathcal{N}$ is another smooth manifold
and  $\phi: \zva\rightarrow \mathcal{N}$   a smooth map,
the  \emph{pull-back} of $\phi$ is the map 
$\phi^*:\mathcal{F}(\mathcal{N}) \rightarrow\mathcal{F}(\zva)$ 
defined by  
$\phi^*f = f\circ\phi$ i.e. 
$ (\phi^* f)(z)= f(\phi(z))$ for all $z\in \zva$.
Through $\phi^*$,
$\FR(\mathcal{N})$ can  be viewed as a $\FR(\zva)$ module. 

A derivation $\vf: \FR(\zva)\rightarrow \FR(\zva)$ on $\zva$ 
induces a derivation$\vf|_z: \FR(\zva) \rightarrow \Ri$  at $z$ defined by 
$\vf|_z(f) = \vf(f)(z)$. The set of derivations at a point $z\in \zva$ 
is the tangent space of $\zva$ at $z$. Vector fields on $\zva$ 
can be understood as derivations. 

The \emph{push-forward} or \emph{differential} of $\phi$ is 
defined by 
\[ (\phi_*\vf)(f) (\phi(x)) = \vf(\phi^*f)(x) \]
The coordinate expression for $\phi_* \vf$  
is given by the chain rule. 
Yet this \emph{star} formalism allows us to write formulae in a compact 
way and we shall use it extensively. 

\further{\begin{compy}
If $x=(x_1, \ldots, x_m)$ is a coordinate system on $\zva$ and
$y=(y_1, \ldots, y_n)$ we can write
$\vf = \sum_{i=1}^{m} \alpha_i \, \pdif{}{x_i}$ 
where $\alpha_i \in \FR(\zva)$.
Then $\phi^* \vf = \sum_{j=1}^{n} \beta_j \, \pdif{}{y_j}$ 
with
$\beta_j = \sum_{i=1}^m  \pdif{\phi_j}{x_i}\,\alpha_i $,
where $\phi_j = \phi^* y_j$. 
In matrix form $\beta = \left(  \pdif{\phi}{x}\right) \alpha$.
\end{compy}
}

\subsubsection*{Local action on a manifold}

We consider a connected Lie group \gva of dimension $\dimg$.
The multiplication of two elements 
$\lambda,\mu \in \gva$ is denoted as $\lambda \cdot \mu$.
An action of $\gva$ on a manifold $\zva$ is defined by a 
map 
$\ac : \gva \times \zva \rightarrow \zva$  
that satisfies $\ac(\lambda, \ac(\mu, z))=\ac(\lambda \cdot \mu, z)$.
We shall implicitly consider local actions, that is $\ac$ is 
defined only on an open subset  of $\gva \times \zva$ that contains 
$\{e\}\times\zva$.
We assume that $\zva$ is made of a single coordinate 
chart. If  $(z_1,\ldots,z_k)$ are the coordinate functions
 then $\ac^*z_i :  \gva\times\zva \rightarrow \Ri$ 
represents the $i$th component of the map $\ac$.

There is a fine interplay of right and left invariant 
vector fields in the paper. We thus detail what we mean there now.
Given a group action 
$\ac:\gva\times \zva \rightarrow \zva$ 
define, for $\lambda\in \gva$, 
$\ac[{\lambda}]:  \zva \rightarrow \zva$
by $\ac[\lambda](z)=\ac(\lambda,z)$ for $z\in \zva$.
A vector field $X$ on $\zva$  is
$\gva$-\emph{invariant} if  $\ac[{\lambda *}] X =  X$ 
for all $\lambda\in \gva$,  that is
\[ \forall f \in \FR(\zva),\; \forall z\in \zva,\;
     X ( f\circ \ac[\lambda] )(z) = X(f)(\ac[\lambda](z)) .\]

A vector field on $\gva$ is right invariant if it is invariant under
the action of $\gva$ on itself by right multiplication. 
In other words, if $r_\mu: \gva \rightarrow \gva$ 
is the right multiplication by $\mu^{-1}$,
$r_\mu(\lambda)=\lambda\cdot \mu^{-1}$,  a vector field $\liea$ on 
$\gva$ is right invariant if
\[ \liea(f\circ r_\mu)(\lambda) = \liea(f) (\lambda \cdot \mu^{-1}), \quad 
\forall f\in \FR(\gva). \]

\further{\begin{compy}
Assume $\gva$ has coordinate 
functions $\lambda=(\lambda_1, \ldots, \lambda_s)$.
Let $r_\mu: \gva \rightarrow \gva$ 
be the right multiplication by $\mu\in \gva$ and
write $r_\mu(\lambda) =(r_\mu^1(\lambda), \ldots , r_\mu^s(\lambda))$. 
We look for the maps $\alpha_i: \gva\rightarrow \Ri$ such that
$\liea = \sum_{i=1}^s \alpha_i \pdif{}{\lambda_i}$ 
is a right invariant vector field on $\gva$.

We can write $r_{\mu *} v = \sum_{a=1}^s \beta_i \pdif{}{\lambda_i}$,
where  
$\beta_i: \gva\times\gva \rightarrow \Ri$ are given 
by 
$\beta_i = \sum_{j} \pdif{r^i_\mu}{\lambda_j} \, \alpha_j$ 
or in matrix form
$\beta = \left( \pdif{r_\mu}{\lambda}\right) \, \alpha$.

The invariance property 
$ \liea(f)(\lambda\cdot \mu)= (r_{\mu*}\liea) (f)(\lambda)$  
applied to the coordinate functions 
provide the vector identity: 
\( \alpha(\lambda \cdot \mu^{-1}) = \beta(\lambda, \mu).\) 
In particular for $\lambda=e$ we have
\( \alpha( \mu^{-1}) = \left( \pdif{r_\mu}{\lambda}\right) \alpha(e).\) 
By fixing arbitrarily a vector $\alpha(e)$ 
the formula provides a right invariant vector field. 
\end{compy}
}


For a right invariant vector field on $\gva$,
the \emph{exponential} map $e^{\liea} : \Ri \rightarrow \gva$ 
is the flow of $\liea$ 
such that $e^{\liea}(0)$ is the identity. 
We write $e^{t\liea}$ for $e^{\liea}(t)$.
The defining equation for $e^{\liea}$ is 
 \[ \liea(f)(\lambda)
   =\left.\dt{}\right|_{t=0} f\left(e^{t\liea} \cdot \lambda\right) .\]

Similarly the associated \emph{infinitesimal generator}  $\vf$ 
of the action $\ac$ of $\gva$ on $\zva$ is the vector field
on $\zva$  defined by
\begin{equation} \label{infinitesimal}
\vf(f)(z)= \left. \dt{}\right|_{t=0} f(\ac(e^{t\liea},z)), 
\quad \forall f \in \FR(\zva) .
\end{equation}

Note that $\liea$ is the infinitesimal generator for 
the action of $\gva$ on $\gva$ by left multiplication. 
The infinitesimal generator associated to $\liea$ for the action 
of $\gva$ on $\gva$ by right multiplication,
\( r:\gva \times \gva \rightarrow \gva\), 
\(r(\lambda, \mu)=\mu\cdot \lambda^{-1}\) 
is
\begin{equation}\label{barnougat}
\lliea(f)(\lambda) = 
\left. \dt{}\right|_{t=0} f(\lambda \cdot e^{-t\liea}). 
\end{equation}
We can observe that $\lliea$  is a left invariant 
vector field on $\gva$.

A right invariant vector field on $\gva$ is completely determined by
 its value at identity.
We can thus find a basis $\liea=(\liea[1],\ldots, \liea[r])$ 
for the derivations on $\FR(\gva)$ made of right invariant vector fields.
The associated left invariant vector fields
$\lliea=(\lliea[1], \ldots, \lliea[r])$ 
then also form a basis of derivations on $\FR(\gva)$
 \cite[Chapter 2]{olver:purple}.

The following property is used for the proof of \theor{magnolia} 
and \ref{recurrences}. 
What is used more precisely in \theor{recurrences} is the fact that
 $\liea(\ac^*f)|_{e}=\vf(f)$. This can also be deduced from
Theorem 3.10 by \cite{fels99}. 
In our notations this latter reads as: 
$\liea(\ac^*z_i)=\ac^* \vf(z_i)$. 

\begin{propy} \label{inf2inf}
Let $\liea$ be a right invariant vector field on $\gva$,
 $\lliea$ the associated  infinitesimal generator 
for the action  of $\gva$ on $\gva$ by right multiplication
and $\vf$ the associated infinitesimal generator of the action
 $\ac$ of $\gva$ on $\zva$.

When both $\lliea$ and $\vf$ are considered as derivations on 
$\FR(\gva\times\zva)$ then 
\[ \lliea ( \ac^*f ) + \vf(\ac^*f) =0
   \quad   \hb{and} \quad 
    \vf(\ac^*f)(e,z) = \vf(f)(z), 
    \quad 
   \forall f\in \FR(\zva) .\]
As a particular case we have 
$\lliea(f)(e) = -\liea(f)(e)$.
\end{propy}

\begin{proof}
$\lliea$ is a linear combination of derivations
 with respect to the group parameters, 
i.e. the coordinate functions on $\gva$,
while $\vf$ is  a combination of derivations with respect to 
the coordinate functions on $\zva$.
By \Ref{infinitesimal}  and \Ref{barnougat} we have 
  \[\vf(\ac^*f)(\lambda,z) = 
     \left.\dt{}\right|_{t=0}(\ac^*f)(\lambda, \ac(e^{t\liea},z))\]
and
   \[\lliea(\ac^*f) (\lambda, z) =  \left.\dt{}\right|_{t=0} 
  (\ac^*f)(\lambda\cdot e^{-t\liea}, z) = -\left.\dt{}\right|_{t=0} 
  (\ac^*f)(\lambda\cdot e^{t\liea}, z) . \]
The conclusion follows from the group action property that imposes:
\[ (\ac^*f)(\lambda, \ac(e^{t\liea},z))
    = f(\ac(\lambda, \ac(e^{t\liea},z))
    = f(\ac(\lambda \cdot e^{t\liea}, z))
    =(\ac^*f)(\lambda \cdot e^{t\liea},z).
\]
\end{proof}

\begin{exy} \label{sl1:gp} \label{sl1:inf} 
We consider the group $\gva = \Ri_{>0}^* \ltimes \Ri$  with multiplication
$(\lambda_1,\lambda_2)\cdot(\mu_1,\mu_2)^{-1}= 
 (\frac{\lambda_1}{\mu_1},-\lambda_1\,\frac{\mu_2}{\mu_1}+\lambda_2)$.

A basis of right invariant vector fields is given by 
 \cite[Example 2.46]{olver:purple}
\[ \liea[1]=\lambda_1 \pdif{}{\lambda_1}+\lambda_2\, \pdif{}{\lambda_2}, 
   \quad \liea[2]=  \pdif{}{\lambda_2} .\]
The associated left invariant vector fields, i.e. the 
infinitesimal generators 
for the action of $\gva$ on $\gva$ by right multiplication, are: 
\[ \lliea[1]= -\lambda_1 \pdif{}{\lambda_1}, 
   \quad \lliea[2]=  -\lambda_1 \pdif{}{\lambda_2} .\]

If we consider the action $\ac$ of $\gva$ on $\Ri$ given by
\( \ac^* x =\lambda_1\,x+\lambda_2,\)
the associated infinitesimal generators for this action are
\[ \vf_1 = x\pdif{}{x}, \quad  \vf_2 = \pdif{}{x}. \]
Note that 
\( \lliea[i] ( \ac^* x) = - \vf_i ( \ac^*x)\) and $\lliea[i]|_e = - \liea[i]|_e$.
\end{exy}

\subsection{Action prolongations} \label{prolong}

We shall consider now a manifold $\xva\times \uva$.
We assume that $\xva$ and $\uva$ are 
covered by a single coordinate chart with
respectively $\xx=(\xx[1], \ldots, \xx[\dimx])$ and 
$\uu=(\uu_{1}, \ldots, \uu_{\dimu})$ as coordinate functions.
The $x$ are considered as the independent variables 
and the $u$ as dependent variables.
We discuss briefly the prolongation of an 
action of $\gva$ on $\xva\times \uva$ to its jet space following
 \cite{olver:yellow, olver:purple}.

\begin{noty}  \label{tupleware} The $\dimx$-tuple with 
$1$ at the $i^{\hbox{th}}$ position and $0$ otherwise is denoted
by $\epsilon_i$ .
For $\alpha=(\seq{\alpha}{\dimx})\in \Ni^m$
we note $|\alpha|=\alpha_1+\ldots+\alpha_\dimx$.
If $\D_1,\ldots,\D_\dimx$ are derivations we write
$\D[\alpha]$ for $\D[\alpha_1]_1\ldots\D[\alpha_\dimx]_\dimx$.
Similarly $u_\alpha$ stands for 
$\frac{\partial^{|\alpha|} u}{\partial\; x^\alpha}
= \frac{\partial^{|\alpha|} u}{\partial x_1^{\alpha_1}\ldots\partial x_\dimx^{\alpha_\dimx}}$.

\end{noty}

\subsubsection*{Total derivations} 

The $k$-th order jet space is noted $\jva[k](\xva,\uva)$,
 or $\jva[k]$ for short, while the infinite jet space is $\jva$. 
Besides $\xx$ and $\uu$ the  coordinate functions of  $\jva[k]$
are  $\uu_\alpha$ for $\uu$ in $\{\uu_1, \ldots,\uu_\dimu\}$
and  $\alpha\in \Ni^m$
 with $|\alpha|\leq k$.

The \emph{total derivations with respect to the independent variables} 
are the derivations on $J$ defined by
\begin{equation} \label{totalder}
 \der[i] = \pdif{ }{x_i} 
+ \sum_{\uu\in \uva,\, \alpha\in \Ni^m }
\uu_{\alpha+\epsilon_i} \pdif{ }{\uu_{\alpha}}, \hbox{ for } 1\leq i\leq \dimx .
\end{equation}
In other words, $\der[i]$ is such that
for any $\uu\in\uva$  and $\alpha\in \Ni^m$,
$\der[i](\uu_{\alpha})=\uu_{\alpha+\epsilon_i}$, 
while $\der[i]x_j=1$ or $0$ according to whether $i=j$ or not.

Pragmatically the set of \emph{total derivations}
is the free $\F$-module with basis  $\der=\{\seq{\der}{\dimx}\}$. 
Geometrically one defines total derivations as
the derivations of $\F$ that annihilate the contact forms 
\citep{olver:purple}.
Alternatively they correspond to the formal derivations in
\citep{kumpera74,kumpera75a,kumpera75b,munoz03}.
A total derivation $\D$ is of order $l$ if
 for all $f\in \F[l+k]$, $k\geq0$, $\D(f) \in \F[l+k+1]$.
The total derivations of order $l$ form a
$\F[l]$-module.

\subsubsection*{Prolongation of  vector fields}

Vector fields on $\jva[k]$
form a free $\F[k]$-module a basis of which 
is given by 
$\{ \pdif{}{x}\;|\; x\in \xva\} \cup 
\{ \pdif{}{u_{\alpha}}\;|\; u\in \uva, |\alpha|\leq k\}$.

\begin{defy} 
Let $\vf[0]$ be a vector field on $\jva[0]$.
The $k$-th prolongation $\vf[k]$, $k\geq 0$, 
is  the unique vector field of $\F[k]$
defined recursively  by the conditions
\[ \vf[k+1]|_{\F[k]}= \vf[k], 
\; \hb{and}  \;
 \vf[k+1]\circ \der[i] - \der[i] \circ \vf[k] 
\hb{is a total derivation for all} 1\leq i\leq \dimx . \]
\end{defy}

This definition is to be compared with 
 \citep[Proposition 4.33]{olver:purple} 
given in terms of contact forms. 
The explicit form of the prolongations are given in
Chapter 4 of  \cite{olver:purple}. 

\begin{propy} \label{caspar}
The prolongations of a vector field 
$\vf[0]=\sum_{i=1}^n \xi_{i}\, \pdif{}{x_i}
 +\sum_{j=1}^\dimu\eta_{j}\, \pdif{}{\uu_{j}}$ on $\jva[0]$ 
are  the appropriate restrictions of the vector field
\[ \vf = \sum_{i=1}^n \xi_{i}\, \der[i]
 + \sum_{1\leq j \leq \dimu,\,\alpha\in\Ni^m}
   \D[\alpha](\zeta_{j})\, \pdif{}{ {\uu_{j}}_{\alpha}} 
\quad \hb{where} \quad
 \zeta_{j}= \eta_{j}-\sum_{i=1}^m \xi_{i}\, \der[i]({\uu_{j}}).\]
Furthermore
\(\ds \der[j] \circ \vf -\vf\circ \der[j] = \sum_{i=1}^m \der[j](\xi_i) \der[i], \quad \forall j \in \{1,\ldots, \dimx\} .\)
\end{propy}

\further{\begin{proof}
Observe that we can write 
$\vf[0] = \sum_{i=1}^m \xi_i \der[i]^{(0)} 
 +\sum_{u\in \uva} \zeta_{u} \,\pdif{}{u} $
and that
$ \der[j] \circ \vf[0]- \vf[0]\circ \der[j]
= \sum_{i=1}^m \der[j](\xi_i) \der[i] +
   \sum_{u\in \uva} \der[j](\zeta_u) \, \pdif{}{u} .$
The result follows by induction.
\end{proof}}

\subsubsection*{Action prolongations}
\label{prolongaction}

Consider a connected Lie group $\gva$ 
of dimension $\dimg$ acting on 
$\jva[0]=\xva\times \uva$.

An action of $\gva$  on $\jva[0]=\xva\times\uva$ 
can be prolonged in a unique way to an action 
$\gva \times   \jva[\kappa]\rightarrow \jva[\kappa]$
that defines a contact transformation for each $\lambda\in \gva$.
We shall write $\ac$ as well for the action on any $\jva[k]$. 
The explicit expressions for $\ac^*u_{\alpha}$ is obtained as
follows \citep[Chapter 4]{olver:yellow}.

In order to obtain compact formulae
 we introduce vectorial notations.
 $\D$ denotes the vector of total derivations 
$\der=(\der[1],\ldots \der[\dimx])^T$ on $\F$.
Define the vector $\pD=(\pD_1, \ldots, \pD_\dimx)^T$ of
 derivations on $\FR(\gva\times\jva)$ as  
\begin{equation} \label{elora}
 \pD = A^{-1} \D \; \hb{ where }\; A=\left(\D_i(\ac^*x_j)\right)_{ij}.
\end{equation} 
The total derivations $\D$ are here implicitly 
extended to be derivations on functions of $\gva\times\jva$.
The derivations $\pD$ commute and are such that
$\pD_i(g^* x_j) = \delta_{ij}$ 
and 
$\ac^* u_\alpha = \pD[\alpha] (\ac^* u)$ \cite[Chapter 4]{olver:purple}.
The prolongations are then given by:
\begin{equation} \label{chesapeake}
 \ac^{*} ( \D f)=\pD(\ac^*f), \; \forall f\in \F .
\end{equation} 

If $\vf[0]=(\vf[0]_1, \ldots, \vf[0]_\dimg)$ are the infinitesimal
 generators for the action of $\ac$ on $\jva[0]$ 
then their $k$-th prolongations
$\vf[k]=(\vf[k]_1, \ldots, \vf[k]_\dimg)$ are the infinitesimal
generators for the action of $\ac$ on $\jva[k]$.

\begin{exy} \label{sl1:prolong} 
We consider the group of \exr{sl1:gp}, 
$\gva = \Ri_{>0}^* \ltimes \Ri$ 
and extend trivially its action 
on $\xva^1\times \uva^1$ as follows:
\[ \ac^* x = \lambda_1\,x+\lambda_2, \quad \ac^* u= u.\]
The derivation $\pD=\frac{1}{\lambda_1} \D$ 
allows to compute the prolongations of the action: 
\( \ac^* u_k = \frac{u_k}{\lambda_1^k} .\)
The infinitesimal generators of the action were given 
in \exr{sl1:inf}. Their prolongations are:
\[ \vf_1 = x\, \D - \sum_{k\geq 0} \D[k](x\, u_1) \pdif{}{u_k} 
   = x\, \pdif{}{x}-k\, u_k \pdif{}{u_k},\quad
   \vf_2 = \pdif{}{x} .
\]
\end{exy}

\section{Local and differential invariants} 
   \label{amf}   

We first define the  normalized invariants 
 in the context of a 
group action on a manifold $\zva$. 
We then generalize those concepts to differential invariants.
The material of this section is essentially borrowed from
 \cite{fels99} and  \cite{hubert07b}, following closely this latter.  
We refer the readers to those papers for more details 
 and a substantial set of examples.

\subsection{Normalized invariants}  \label{acni}

We consider the action
$\ac:\gva\times\zva\rightarrow\zva$ 
of the $\dimg$-dimensional Lie group $\gva$ on the smooth manifold 
$\zva$.
\begin{defy}\label{linv} 
A smooth function $f$, defined on an open subset of $\zva$,
 is a \emph{local invariant}
if $\vf(f)=0$ for any infinitesimal generator 
$\vf$ of the action $\ac$ of $\gva$ on $\zva$. 
The set of local invariants is denoted $\FRG(\zva)$.
\end{defy}

This is equivalent to say that, for $z$ in the  definition set of $f$,
 $\ac[\lambda]^*f(z)=f(z)$  for all $\lambda$ in a neighbourhood of
 the identity in $\gva$.

The orbit of a point $z\in \zva$ is the set of points 
$\ova_z=\{\ac(\lambda, z) | \lambda\in \gva\}$.
The action is semi-regular if all the orbits have the same dimension,
 say $\dimo$.
For those  a maximally independent set of 
 local invariants 
is classically shown to exist by Frobenius theorem
 \cite[Theorem 2.23 and 2.34]{olver:purple}.
Alternatively, a geometric method was described for free action  
based on a \emph{\mfm} by \citet{fels99}
and extended to  semi-regular actions 
with the sole use of a cross-section by
 \citet{hubert07b}.

\begin{defy} \label{lsection}
An embedded submanifold $\sva$ of $\zva$ is a \emph{local cross-section}
 to the orbits if there is an open set $\ncs$ of $\zva$
 such that
\begin{itemize}
\item[-] $\sva$ intersects $\ova_{z}^0\cap\ncs$ at a unique
      point $\forall z\in\ncs$, where $\ova_{z}^0$ is the 
      connected component of  $\ova_{z}\cap
      \ncs$, containing $z$.
\item[-] for all $z \in \sva\cap \ncs$,
   $\ova_{z}^0$ and $\sva$ are transversal and of complementary
   dimensions.
\end{itemize}
\end{defy}

Most of the results in this paper restrict to $\ncs$.
We shall thus assume, with no loss, that $\ncs=\zva$.

An embedded submanifold  of codimension  $\dimo$ can be locally 
defined as the zero set of a map $P:\zva \rightarrow \Ri^\dimo$
where the components $(p_1, \ldots, p_\dimo)$  are 
independent functions along $\sva$. 
The transversality and dimension condition in the definition 
induce the following necessary condition for $P$ to define a 
local cross-section $\sva$:
\begin{equation}\label{tr} \mbox{ the rank of the } \dimg\times\dimo
   \mbox{ matrix  }
   \left(\vf_i(p_j)\right)_{i=1..\dimg}^{j=1..\dimo}
   \mbox{ equals to } \dimo \mbox{ on } \sva .\end{equation}

When  $\gva$ acts semi-regularly on $\zva$ there 
is a lot of freedom in choosing a cross-section.
In particular we can always choose a coordinate 
cross-section \cite[Theorem 5.6]{hubert07b}.

A cross-section on $\zva$  defines an invariantization
process that is a projection from $\FR(\zva)$ to $\FRG(\zva)$.

\begin{defy} \label{iota} Let $\sva$  be a local cross-section to
 the orbits of the action $\ac : \gva \times \zva\rightarrow \zva$. 
 Let $f$ be a smooth function on $\zva$. 
The \emph{invariantization} $\icv f$ of $f$  is the function
 defined by  $\icv f(z)=f (z_0)$ 
for each  $z\in \zva$,  where $z_0=\ova_{z}^0\cap\sva$. 
\end{defy}

The invariantization of the coordinate functions on $\zva$ 
are  the \emph{\cnis[]}. \citet[Definition 4.9]{fels99} explain
how invariantization actually  ties in with 
the normalization procedure in Cartan's work.
The following theorem
\citep[Theorem 1.8]{hubert07b} entails  that
normalized invariants  form a  generating set that is
 equipped with a trivial rewriting process. 

\begin{theoy} \label{invariantizationII} \label{susana}
Let a Lie group $\gva$ act semi-regularly on a manifold $\zva$,
and let $\sva$ be a local cross-section to the orbits.
Then the invariantization $\icv f$ of $f:\zva\rightarrow \Ri$ 
is the unique local invariant 
whose restriction to $\sva$ is equal to the restriction of $f$ to
$\sva$. In other words $\icv f|_{\sva}=f|_{\sva}$.
\end{theoy}

Contained in this theorem as well is the fact that two 
local invariants are equal 
if and only if they have the same restriction on $\sva$. 
In particular if $f\in \FRG(\zva)$ then $\inv f = f$.
Now, by comparing the values of the functions involved at the cross-section, 
it is furthermore easy to check that:

 \begin{coly} \label{thomasa}
For $f\in \FR(\zva)$,
 $\inv f (z_1,\ldots, z_n) = f(\inv z_1, \ldots, \inv z_n)$.
\end{coly}

Thus for $f\in  \FRG(\zva)$ we have 
$f(z_1, \ldots, z_n) =  f(\inv z_1, \ldots, \inv z_n)$.
Therefore the normalized invariants
$\{\inv z_1, \ldots, \inv z_n\}$ form a generating set of local invariants: 
any local invariant can be written as a function of those. 
The rewriting is furthermore a simple replacement: we substitute
 the coordinate functions by their invariantizations.

The normalized invariants are nonetheless not functionally independent.
Characterizing the functions that vanish on $(\inv z_1, \ldots, \inv z_n)$ 
amounts to characterize the functions the invariantization of which is zero.
The functions that cut out the cross-section are an example of those.

\begin{propy}  \label{colibria}
Assume the cross-section $\sva$ is the zero set of the map
$P=(\seq{p}{\dimo}): \zva \rightarrow \Ri^{\dimo}$
 which is of maximal rank $\dimo$ along $\sva$.
The invariantization of  $f\in \FR(\zva)$ is zero if and only if,
in a neighbourhood of each point of $\sva$, 
there exist $a_1,\ldots,a_\dimo \in\FR(\zva)$ 
such that  $f = \sum_{i=1}^\dimo a_i \, p_i$. 
\end{propy}

\begin{proof}
Taylor's formula with integral remainder shows the following 
\cite[Paragraph 2.5]{bourbaki67}. For a smooth function $f$ on an open set 
$ I_1\times \ldots \times  I_\dimo\times U\subset \Ri^{\dimo}\times\Ri^l$, 
where the $I_i$ are intervals of $\Ri$ that contain zero,
there are smooth functions $f_0$ on $U$, and 
$f_i$ on $I_1\times \ldots \times  I_i \times U$, $1\leq i\leq \dimo$
such that
$f(t_1, \ldots, t_\dimo, x)=f_0(x)+ \sum_{j=1}^l t_j\, f_j(t_1,\ldots, t_j,x)$
where $f_0(x)=f(0,\ldots, 0, x)$.

Assume that $\inv f=0 \equi f|_{\sva}=0 $.
Since $(\seq{p}{\dimo})$ is of rank $\dimo$ along $\sva$ we can find, 
in the neighbourhood of each point of $\sva$,
$x_{\dimo+1},\ldots, x_{n} \in \FR(\zva)$ such that
$(\seq{p}{\dimo},\seq[{\dimo+1}]{x}{{n}})$
is a coordinate system.
In this coordinate system we have 
$f(0,\ldots, 0, \seq[{\dimo+1}]{x}{{n}})=0$.
The result therefore follows from the above Taylor formula.
\end{proof}

When $\gva$ is an algebraic group and $\ac$ a rational action,
the \cnis  $(\icv z_1, \ldots, \icv z_n)$ are algebraic functions 
and their defining ideal can be computed effectively
\cite[Theorem 3.6]{hubert07b}. 
The method of \citet{fels99} proceed through the moving frame.

\subsection{Moving frames}
\label{mfmapsec}

Invariantization was first defined by \citet{fels99}
 in terms of an $\gva$-equivariant map $\rho:\zva \rightarrow \gva$
called a \emph{moving frame} in reference 
to the \emph{rep\`ere mobile} of \citet{cartan35,cartan37} 
of which they offer a new interpretation.  
As noted already by 
\citet{griffiths74,green78,jensen77,ivey03},
the geometric idea of classical moving frames,
like the Frenet frame for space curves in Euclidean geometry, 
can indeed be understood  as maps to the group.

 An action of a Lie group $\gva$ on a manifold
$\zva$ is \emph{locally free} if for every point $z\in \zva$ its
isotropy group $\gva_{z}=\{\lambda\in \gva\,|\,\lambda\cdot z=z\}$ is
discrete.
Local freeness implies semi-regularity with
the dimension of each orbit being equal to the dimension of the
group.  \citet[Theorem~4.4]{fels99} established the
existence of moving frames for actions with this property.  
It can indeed then be defined by a cross-section to the orbits.

\sil{
\begin{theoy}\label{lmf}
A Lie group $\gva$ acts  locally freely on $\zva$ if and only if
every point of $\zva$ has an open neighborhood $\ncs$ such that there exists a
 map $\rho\colon \ncs\to \gva$
that makes the following diagram commute.
Here right multiplication is chosen for the action of $\gva$ on itself, and
 $\lambda$ is taken in a suitable neighborhood
of the identity in $\gva$.
\[ \xymatrix{
   \zva \ar[d]_{\rho}  \ar[r]^{\lambda} & \zva\ar[d]^{\rho}\\
   \gva \ar[r]_{{\lambda}} & \gva }\]
\end{theoy}

In other words, a \emph{\mfm} is a locally $\gva$-equivariant map,
 that is  $\rho(\lambda\cdot z)= \rho(z) \cdot \lambda^{-1}$ 
for $\lambda$ sufficiently close to the identity.
As before, we shall restrict our attention to $\ncs$ 
and therefore we assume it is equal to $\zva$.}

If the action is locally free and 
$\sva$ is a local cross-section on $\zva$, then the equation
   \begin{equation}\label{mfdef}
        \ac( \rho(z), z )\in \sva \hb{for} z\in\zva 
    \;\hb{and}\; \rho(z)=e, \forall z\in \sva
    \end{equation}
uniquely defines a smooth map $\rho:\zva \rightarrow \gva$ 
in a sufficiently small neighborhood of any point of the cross-section. 
This map is seen to be equivariant: 
$\rho(\lambda\cdot z)= \rho(z) \cdot \lambda^{-1}$
for $\lambda$ sufficiently close to the identity.

If $\sva$ is the zero set of the map 
$P=(p_1, \ldots,  p_\dimg):\zva\rightarrow \Ri^\dimg$ then 
$p_1(\ac(\rho, z) ) =0, \ldots, p_\dimg(\ac(\rho, z) ) =0$ 
are implicit equations for the \mfm. If we can solve 
those,  $\rho$ provides an explicit construction for the 
invariantization process. 
To make that explicit let us introduce the following maps. 
\begin{equation} \label{sigmapi}
 \sigma : \begin{array}[t]{ccc} \zva & \rightarrow & \gva \times \zva \\
                z & \mapsto & (\rho(z), \, z) \end{array}
 \quad \hb{and}\quad 
 \pi=  \ac \circ \sigma:  
      \begin{array}[t]{ccc} \zva & \rightarrow &  \zva \\
                z & \mapsto & \ac(\rho(z), \, z)  
\end{array}
\end{equation}
Proposition 1.16 of \cite{hubert07b} can be restated as:
\begin{propy}\label{foinv} 
\(\icv f = \pi^* f,\) that is 
 \( \icv f(z) = f( \ac(\rho(z), z)) \) for all $z\in \zva$.
\end{propy}

\subsection{Differential invariants} \label{diffinv}

We consider an action $ \ac $ of $\gva$ on $\jva[0]=\xva\times\uva$ 
and its prolongations to the jet spaces $\jva[k]$.
The prolongations of the  infinitesimal generators on $\jva[k]$
are denoted 
$\vf[k]=(\vf[k]_1, \ldots, \vf[k]_\dimg)$
while their prolongations to $\jva$ are denoted
$\vf=(\vf_1, \ldots, \vf_\dimg)$.

\begin{defy} A differential invariant of order $k$  
is a function $f$ of $\F[k]$ such that
$ \vf[k]_1(f)=0, \ldots, \vf[k]_\dimg(f)=0$.
\end{defy}

A differential invariant of order $k$ is thus 
a local invariant of the action prolonged to $\jva[k]$. 
The ring of differential invariants of order $k$ is accordingly 
denoted by $\FG[k]$. The ring of differential invariants of any order is $\FG$.

The maximal dimension of the orbits 
can only increase as the action is
prolonged to higher order jets. 
It can not go beyond the dimension of the group though. 
The stabilization order is  the order at which the maximal 
dimension of the orbits becomes stationary. 
If the action on $\jva[0]$ is locally effective on subsets
 \cite[Definition 2.2]{fels99}, 
i.e.  the global isotropy group of any open set is discrete,
then, for $s$ greater than the stabilization order, 
the action on $\jva[s]$ is locally 
free on an open subset of $\jva[s]$ \cite[Theorem 5.11]{olver:purple}. 
We shall make this assumption of an action that acts locally 
effectively on subsets.
The dimension of the orbits in $\jva[s]$ is then $\dimg$, 
the dimension of the group.

For any $k$, a cross-section to the orbits of $\ac$ in $\jva[k]$ 
defines an invariantization and a set of \cnis on an open set of $\jva[k]$.
As previously we tacitly restrict to this open set though
we keep the global notation $\jva[k]$. 
Let $s$ be equal to or bigger than the  stabilization order and
 $\sva[s]$  a cross-section to the orbits in  $\jva[s]$.
Its pre-image $\sva[s+k]$ in $\jva[s+k]$ by the projection map
$\pi_s^{s+k} : \jva[s+k]\rightarrow \jva[s]$
is a cross-section to the orbits in $\jva[s+k]$. 
It defines an invariantization 
\( \inv :\F[s+k] \rightarrow \FG[s+k] \).
The \emph{normalized invariants of order $s+k$} are 
the invariantizations of the coordinate functions on $\jva[s+k]$.
We note the set of those:
\[
 \ninv[s+k] = \{ \inv x_1, \ldots, \inv x_\dimx\} 
  \cup \{ \inv u_\alpha \, |\, u\in \uva, \; |\alpha|\leq s+k\} .
\]

We can immediately extend \theor{susana} and its \colr{thomasa} to show that 
$\ninv[s+k]$ is a  generating set of differential invariants of order $s+k$ endowed with a trivial rewriting.

\begin{theoy}  \label{susan}
Let $s$ be equal to or greater than  the stabilization order and 
let $\sva[s]$ be a cross-section in $\jva[s]$.
For $f\in \F[s+k]$, $k\in \Ni$,
 $\icv f$ is the unique differential invariant (of order $s+k$) 
whose restriction to $\sva[s+k]$ is equal to the restriction of $f$ to
$\sva[s+k]$.
\end{theoy}

\begin{coly} \label{thomas}
For $f\in \F[s+k]$, $\inv f (x, u_\alpha) = f(\inv x,  \inv u_\alpha)$.
\end{coly}

In particular, if $f\in \FG[s+k]$ then $\inv f = f$ 
and $f (x, u_\alpha) = f(\inv x,  \inv u_\alpha)$.

We furthermore know the functional relationships among the elements
 in $\ninv[s+k]$. 
They are given by the functions the invariantization of which is zero. 
Those are essentially characterized by \propr{colibria}.

\begin{propy}  \label{colibri}
Let $s$ be equal to or greater than the stabilization order.
Consider the cross-section $\sva[s]$  in $\jva[s]$ 
that we assume given as the zero set of 
 $P=(\seq{p}{\dimg}): \jva[s] \rightarrow \Ri^{\dimg}$, a map of 
maximal rank $\dimg$ along $\sva[s]$. 
The invariantization of  $f\in \F[s+k]$, for $k\in \Ni$, 
is zero iff, in the neighbourhood of each point of $\sva[s+k]$,
there exists $a_1,\ldots,a_\dimg \in\F[s+k]$ 
such that  $f = \sum_{i=1}^\dimg a_i \, p_i$. 
\end{propy}

\begin{exy} 
\label{sl1:invariantization} 
We carry on with \exr{sl1:prolong}.

We can choose $P=(x,u_1-1)$ as cross-section in $\jva[1]$.
This already implies that  $\inv x = 0, \inv u_0 = u_0, \inv u_1 = 1$.
The associated moving frame $\rho: \jva[1] \rightarrow \gva$ 
is then defined by  $\rho^* \lambda_1 = u_1, \rho^* \lambda_2 = -x\, u_1$
so that 
  $\inv u_i =  \frac{u_i}{u_1^i}$ 
since 
  $\ac^*u_i = \frac{u_i}{\lambda_1^i}$. 
\end{exy}

\begin{exy}  \label{sl2:invariantization}
We consider the action of $\gva=\Ri_{>0}^*\ltimes \Ri^2$ on 
 $\jva[0]=\xva^2\times\uva^1$, with coordinate $(x,y,u)$, given by:
\[ \ac^* x_1 = \lambda_1 \, x_1+\lambda_2, \quad 
   \ac^* x_2 = \lambda_1 \, x_2+\lambda_3, \quad
   \ac^* u =  u . \]
The derivations 
$\pD_1 = \frac{1}{\lambda_1} \der[1]$ and 
$\pD_2=\frac{1}{\lambda_1}\der[2]$ allow to compute its prolongations:
\[ \ac^* u_{ij}= \frac{u_{ij}}{\lambda_1^{i+j}} .\]

The action is locally free on $\jva[1] \setminus \mathcal{S}$ 
where $\mathcal{S}$ are the points where both $u_{10}$ and $u_{01}$ are zero.
The moving frame associated with  the cross-section defined by
 $P=(x_1,x_2,u_{10}-1)$ is 
$ \rho^*\lambda_1={u_{10}},$
$\rho^*\lambda_2=-x_1\, u_{10},$
$\rho^*\lambda_3=-x_2\, u_{10}.$
It is defined only on a proper subset of $\jva[1]\setminus \mathcal{S}$,
as are the normalized invariants: 
$ \inv u_{ij} = \frac{u_{ij}}{u_{10}^{i+j}}$

On the other hand, if  
we choose the cross-section defined by 
$$P=\left(\;x_1,\;x_2,\;\frac{1}{2}-\frac{1}{2}(u_{10}^2+u_{01}^2)\right)$$
the associated moving frame is well defined on the whole of 
$\jva[1]\setminus \mathcal{S}$:
\[  \rho^*\lambda_1={\sqrt{u^2_{10}+u_{01}^2}}, 
\quad \rho^*\lambda_2=- x_1\,\sqrt{u^2_{10}+u_{01}^2} ,
\quad \rho^*\lambda_3=-x_2\,\sqrt{u^2_{10}+u_{01}^2} .
\]
as are the normalized invariants:
\[ \inv x_1 =0,\quad \inv x_2 = 0, \hb{and}
 \inv u_{ij} =\frac{u_{ij}}{{(u^2_{10}+u_{01}^2)}^{\frac{i+j}{2}} }. \]
This shows that a nonlinear cross-section might have some 
desirable properties.
\end{exy}

\section{Invariant derivations} 
    \label{invder}  

An invariant derivation is a total derivation that 
commutes with the infinitesimal generators. It maps 
differential invariants of order $k$ to differential 
invariant of order $k+1$, for $k$ large enough. 
Classically a basis of commuting 
invariant derivations is constructed with the use of
 sufficiently many differential invariants 
\citep{olver:purple,ovsiannikov78,
 kumpera74,kumpera75a,kumpera75b,munoz03}.
The novel construction proposed by \citet{fels99} is based 
on a \mfm[]. The constructed invariant derivations do not 
commute in general. Their principal benefit
is that they bring an explicit formula for the 
derivation of  normalized invariants. 
This has been known as the \emph{recurrence formulae} 
\citep[Section 13]{fels99}.
They are the key to most results about generation 
and syzygies in this paper.
All the algebraic and algorithmic treatments of differential 
invariants and their applications
\citep{mansfield01,olver07a,hubert07c,hubert08a}
 come as an exploitation of those formulae.

In  \theor{recurrences} we present the derivation formulae 
for any invariantized functions. 
For the proof we take the dual approach of the one of
\cite{fels99} which is therefore close in essence to the 
one presented by 
\cite{mansfield:bk}, based on the application of the chain rule. 

We always consider the action $\ac$ of  a connected 
$\dimg$-dimensional 
 Lie group $\gva$  on $\jva[0]=\xva\times\uva$ and its prolongations.
We make use of a basis of right invariant vector fields 
 $\liea=(\liea[1], \ldots, \liea[r])$ on $\gva$, 
and the associated infinitesimal generators: 
\begin{itemize} 
\item $\vf=(\vf_1, \ldots, \vf_r)^T$ is the 
vector of infinitesimal generators for the action  $\ac$ of $\gva$ on $\jva$
\item $\lliea = (\lliea[1], \ldots, \lliea[r])^T$ is the vector 
of infinitesimal generators for  the action of 
  $\gva$ on itself by right multiplication.
\end{itemize}

\subsection{Infinitesimal criterion}

Recall from \secr{prolong} that total derivations are the derivations 
on $\jva$ that belong to the $\F$-module with basis
 $(\der[1],\ldots,\der[\dimx])$, the total derivations with respect to 
the independent variables $\seq{x}{\dimx}$.

\begin{defy} \label{ider:def} 
An invariant derivation $\iD$ is a total derivation that commutes with 
any infinitesimal generator  $\vf$ of the group action:
 $\iD \circ \vf = \vf\circ \iD$.
\end{defy}

As an immediate consequence of this definition we see that
if $f$ is a differential invariant 
and $\iD$ an invariant derivation
then $\iD(f)$ is an differential invariant.

\begin{propy} \label{ider:crit}
Let  $A=(a_{ij})$ be an invertible $\dimx \times \dimx$ matrix
 with entries in $\F$.
A vector of total derivations
$\iD =\left( \ider[1],\ldots, \ider[\dimx]\right)^T$ defined by
\( \iD = A^{-1} \, \D\)
is a vector of invariant derivations  if and only if,
 for all infinitesimal generator $\vf$ of the action,
\[  \vf(a_{ij})  +\sum_{k=1}^\dimx \der[i](\xi_k)\, a_{kj} = 0 ,
 \quad  \hb{where} \xi_k = \vf(x_k),
\quad  1\leq i,j\leq \dimx . \]
\end{propy}

\begin{proof} 
For all $i$ we have $\der[i]= \sum_{j=1}^m a_{ij} \, \ider[j]$.
By expanding the equality 
$[\der[i],\vf]= \sum_{k=1}^\dimx \der[i](\xi_k)\,\der[k]$ (\propr{caspar}) 
we obtain,  for all $i$,
\[ \sum_{j=1}^\dimx a_{ij}\, [\ider[j], V] = 
\sum_{j=1}^\dimx \left( V(a_{ij})
  +\sum_{k=1}^\dimx \der[i](\xi_k) a_{kj} \right) \ider[j] \]
Since $A$ is  of non-zero determinant
\(  [\ider[j] , V] =0\) for all $j$ if and only if
\( V(a_{ij})  +\sum_{k=1}^\dimx \der[i](\xi_k) a_{kj} = 0, \forall i,j . \)
\end{proof}

As illustration, a classical construction 
 of invariant derivations is given by the following proposition 
\citep{kumpera74,kumpera75a,kumpera75b,olver:purple,ovsiannikov78,munoz03}:
\begin{propy} \label{iderclassical}
If $f_1, \ldots, f_m$ are differential invariants such that the matrix
$A = (\der[i](f_j))_{i,j}$ is invertible 
 then the derivations  \( \iD = A^{-1} \D \)
are invariant derivations.
\end{propy}

\begin{proof}
If $a_{ij} = \der[i](f_j)$ then, by \propr{caspar},
\[ V(a_{ij}) = V( \der[j](f_i) ) = 
  \der[j](V(f_i)) -\sum_{k} \der[j](\xi_k) \,\der[k](f_i)
  =\der[j](V(f_i)) -\sum_{k} \der[j](\xi_k) \,a_{ik}. \]
By hypothesis $V(f_i)=0$ so that the result follows from \propr{ider:crit}.
\end{proof}

The above derivations commute. They can be understood as derivations 
with respect to the new \emph{independent variables} $\seq{f}{\dimx}$.

As a side remark, note that  \defr{ider:def}  is  dual 
 to the infinitesimal  condition for a 1-form to be contact invariant 
\citep[Theorem 2.91]{olver:purple}. 
The  invariant derivations of \propr{iderclassical} 
are dual to the contact invariant 1-forms  
$\dd_H f_1, \ldots, \dd_H f_m$.

\subsection{Moving frame construction of invariant derivations} 
\label{invderdef}

Assume that there exists on $\jva[s]$  a
\mfm  $\rho: \jva[s]\rightarrow \gva$. 
As in \secr{amf} we  construct the additional maps
\begin{equation} \label{dsigmapi}
 \sigma : \begin{array}[t]{ccc} \jva[s+k] & 
          \rightarrow & \gva \times \jva[s+k] \\
                z & \mapsto & (\rho(z), \, z) \end{array}
 \quad \hb{and}\quad 
 \pi= \ac\circ \sigma:  
      \begin{array}[t]{ccc} \jva[s+k] & \rightarrow & \jva[s+k] \\
                z & \mapsto & g(\rho(z), \, z)  
\end{array}
\end{equation}

\begin{theoy} \label{magnolia}
The vector of derivations $\iD = (\sigma^*A)^{-1} \, \D$, 
where $A$ is the $\dimx\times\dimx$ matrix 
$\left(  \der[i]( g^*x_j) \right)_{ij}$,
is a vector of invariant derivations.
\end{theoy}

The matrix $A$ has entries in $\FR(\gva\times \jva[1])$.
Its pull back $\sigma^* A$ has entries in $\F[s]$.
The above result is proved by checking that 
the formula of \propr{caspar} holds. 

\begin{proof}
The equivariance of $\rho$ implies 
$  \rho(g(e^{t\liea},z))=\rho(z)\cdot e^{-t\liea} $
so that 
\( \rho_* \vf = \lliea  \). 
Thus  $\sigma_* V = \lliea+\vf$
that is
\( \sigma_* V(a_{ij})= 
 \lliea( \der[i](g^*x_j) ) + V( \der[i](g^*x_j) ) \).
As derivations on $\FR(\gva\times\jva[s])$, 
$\der[i]$ and $\lliea$ commute while 
the commutator of $\der[i]$ and $\vf$ is given by \propr{caspar}. 
It follows that
\(\sigma_* V(a_{ij}) =  \der[i]( \lliea(g^*x_j))
+\der[i]( \vf(g^*x_j))-\sum_{k=1}^m\der[i](\xi_k)\,\der[k]( g^*x_j) 
.\)
By \propr{inf2inf} the two first terms cancel
and since 
\( V( \sigma^* a_{ij}) =\sigma^* (\sigma_* V)(a_{ij})\)
we have 
\( V( \sigma^* a_{ij})
=-\sum_{k=1}^m \der[i](\xi_k)\,\sigma^*a_{kj}.\)
We can conclude with \propr{ider:crit}.
\end{proof}

\begin{exy} \label{sl1:invder} 
We carry on with  \exr{sl1:prolong} and \ref{sl1:invariantization}.

We found that the equivariant map
associated to $P=(x,u_1-1)$  is given by
$\rho^* \lambda_1 = u_1, \rho^* \lambda_2 = -x\, u_1$.
In addition  $\pD = \frac{1}{\lambda_1} \D$ 
while
$\vf_1 = x\, \pdif{}{u}- \sum_{k\geq0}k\, u_k \pdif{}{u_k}$
and
$ \vf_2 = \pdif{}{x}$. 

Accordingly define  $\iD = \frac{1}{u_1} \D$. We can then verify that
\( [\vf_1, \iD] =0\) and   \( [\vf_2, \iD] = 0. \)
The application of $\iD$ to a differential invariant thus
produces a differential invariant. 
For instance 
\[\iD \left( \frac{u_i}{u_1^i} \right) 
= \frac{u_{i+1}}{u_1^{i+1}} - \frac{u_i}{u_1^{i+2}} u_2 
= \frac{u_{i+1}}{u_1^{i+1}} - \frac{u_i}{u_1^{i}} \frac{u_2}{u_1^2}
.\]
Remembering that $\inv u_i=\frac{u_i}{u_1^i}$ we  can observe that 
$\iD (\inv u_i) =   \inv u_{i+1}-\inv u_2\, \inv u_i$.
This shows that $\iD (\inv u_i) \neq   \inv u_{i+1}$ in general.
The relationship between  these two quantities is the subject 
of \theor{recurrences} below. We shall furthermore observe that nonetheless 
$\iD( \frac{u_i}{u_1^i}) = \inv ( \D( \frac{u_i}{u_1^i}) )$ 
(\colr{iDoninv}).
\end{exy}

\subsection{Derivation of invariantized functions.}

An essential property of the invariant derivations of 
\theor{magnolia} is that we can write explicitly their action on the
invariantized functions.
\theor{recurrences} below is a general form for 
 the recurrence formulae
of  \citet[Equation 13.7]{fels99}.  

Assume that the action of $\ac$ on $\jva[s]$ is locally free and 
that $P =(\seq{p}{\dimg})$ 
defines the cross-section $\sva$.
Let $\rho:\jva[s]\rightarrow \gva$ be the associated \mfm[].
We construct the vector of invariant derivations 
$\iD=(\iD_1,\ldots, \iD_\dimx)$ as in \theor{magnolia}.

Denote by $\D(P)$  the $\dimx \times \dimg$ matrix 
$\left( \der[i]( p_j)\right)_{i,j}$
with entries in $\F[s+1]$ 
while $\vf(P)$ is the $\dimg\times \dimg$ matrix 
$\left( \vf_i(p_j)\right)_{i,j}$
with entries in $\F[s]$.
As $\sva$ is transverse to the orbits of the action of $\gva$ on $\jva[s]$,
the matrix $\vf(\sid)$ has non zero determinant 
along $\sva$ and therefore  in a neighborhood of each of its points.

\begin{theoy} \label{recurrences}
Let $P=(p_1, \ldots, p_\dimg)$ define a cross-section $\sva$
to the orbits in $\jva[s]$, where $s$ is equal to or greater 
than the stabilization order.
Consider $\rho:\jva[s]\rightarrow \gva$ the associated
 \mfm and $\inv:\F \rightarrow \FG$ 
the associated invariantization.
Consider $\iD = (\iD_1, \ldots, \iD_\dimx)^T$ the vector of 
invariant derivations constructed in \theor{magnolia}.
Let $K$ be the $\dimx\times\dimg$ matrix obtained by invariantizing the entries of
\(\D(P)\, \vf(P)^{-1} \).
Then
\[ \iD( \inv f ) = \inv( \D f) - K \, \inv(\vf(f) ). \]
\sil{satisfies
\( \rho^* \omega = -K^T \, \idx \mod \Theta \).}
\end{theoy}

\begin{proof}
From the definition of $\sigma: z \mapsto (\,\rho(z),z\,)$ 
and the chain rule we have
\begin{equation} \label{ostereich}
 \iD( \inv f) (z) =\iD( \sigma^* g^* f) (z) = 
   \iD(g^* f) ( \rho(z),z) +(\rho_* \iD)( g^* f) (\rho(z), z)  .
\end{equation}

Recall the definition of $\pD$ in \secr{prolongaction} that
satisfies $\pD_j(g^*f)= g^*( \der[j]f)$  for all $f\in \F$.
We have 
\( \iD(g^* f)(\rho(z),z)
      =(\sigma^*\pD(g^*f)) (z)
      = \sigma^*g^* (\der f) (z)=\inv (\der f)(z) \)
and \Ref{ostereich} becomes
\begin{equation} \label{chicoutimi}
 \iD( \inv f) (z) =
  \inv(\D f)(z)  + \sigma^*(\rho_* \iD)( g^* f) ( z)  .
\end{equation}

Since  $\lliea=(\lliea[1], \ldots, \lliea[\dimg])$ 
form a basis for the derivations on $\gva$
there is a matrix\footnote{
  With $\iD$ known explicitly,
  we can write $\tilde{K}$ explicitly in terms of coordinates 
  $\lambda=(\lambda_1, \ldots, \lambda_\dimg)$.
  $\tilde{K}$ is the matrix obtained by multiplying the matrix
  $\iD(\rho) = \left(\iD_j(\rho^* \lambda_i)\right)$
  with the inverse of 
  $\lliea ( \lambda ) = \left( \lliea[i]( \lambda_j) \right)$.
  Yet $\sigma^* \tilde{K}$ needs not have differential invariants 
  as entries and we shall seek $\inv (\sigma^* \tilde{K})$ 
  in a more direct way. See \exr{sl1:K}. }
 ${\tilde{K}}$ with entries
 in $\FR( \gva\times\jva[s])$ 
 such that $\rho_*\iD = \tilde{K} \, \lliea$.

We can write \Ref{chicoutimi} as
\( \iD(\inv f) (z)=\inv(\D f)(z)
   +\sigma^*\left(\tilde{K} \lliea(g^* f)\right)( z)\) 
so that, by \propr{inf2inf},
\begin{equation} \label{chicoutima}
  \iD(\inv f) (z) = \inv(\D f)(z) - 
 \sigma^* \left( \tilde{K} \vf ( g^* f)\right) (z) .
\end{equation}

This latter equation shows that 
\( \sigma^* \left( \tilde{K} \vf ( g^* f)\right)
  = \inv(\D f) - \iD(\inv f)  \)
 is a differential invariant.
As such it is equal to its invariantization and thus 
\[  \sigma^* \left( \tilde{K} \vf ( g^* f)\right) 
= \inv (\sigma^* \tilde{K}) \, \inv (\sigma^* \vf(g^*f) ) .\]

For all $z\in \sva$, $\rho(z)= e$ and therefore
 $\sigma^* \vf(g^*f)$ and $\vf(f)$ agree on $\sva$:
for all $z\in\sva$,
\(\sigma^* \vf(g^*f)\,(z) =\vf(g^*f) \,(e,z)=\vf(f) (z) \)  
by \propr{inf2inf}.
It follows that $ \inv (\sigma^* \vf(g^*f) )= \inv ( \vf(f) )$
so that \Ref{chicoutima} becomes
\begin{equation} \label{multiplesclerosis}
 \iD(\inv f) (z) = \inv(\D f)(z)-\inv (\sigma^* \tilde{K})\,\inv ( \vf(f) ).
\end{equation}

To find the matrix $K= \inv (\sigma^* \tilde{K})$ 
we use the fact that  $\inv p_i =0$ for all $1\leq i\leq r$. 
Applying $\iD$ and \Ref{multiplesclerosis} to this equality we obtain:
\(  \inv (\D p_i) = K \,  \inv( \vf ( p_i) ) \)
so that \( \inv (\D(P)) = K \,  \inv( \vf ( P) ) \).
The transversality of  $\sva$ imposes that
$\vf(P)$ is invertible along $\sva$, and thus so is $\inv(\vf(P))$. 

We thus have proved that 
\(\iD( \inv f ) = \inv( \D f) - K \, \inv(\vf(f) )\)
where 
\( K = \inv (\sigma^* \tilde{K}) = \inv (\D(P) V(P)^{-1}) .\)
\end{proof}

\sil{On the other hand, 
since $\rho^*\omega$ is a vector of invariant form while 
$\idx$ is a vector of contact invariant forms whose entries 
span the horizontal coframe there is a matrix $A$ with 
differential invariants as entries such that
\( \rho^* \omega = A \idx .\) We prove that $A=-K^T$.

Since $\idx$ and $\iD$ are dual we have 
$A_{ai}(z)  
      = \langle \rho^* \omega_a , \iD[i] \rangle 
      = \langle  \omega_a , \rho_*\iD[i] \rangle$
which can be written in matrix form as
$A^T = \omega^T  (\rho_*\iD) $ 
and thus, taking forms and vector field where they should be taken, 
$A^T=\omega|_{\rho(z)}^T \tilde{K}(\rho(z),z) \lliea|\rho(z)$.
In particular for $z\in \sva$ we have
$A(z)^T=\omega_e^T \tilde{K}(e,z) \lliea|_e$
By \propr{inf2inf} $\lliea|_e = -\liea|_e$ and $\liea$ is dual to $\omega$.
So $A(z)^T= - \tilde{K}(e,z)$ for all $z\in \sva$. Therefore 
$A = -\inv (\sigma^*\tilde{K})^T$ and we showed that this latter is $K$.}

If $f$ is a differential invariant, $\iD(f)$ 
is also a differential invariant, while $\D(f)$ need not be.  
But if we invariantize this latter though we find nothing else 
than $\iD(f)$ .
This follows immediately from  the above way of writing the
 \emph{recurrence formulae} yet 
we have not seen the following corollary in previous papers on the subject. 

\begin{coly} \label{iDoninv}
If $f$ is a differential invariant then $\iD(f) = \inv (\D(f))$ .
\end{coly}

\begin{proof}
If $f$ is a differential invariant then $\inv f = f $ and $\vf(f)=0$. 
The result thus follows from the above theorem.
\end{proof}

By deriving a recurrence formula for forms,  \cite[Section 13]{fels99}
derived explicitly the  commutators of the invariant derivations . 
It can actually be derived directly from 
\theor{recurrences} through the use of 
\emph{formal invariant derivations} \citep{hubert08a}.

\begin{propy} \label{commuterules}
 For all $1\leq i,j\leq \dimx$,
\(\ds \left[\iD_i, \iD_j\right] = \sum_{k=1}^\dimx\,  \cminv[ijk] \,\iD_k\)
where
$$\Lambda_{ijk}= \sum_{c=1}^\dimg K_{ic}\, \inv(\D_j(\xi_{ck}))
             - K_{jc} \, \inv(\D_i(\xi_{ck}))  \; \in \;\FG[s+1],$$
$K=\inv\left(D(P) \, V(P)^{-1}\right)$, and
$\xi_{ck}=\vf_c(x_k)$.
\end{propy}

\begin{exy}  \label{sl1:K} We carry on with 
\exr{sl1:prolong}, \ref{sl1:invariantization}, and \ref{sl1:invder}.

We chose  $P=(x,u_1-1)$ and showed that $\iD = \frac{1}{u_1}\D$ 
while $\inv u_i=\frac{u_i}{u_1^i}$.
We  computed 
\[\iD( \inv u_i) = \frac{u_{i+1}}{u_1^{i+1}} 
         - i\,\frac{u_i}{u_1^{i}} \frac{u_2}{u_1^2}
         = \inv u_{i+1} - i \,\inv u_2 \, \inv u_i .\]
We have 
$\D(P)= (\begin{array}{ll} 1 &  u_2\end{array})$
and
$\vf(P)= \left( \begin{array}{cc} x & -u_1 \\ 1 & 0 \end{array} \right)$.
The matrix $K$ of  \theor{recurrences} is thus
$K = \inv \left( D(P) \, V(P)^{-1}\right) 
= (\begin{array}{ll} -\inv u_2 & 1 \end{array})$
and the formula is verified:
\[ \iD ( \inv u_i)= \inv u_{i+1}
      - (\begin{array}{ll}-\inv u_2 & 1 \end{array}) \, 
       \left(  \begin{array}{c} \inv \vf_1(u_i) \\   
         \inv \vf_2(u_i)\end{array} \right)
\hb{ since }
\inv \vf(u_i)= \left(\begin{array}{cc} -i\,u_i & 0\end{array}\right)^T.\]

What we shall do next is illustrate the proof by exhibiting 
the matrix $\tilde{K}$ that arises there. It is defined by
 $\rho_* \iD=\tilde{K} \,\lliea$ 
and the fact that
$\sigma^* \tilde{K} \vf(g^* f)$ is an invariant 
for any $f\in \F$.

We have 
 \( \lliea[1]= -\lambda_1 \pdif{}{\lambda_1}, 
   \quad \lliea[2]=  -\lambda_1 \pdif{}{\lambda_2} \)
and saw that 
$\rho^*\lambda_1 = u_1$ and $\rho^*\lambda_2= -xu_1$.
Thus 
\[ \rho_* \iD  = 
   \left( \begin{array}{cc}
   \iD( \rho^* \lambda_1) &  \iD (\rho^* \lambda_2)
   \end{array} \right)
 \left( \begin{array}{c} \pdif{}{\lambda_1} 
     \\ \pdif{}{\lambda_2} \end{array} \right)
  =  \left( \begin{array}{cc}
           -\frac{u_2}{u_1}\frac{1}{\lambda_1} & 
            \frac{u_1+x\, u_2}{u_1} \, \frac{1}{\lambda_1} 
           \end{array}\right)  
    \left(\begin{array}{c} \lliea[1] \\ \lliea[2]\end{array}\right) .\]
So here 
$\sigma^* \tilde{K}=   
   \left( -\frac{u_2}{u_1^2}, \frac{u_1+xu_2}{u_1^2} \right)$.
We indeed have that $\inv \sigma^* \tilde{K}= K$ as used in the proof.
We verify here that $\sigma^*\left( \tilde{K} \vf(g^*f)\right)$ 
is a vector of differential invariants.
We have 
\[ \vf(g^* x) =
  \left(\begin{array}{c} \lambda_1 \, x \\ \lambda_1\end{array}\right) ,
\;
\vf(g^* u_i) =
\left(\begin{array}{c}  -i\,\frac{u_i}{\lambda_1^i} \\ 0\end{array}\right) 
\]
so that
\( \sigma^*\tilde{K}\vf(g^*x)=1 \) and
\(\sigma^*\tilde{K}\vf(g^*u_i)= i\,\frac{u_2}{u_1^2} \, \frac{u_1}{u_1^i} 
 = i\,\inv u_2 \, \inv u_i .
\)
\end{exy}

\begin{exy}  \label{sl2:K} 
We carry on with  \exr{sl2:invariantization}.

 We chose 
$$P=\left(x_1,x_2,\frac{1}{2}-\frac{1}{2}(u_{10}^2+u_{01}^2)\right).$$

On one hand the prolongations of the  infinitesimal generators to $\jva$  are
\[ 
 \vf_1= \pd{x_1}, \quad \vf_2=\pd{x_2}, \quad 
 \vf_3=x_1\,\pd{x_1}+x_2\,\pd{x_2}-\sum_{i,j\geq 0}(i+j)\,u_{ij}\,\pd{u_{ij}}
\]
so that
$$\vf(P) = \left(\begin{array}{ccc} 
   1&0&0\\ 0&1&0 \\x_1 & x_2 &u_{10}^2+u_{01}^2
\end{array}\right) 
\hb{while}
\D(P)= 
   \left(\begin{array}{ccc} 
    1&0& v \\ 0&1& w
   \end{array}\right)
$$
where  
$$v=-(u_{10}u_{20}+u_{01}u_{11}) \hb{and} w=-( u_{10}u_{11}+u_{01}u_{02}).$$
Since
\( \inv x_1=0,\,\inv x_2=0\)  and  $\inv( u_{10}^2+ u_{01}^2)=1$,
$\inv \vf(P)$ is the identity matrix   so that
\[ K= \inv (\D(P)\vf(P)^{-1})=
  \left(\begin{array}{ccc} 1&0&\inv v \\ 0&1&\inv w
   \end{array}\right). \]

On the other hand the 
normalized invariants and  invariant derivations are 
\[ \inv u_{ij}=\frac{u_{ij}}{{(u^2_{10}+u_{01}^2)}^{\frac{i+j}{2}}},\; \forall i,j;
\qquad 
 \ider[i]=\frac{1}{\sqrt{u^2_{10}+u_{01}^2}} \der[i],\; i=1,2.
\]
We can thus check that
\[ \left( \begin{array}{c}
    \ider[1](\inv u_{ij})\\ 
    \ider[2](\inv u_{ij})
\end{array}\right)
=   \left( \begin{array}{c}  
   \inv(  u_{i+1,j})\\ 
    \inv (u_{i,j+1} ) \end{array}\right)
 - K \,  \left( \begin{array}{c}
    0\\ 0\\ -(i+j)\,\inv  u_{ij} 
\end{array}\right) ,
\]
as predicted by \theor{recurrences},
and that 
$ [\ider[2],\ider[1] ] =  \inv w\,\ider[1]-\inv v \, \ider[2], $
as predicted by \propr{commuterules}.
\end{exy}

\section{Finite generation and rewriting} 
    \label{geni} 

The recurrence formulae, \theor{recurrences},
together with the replacement theorem,
\theor{thomas},
show that any differential invariant can be written in terms of the
normalized invariants of order $s+1$, where $s$ is the order of the
moving frame, and their invariant derivatives. The rewriting is effective.

In the case of a cross-section of minimal order,
we  exhibit another  generating  set of
differential invariants  with bounded cardinality. 
This bound is  $\dimx\dimg$ in the case of an action transitive on $\jva[0]$. 
When  in addition we choose a coordinate cross-section, this
set consists of normalized invariants and we retrieve the result of
\citet{olver07}. This was incorrectly stated for any cross-section
by \citet[Theorem 13.3]{fels99}.

\subsection{Rewriting in terms of normalized invariants of order $s+1$} 
 \label{cni}

Let $s$ be  equal  to or greater then the  stabilization order and
let $\sva$ be a cross-section to the orbits in  $\jva[s]$ 
defined by $P=(p_1, \ldots, p_r)$ with $p_i\in \F[s]$.
Recall from \secr{diffinv} that
\[ \ninv[s+k] = \{ \inv x_1, \ldots, \inv x_\dimx\} 
  \cup \{ \inv u_\alpha \, |\, u\in \uva, \; |\alpha|\leq s+k\},
\]
where $\inv :\F[s+k] \rightarrow \FG[s+k]$ 
is the invariantization associated to $\sva$, 
forms a generating set of local invariants for 
the action of $\ac$ on $\jva[s+k]$.
Those invariants have additional very desirable properties: 
we can trivially rewrite any differential invariants of order $s+k$ 
in terms of them.
Yet it is even more desirable to describe the differential 
invariants of all order in finite terms.

\sil{\begin{defy} Let $\iD=(\iD_1,\ldots, \iD_m)$ 
be the set of invariant derivations of \theor{magnolia}.
A set of differential invariants $\mathcal{I}$ 
is generating  if all differential invariants 
can be written as a function of the elements of $\mathcal{I}$ 
and their monotone derivatives.
\end{defy}}

\theor{recurrences} implies in particular that
\[ 
\inv( \der[i] u_{\alpha}) = \ider[i]( \inv u_\alpha ) +
\sum_{a=1}^r K_{ia} \, \inv(\vf_a( u_\alpha) )  
\]
where $K=\inv(\D(P)\vf(P)^{-1})$ 
has entries that are function of $\ninv[s+1]$.
It is then an easy inductive argument to show that any
$\inv u_\alpha$ can be written as a function of $\ninv[s+1]$ 
and their derivatives of order $\max(0,|\alpha|-s-1)$.
Combining with the replacement property,
\theor{thomas}, we have a  constructive way of 
rewriting any differential invariants 
in terms of the elements of $\ninv[s+1]$ and their derivatives:
A differential invariant  of order $k$ 
is first trivially rewritten in terms of $\ninv[k]$ by \theor{thomas}.
If $k\leq s+1$ we are done. 
Otherwise, any element $\inv u_{\alpha}$ of $\ninv[k]$ 
with $|\alpha|=k$ is a 
$\inv (\D_i u_\beta)$, for some $1\leq i \leq \dimx$ and  $|\beta|=k-1$. 
We can thus write it as:
\[ \inv u_\alpha = \inv (\D_i u_\beta) = 
   \ider[i] (\inv u_\beta) + 
   \sum_{a} K_{ia} \, \inv\left(\vf_a( u_\beta) \right). \]
This involves only elements of $\ninv[k-1]$ 
and their derivatives. Carrying on recursively we can rewrite 
everything in terms of the elements of $\ninv[s+1]$ and their derivatives.

This leads to the following result that will be refined in \secr{syzygies}.
Indeed the rewriting is  not unique: 
at each step there might be several 
choices of pairs $(i,\beta)$ such that 
$u_\alpha = \D_i u_\beta$. 

\begin{theoy} \label{dolce}
Any differential invariant of order $s+k$ can be written in terms of the elements of $\ninv[s+1]$ and their derivatives of order $k-1$ and less.
\end{theoy}

\subsection{Case of minimal order cross-section}

A natural question is to determine a smaller set of 
differential invariants that is generating. 
 \citet{olver07} proved
that when choosing a coordinate cross-section of \emph{minimal order}
the normalized invariants corresponding to 
 the derivatives of the coordinates functions which are set to constant 
 form a generating set of differential invariants. 
Here we generalize the result to noncoordinate cross-sections.
The proof is based on  the same idea.

Let $s$ be equal to or greater  than  the stabilization order. 
A local cross-section $\sva$ in $\jva[s]$ is of 
\emph{minimal order} if its projection on $\jva[k]$, for all $k\leq s$, 
is a local cross-section to the orbits of the action of $\ac$ on $\jva[k]$
\citep{olver07}.
Assume $P=(p_1,\ldots, p_\dimg)$ defines 
a cross-section $\sva$ of minimal order. 
Without loss of generality we can assume that 
$P_k=(p_1, \ldots, p_{\dimg_k})$
where $\dimg_k$ 
is the dimension of the orbits of the action of $\ac$ on $\jva[k]$,
defines the projection of $\sva$ on $\jva[k]$. 

\begin{theoy} \label{minorder}
If $P= (p_1,\ldots, p_\dimg)$ defines 
a cross-section for the action of $\ac$ on $\jva$ such that
$P_k=(p_1, \ldots, p_{\dimg_k})$ defines a cross-section 
for the action of $\ac$ on $\jva[k]$, for all $k$, then 
$\moinv=
\{\inv(\der[i](p_j))\,|\, 1\leq i\leq \dimx, \; 1\leq j\leq \dimg \}$ 
together  with $\ninv[0]$ 
form a generating set of differential invariants.
\end{theoy}

\begin{proof}
The minimal order condition imposes that
the $\dimg\times \dimg_k$ matrix $\vf(P_k)$ 
has maximal rank $r_k$ on $\sva$, and therefore on 
an open neighborhood of each point of $\sva$.
As $\vf[k]$ has rank $r_k$, for any $f$ 
in $\F[k]$, $\vf(f)$ is linearly dependent on 
$\vf(p_1), \ldots, \vf(p_{r_k})$.
In a neighborhood of each point of $\sva[k]$ there is thus a relation
\[ \vf(f) = \sum_{i=1}^{\dimg_k} a_i\, \vf(p_i) , 
   \hbox{ where } a_i \in \F[k] .\]
On one hand, by  \theor{recurrences}, we have
\( \inv( \D f)= \iD( \inv f )  + K \, \inv(V(f) ) \) 
so that \( \inv( \D f)= \iD( \inv f )  
  +  \sum_{i=1}^{r_k} \,\inv(a_i)\,K\,\inv(\vf(p_i)) \).
On the other hand  $\inv(p_i)=0$ so that
 \( \inv(\D p_i) = K \inv(\vf(p_i)) .\) 
It follows that
\[  \inv( \D f) = \iD( \inv f )  +
    \sum_{i=1}^{r_k} \,\inv(a_i)\, \inv(\D p_i) . \]

Note that $\inv(a_i)$ can be written in terms of the 
$\inv(u_\beta)$ with $|\beta |\leq k$.
So the formula implies that any $\inv u_\alpha$, 
with $|\alpha|=k+1$, can be written in terms of
$\{\,\inv(\D p_i)\;| \; 1\leq i\leq r_k \,\}$ and  
$\{\,\inv(u_\beta)\; | \;|\beta |\leq k\,\}$ 
together with 
their derivatives with respect to the invariant derivations $\iD$.
By induction, it follows that any  $\inv u_\alpha$
can be written in terms of the zero-th order normalized
 invariants together with the elements of $\moinv$ and their derivatives. 
\end{proof}

In the case of a coordinate cross-section
$\moinv$ is a subset of the normalized invariants $\ninv[s+1]$ that 
\citet{olver07} named the \emph{edge invariants} 
for the  representation of 
the derivatives of a  dependent function on a lattice. 
 We shall extend this name in the case of non coordinate 
cross-section though the pictorial representation is no longer valid. 

Minimality is necessary for the edge invariants to be generating 
in general.
\cite{olver07a} exhibits a choice of non minimal (coordinate) 
cross-section for which the edge invariants are not generating. 
We review this example in \secr{curves}.

A consequence of \theor{minorder} is that we can bound the number 
of differential invariants necessary to form a generating set. 
The bound is  $\dimx\, \dimg+ d_0$, where
$d_0=m+n-r_0$ is the codimension of the orbits of the action 
of $\ac$ on $\jva[0]$.
Transitive actions on $\jva[0]$ are of particular interest.
There $d_0=0$  and the  bound is simply $\dimx \,\dimg$.
\cite{hubert07m}  exhibits a generating set of such cardinality 
even in the case of non minimal cross-section.

\begin{exy} \label{sl2:minorder}
Consider  \exr{sl2:K} again. The chosen cross-section is of minimal order.
Specializing \theor{recurrences} we obtained
$$
 \left( \begin{array}{c}  
   \inv(  u_{i+1,j})\\ 
    \inv (u_{i,j+1} ) \end{array}\right)
=
\left( \begin{array}{c}
    \ider[1](\inv u_{ij})\\ 
    \ider[2](\inv u_{ij})
\end{array}\right)
- (i+j) \, \inv  u_{ij} 
\left(\begin{array}{c}
     \inv v \\ \inv w
   \end{array}\right)
$$
from which it is clear that all the  normalized invariants can be inductively written in terms of $\inv u_{00},\inv v$ and $\inv w$, i.e the non constant elements of $\ninv[0]\cup\moinv$, and their derivatives.
\end{exy}

\section{Syzygies}
   \label{syzygies}   

Loosely speaking, a \emph{differential syzygy} is a relationship
among  a (generating) set of differential invariants 
and their derivatives. 
A set of differential syzygies is complete if 
any other syzygies is inferred by those and their derivatives.
In this section we formalize a 
definition of syzygies by introducing the appropriate 
differential algebra. We then show the completeness of a finite
set of differential syzygies on the normalized invariants of order $s+1$.

\citet[Theorem 13.2]{fels99} claimed a complete set of  syzygies
for  edge invariants, in the case of coordinate cross-section.
It  has so far remained unproven\footnote{An necessary amendment of the
 statement is that $K$ might be taken as the empty set in
(iii).}. As we are finishing this paper \cite{olver07b} 
announce a syzygy theorem for pseudo-groups.
The \emph{symbol module} of the infinitesimal determining system 
takes there a prominent place:
on one hand it dictates the coordinate cross-section to be used and, 
on the other hand,  its (algebraic) syzygies prescribe the syzygies on 
the differential invariants. 
Let us note here two immediate advantages of our result for 
 Lie group actions: we do not need to have any side algebraic computations 
(over a ring of functions) nor are we
restricted in our choice of cross-section.
In particular we are neither restricted to 
 minimal order nor coordinate cross-section. 
Even if those latter are often the best choice, 
there are needs for more options. 
Such is the case in the symmetry reduction considered by 
\cite{mansfield01}.
Also in \exr{sl2:invariantization} the nonlinear cross-section is defined 
for the whole open set where the action is regular, 
while a linear cross-section is only defined for a subset.

The commutation rules, \theor{commuterules}, imply infinitely many 
relationships on derivatives of normalized invariants.
\cite{fels99}, as well as \cite{olver07b},  considered those as syzygies.
Our approach is in the line of \citet{hubert05}.
We encapsulate those relationships  
in a  recursive definition of the derivations to 
work exclusively with \emph{monotone derivatives}.
The differential algebra of monotone derivatives 
that arises there is a  generalization  of the 
differential polynomial rings 
considered by \citet{ritt} and \citet{kolchin}
to model nonlinear differential equations.
Of great importance is the fact that it is endowed with 
a proper differential elimination theory \citep{hubert05}.
This generalization is effective and has been implemented 
\citep{ncdiffalg}.

Refining the discussion of \secr{geni},
we first observe that any differential 
invariant can be written in terms of the monotone derivatives 
of the normalized invariants of order $s+1$. 
The rewriting is nonetheless not unique in general.
The syzygies can be understood as 
the relationships among the monotone 
 derivatives that govern this indeterminacy.

For the normalized invariants of order $s+1$
we introduce the concept of \emph{normal derivatives}.
They provide a 
canonical rewriting of any differential invariant.
The set of differential relationships that allows one to rewrite any monotone 
derivative in terms of normal derivatives is
then shown to be a complete set of  syzygies for the 
normalized invariants of order $s+1$ (\theor{cnisyz}).

To prove these results we formalize the notion of syzygies 
by introducing the algebra of monotone derivatives. 
We endow this algebra with derivations so as to have 
a differential morphism onto the algebra of differential 
invariants. 
The syzygies are the elements of the kernel of this morphism.
It is a differential ideal and \theor{cnisyz} actually
 exhibits a set of generators. 

\subsection{Monotone and normal derivatives}

In \secr{geni} we showed that any differential invariant
can be written in terms of \ninv[s+1] and its derivatives. 
However, this rewriting is not unique. 
We can actually restrict the derivatives to be used in this rewriting,  
first to \emph{monotone derivatives}, then to \emph{normal derivatives}.
Normal derivatives provide a canonical rewriting.

\begin{defy} \label{monotonedef}
An invariant derivation operator $\ider[j_1]\ldots\ider[j_k]$ 
is monotone if $j_1\leq \ldots\leq j_k$. 
Such a monotone derivation operator is noted $\iD[\alpha]$ 
where $\alpha=(\alpha_1, \ldots, \alpha_\dimx)\in \Ni^\dimx$ 
and $\alpha_i$ is the cardinality of $\{j_l \;|\; j_l=i\}$.
\end{defy}

There is an inductive process to rewrite any normalized invariants, 
and therefore any differential invariants,  in terms of the monotone
 derivatives of $\ninv[s+1]$.
For the inductive rewriting of  $\inv u_{\beta}$, for $|\beta|>s+1$,
in terms of the monotone derivatives of $\ninv[s+1]$ we can proceed as
follows: split $\beta$ in
 $\beta=\cre+\bas$ where $|\bas|=s+1$
and then rewrite $\inv u_{\beta}-\iD[\cre](\inv u_{\bas})$ which is of
lower order.
There might be several ways to split $\beta$, each
leading to a different rewriting.
The following definition imposes a single choice of 
splitting\footnote{
The idea is reminiscent of  \emph{involutive division}. Originally
introduced by \citet{riquier10} and \cite{janet29} for the completion of 
partial differential systems, generalizations and algorithmic refinements 
have been worked out by several authors in the past decade 
for polynomial systems as well within the framework of computer algebra.
}.

\begin{noty} \label{sloop}
For $\beta=(\beta_1,\ldots,\beta_\dimx)\in \Ni^\dimx$, we denote
\[ \bas[\beta] = \left\{\begin{array}{l}
  \ds \beta \hb{if} |\beta|\leq s+1 \\ 
 (0,\ldots, 0, \beta_i', \beta_{i+1}, \ldots, \beta_\dimx) \hb{otherwise}\\
 \quad \hb{with}
  i = \max\;\{ j\;|\; \beta_j+\ldots+\beta_\dimx\geq s+1\}\\
\quad 
\hb{and} \beta_i'=  (s+1)-\beta_{i+1}-\ldots-\beta_\dimx 
\end{array}\right.
\]
and $\cre[\beta]=\beta-\bas[\beta]$.
\end{noty}

With those notations, 
$\cre[\beta]=0$ when $|\beta|\leq s+1$
and
 $|\bas[\beta]|$ is always less or equal to $s+1$. 



\begin{defy} The normal derivatives of $\ninv[s+1]$ are the elements
of the set 
$$\ndinv = \ninv[s+1]
  \cup
 \left\{\iD[{\cre[\beta]}](\inv u_{\bas[\beta]})
      \;|\;\beta \in \Ni^\dimx,\;|\beta|> s+1 \right\}.$$
The set $\ndinv[k]$ of the 
normal derivatives of order $k$ is the subset thereof with
 $|\cre[\beta]|\leq k$. 
\end{defy}

We introduce a further notation to deal with tuples that is used
in the coming inductive proofs and in the description 
of a complete set of syzygies  in \theor{cnisyz}.

\begin{noty}
For $\beta \in \Ni^\dimx$, $|\beta|>0$, we define
$\fs[\beta]$ and $\ls[\beta]$ respectively as the first and last
 non zero component  
of $\beta$, i.e.
\[ \fs[\beta]  =\min \,\{ j \,|\, \beta_j\neq 0\}
\quad \hb{and}\quad 
\ls[\beta]  =\max\,\{j\,|\, \beta_j\neq 0\}.
\]
\end{noty}

Note that the splitting of \notr{sloop} is such that 
$\ls[\cre[\beta]]\leq\fs[\bas[\beta]]$ for all $\beta\neq 0$.

\begin{propy} \label{rosemary}
Any differential invariant is a function 
of the normal derivatives $\ndinv$ of $\ninv[s+1]$.
\end{propy}

This result follows from an easy inductive argument on the following
lemma. 

\begin{lemy} \label{kiwi}
For all $\beta\in \Ni^m$, $\beta\neq 0$,
$\inv u_{\beta} -\iD[{\cre[\beta]}](\inv u_{\bas[\beta]})\in\FG[|\beta|-1]$.
\end{lemy}

\begin{proof}
This is trivially true for $|\beta|\leq s+1$ 
since then $\cre[\beta]=(0,\ldots, 0)$.
We proceed by induction for $|\beta|>s+1$. 

Assume the statement is true for all 
$\beta$ with $s+1\leq |\beta|\leq k$. 
Take $\beta$ with $|\beta|=k+1$.
Let $i=\fs[\beta] $ and $\beta'= \beta-\epsilon_i$.
We have $\bas[\beta]'=\bas[\beta]$, 
$\cre[\beta]'=\cre[\beta]-\epsilon_i$
and $\iD[{\cre[\beta]}]= \iD_i \iD[{\cre[\beta]'}]$
so that 
$    \inv u_\beta-\iD[{\cre[\beta]}](\inv u_{\bas[\beta]})
   = \inv(\D_i(u_{\beta'}))-\iD_i \iD[{\cre[\beta]'}](\inv
   u_{\bas[\beta]'})$.
Thus, by \theor{recurrences},
$$\inv u_\beta-\iD[{\cre[\beta]}](\inv u_{\bas[\beta]}) =
 \iD_i \left(\inv u_{\beta'}- \iD[{\cre[\beta]'}](u_{\bas[\beta]'})\right)
      +\sum_{a=1}^r K_{ia} \, \inv\left(\vf_a(u_{\beta'}) \right).$$

The entries of $K$ are functions of $\ninv[s+1]$, 
while the entries of $\inv \left(\vf(u_{\beta'})\right)$ are functions of $\ninv[k]$.
By induction  hypothesis
$\iD[{\cre[\beta]'}](\inv u_{\bas[\beta]'})-\inv u_{\beta'}\in\FG[k-1]$
and thus
$\iD_i \left( \iD[{\cre[\beta]'}](u_{\bas[\beta]'})-\inv u_{\beta'}\right)
\in \FG[k]$. 
\end{proof}

Following the induction on \lemr{kiwi}, 
rewriting any $\inv u_\beta$ in terms of the
normal derivatives of $\ninv[s+1]$ is an effective process.
Now, the normalized invariants $\inv u_\beta$ 
are in one-to-one correspondence with the normal derivatives 
$\iD[\hat{\beta}] (\inv u_{\bar{\beta}})$ of $\ninv[s+1]$.
Extending \propr{colibri}, which bears on normalized invariants,
we show that the rewriting of any differential invariants in terms
 of normal derivatives $\ndinv$ of $\ninv[s+1]$ 
is unique, modulo $P$.

\begin{propy}\label{aioli}
Assume $P=(p_1,\ldots, p_\dimg)$  
are the $\dimg$ independent functions of $\F[s]$ that cut out
 the cross-section $\sva[s]$ to the orbits on $\jva[s]$.
Let $F\in \F[s+k]$ be a function such that 
$F(\inv x, \iD[{\cre}](\inv u_{\bas}))=0$.
Then, in the neighborhood of each point of $\sva$,
there exist $a_1,\ldots,a_\dimg \in\F[s+k]$ 
such that  $F = \sum_{i=1}^\dimg a_i \, p_i$. 
\end{propy}

\begin{proof} 
By \lemr{kiwi}, for $|\beta|\leq s+k$, 
there exists $\zeta_\beta$ in $\F[|\beta|-1]$ such that 
$\iD[{\cre}](\inv u_{\bas}) -\inv u_{\beta}= \inv \zeta_\beta$.
We choose such a family of $\zeta_\beta$ with 
$\zeta_\beta =0$ for $|\beta|\leq s+1$. 
The map $\theta :\F[s+k]\rightarrow \F[s+k]$  then defined by 
$\theta( u_\beta) = u_\beta+\zeta_{\beta}$ is an automorphism of $\F[s+k]$.
It satisfies 
$F(\inv x, \iD[{\cre}]\inv u_{\bas}) = \theta(F) ( \inv x, \inv u_\beta)$
and its restriction to $\F[s+1]$ 
is the identity. In particular  $\theta(p_i)=p_i$. 

If  $F\left(\inv x, \iD[{\cre}](\inv u_{\bas})\right)=0 $ 
then,  by \propr{colibri}, there exist $b_1,\ldots,b_\dimg \in\F[s+k]$ 
such that  $ \theta(F) = \sum_{i=1}^\dimg b_i \, p_i$ 
 in the neighborhood of each point of $\sva$.
Let $a_i \in\F[s+k]$ be such that $b_i = \theta( a_i)$. 
We have $F=\sum_{i=1}^\dimg a_i \, p_i$.
\end{proof}

\subsection{The differential algebra of monotone derivatives}

When we apply the invariant derivation $\iD_i$ to a monotone derivative
 $\iD[\beta] (\inv u_{\alpha})$ we do not obtain a monotone derivative 
unless $i \leq \fs[\beta]$. 
Yet the obtained result can be written in terms of monotone derivatives.
This comes as a result of the general \propr{rosemary}, 
but we could also deduce it from the commutation rules
on the derivations, \propr{commuterules}. This is detailed by 
\cite{hubert05} and leads to an appropriate definition of
differential algebra in the presence of non trivial 
commutation rules for the derivations.\footnote{
The difficulty, and major difference, 
compared  with the case considered for instance
by \citet{kolchin} or \citet{yaffe01}  is that the coefficients 
of the commutation rules are themselves in the polynomial ring to be
defined as opposed as to be in the base field.}

We shall accordingly define a differential algebra 
where the differential indeterminates are in 
one-to-one correspondence with the elements of 
$\ninv[s+1] = \{ \inv x_1, \ldots, \inv x_\dimx\} 
  \cup \{ \inv u_\alpha \, |\, u\in \uva, \; |\alpha|\leq s+1\}$. 
They are noted
$\{ \fix[1]{},\ldots,\fix[\dimx]{}\} \cup \{\fiu{} \,|\, |\alpha|\leq s+1\}$.
The monotone derivatives $\iD[\beta](\inv x_i)$ and $\iD[\beta](\inv u_{\alpha})$
are then represented by the double-scripted indeterminates 
$\fix[i]{\beta}$ and $\fiu[\alpha]{\beta}$.
The correspondence 
is encoded with a natural morphism from this differential algebra to 
$\FG$ given by 
 $\fix[i]{\beta}\mapsto\iD[\beta] (\inv x_i)$ 
and $\fiu[\beta]{\alpha}\mapsto\iD[\beta] (\inv u_{\alpha})$.
We shall then define $\fD_1,\ldots,\fD_\dimx$ acting on the
 $\fix[i]{\beta}$ and $\fiu[\beta]{\alpha}$
so that this becomes a differential morphism, i.e.
 $\fD_j\fix[i]{\beta}\mapsto\iD_j\iD[\beta]( \inv x_i)$ and
$\fD_j \fiu[\beta]{\alpha} \mapsto \iD_j \iD[\beta] (\inv u_{\alpha})$.
The key idea comes from \citet{hubert05}: 
the formal invariant derivations  $\fD_1,\ldots,\fD_\dimx$ 
are given a recursive definition.

We develop here the formalism to incorporate the functional aspect,
 as opposed to the polynomial case developed by \cite{hubert05}.
We thus define first a sequence $(\A[k])_k$ of manifolds\footnote{We 
shall simply think of them as open subsets of $\Ri^l$ for the right $l$.} 
that
correspond to the spaces of the monotone derivatives of $\ninv[s+1]$
of order $k$. 
$\A[0]$ is isomorphic to $\jva[s+1]$ and therefore of dimension
 $N = \dimx +\dimu \binom{m+s+1}{s+1}$.
The coordinate function on $\A[0]$ are noted 
$\{ \fix[1]{0}, \ldots, \fix[\dimx]{0} \} 
       \cup \{\fiu{0} \,|\, |\alpha|\leq s+1\}$.
Then, for each $k$, $\A[k]$ is a submanifold of $\A[k+1]$ and
$\A[k]$ is  of dimension  $\ngen \binom{k+\dimx}{\dimx}$.
A coordinate system is given by
$\{\fix{\beta}\,|\, |\beta|\leq k \} 
       \cup \{\fiu{\beta} \,|\, |\beta|\leq k, \, |\alpha|\leq s+1\}$.
We actually focus on the algebras of smooth functions  $\FA[k]$ and $\FA$, 
where $\A=\bigcup_{k\geq 0} \A[k]$.

We can go back and forth from $\FA$ to $\F$ and this is expressed with
the maps $\phi$ and $\psi$ introduced in the next proposition.
This latter is nothing else than the statement  that  any
differential invariants can be written in terms of 
the monotone derivatives of $\ninv[s+1]$ (\propr{rosemary}).

\begin{propy} \label{casimir} \label{colargole}
On one hand the ring morphism $\phi : \FA[k] \rightarrow \FG[s+k+1]$ defined by 
\[ 
\phi(\fix{\alpha})=\iD[\alpha]( \inv x) \quad \hb{and} \quad
\phi(\fiu[\beta]{\alpha})=\iD[\alpha](\inv u_\beta),
\hb{for all} \alpha\in \Ni^\dimx \hb{and} |\beta|\leq s+1, \]
is surjective.

On the other hand there exists a ring morphism 
$\psi:\F[s+1+k]\rightarrow \FA[k]$ such that 
$\phi \circ \psi ( u_\alpha) = \inv u_\alpha$. 
We can furthermore choose $\psi$ so that
 $\psi(x_i)=\fix[i]{0}$ and $\psi(u_\alpha)=\fiu[\alpha]{0}$,
 for $|\alpha|\leq s+1$.
\end{propy}

In other words, $\psi(u_\alpha)$ is a function that allows one 
to rewrite $\inv u_\alpha$ in terms of the monotone derivatives 
of $\ninv[s+1]$. 

We proceed now to define on $\FA$ the derivations $\seq{\fD}{\dimx}$
that will turn $\phi$ into a differential morphism.

\begin{defy} \label{hepatique} Consider the maps $\phi$ and $\psi$ as in
  \propr{colargole}. We define the \emph{formal invariant derivations}
$\fder[1],\ldots,\fder[\dimx]$ from $\FA[k]$ to $\FA[k+1]$ 
by the following inductive process:
\[ 
\fder[i] ( \fiy{\beta}) = \left\{ \begin{array}{ll}
\fiy{\beta+\epsilon_i}, 
& 
\hb{if}  i \leq \fs[\beta] 
\\
\ds \fder[f]\fder[i] (\fiy{\beta-\epsilon_f})
+\sum_{l=1}^m c_{ifl} \, \fder[l]( \fiy{\beta-\epsilon_f }),
& \hb{where} f=\fs[\beta], \hb{otherwise,}
\end{array}\right.
\] 
where 
\begin{itemize}
\item  $\fiy{}$ ranges over the \emph{differential indeterminates} 
$\{ \fix[1]{},\ldots,\fix[\dimx]{}\} 
\cup 
\{\fiu{} \,|\, |\alpha|\leq s+1\}$
\item  \( c_{ijk} = \psi ( \cminv[ijl] )  \in \FA[1]\), 
for all ${1\leq i,j,l\leq\dimx}$,
where  $\{\cminv[ijl]\}_{1\leq i,j,l\leq\dimx}$ are the commutator invariants
 defined 
in \propr{commuterules}.
\end{itemize}

Endowed with the derivations $(\fder[1],\ldots,\fder[\dimx])$, 
$\FA$ is the differential algebra of monotone derivatives of $\ninv[s+1]$.
\end{defy}

Taking the notation 
$\fD[\alpha]= \fD[\alpha_1]_1\ldots\fD[\alpha_\dimx]_\dimx$
of \defr{monotonedef} we have
$\fD[\alpha](\fiy{0}) = \fiy{\alpha}$ but in general
$\fD[\alpha](\fiy{\beta}) \neq \fiy{\alpha+\beta}$, 
unless $\ls[\alpha]\leq \fs[\beta]$. 
We nonetheless have the following property\footnote{which is expected
for a differential elimination theory.}  that  allows 
to show that $\phi$ is a differential morphism, 
 thus justifying the definition of the formal invariant derivations. 
The proofs of the two next  results are reasonably straightforward inductions
exploiting the definition of the derivations. 

\begin{lemy} \label{cortina}
$\fD[\alpha](\fiy{\beta}) -\fiy{\alpha+\beta} \in \FA[|\alpha+\beta|-1]$,
for any 
$\fiy{} \in \{  \fix[1]{},\ldots,\fix[\dimx]{}\} 
\cup 
\{\fiu{} \,|\, |\alpha|\leq s+1\}$.
\end{lemy}

\begin{proof} 
By definition of the derivations $\fD$, this is true whenever 
$\alpha$ or $\beta$ is zero and when $\ls[\alpha]\leq\fs[\beta]$.
It is in particular true when 
$\ls[\alpha]=1$ or $\fs[\beta]=\dimx$. 
The result is then proved by induction along the well-founded pre-order:
\sil{
\[ (\alpha',\beta')\prec (\alpha,\beta) \equi \left\{ \begin{array}{l}
   \fs[{\beta'}]>\fs[\beta] \\
   \hb{or} \fs[\beta']=\fs[\beta]=f 
    \hb{and} \beta_f'<\beta_f \\
 \hb{or} \fs[\beta']=\fs[\beta]=f 
    \hb{and} \beta_f'=\beta_f \hb{and} \ls[\alpha']<\ls[\alpha] \\
 \hb{or} \fs[\beta']=\fs[\beta]=f 
    \hb{and} \beta_f'=\beta_f \hb{and} \ls[\alpha']=\ls[\alpha]=l
   \hb{and} \alpha_l'<\alpha_l 
\end{array} \right.\]
}
\[ 
(\alpha',\beta')\prec (\alpha,\beta) \equi \left\{ \begin{array}{l}
\beta'\prec_f\beta \hb{or} \\
\fs[\beta']=\fs[\beta]=f 
    \hb{and} \beta_f'=\beta_f \hb{and} \alpha'\prec_l \alpha
\end{array} \right.
\]
where
\[ \beta'\prec_f\beta \equi \left\{ \begin{array}{l}
   \fs[\beta']>\fs[\beta] 
   \hb{or}  \\ \fs[\beta']=\fs[\beta]=f 
    \hb{and} \beta_f'<\beta_f 
\end{array} \right.
\]
and
\[ \alpha'\prec_l \alpha  \equi  \left\{ \begin{array}{l}
 \ls[\alpha']<\ls[\alpha] 
 \hb{or} \\
 \ls[\alpha']=\ls[\alpha]=l
   \hb{and} \alpha_l'<\alpha_l   .
\end{array} \right.
\]
Assume the result is true for all
$(\alpha',\beta')\prec (\alpha,\beta)$.
We only need to scrutinize the case $l=\ls[\alpha]>\fs[\beta]=f$.
By definition of $\fD$ then:
\[ 
\fD[\alpha](\fiy{\beta}) =  
\fD[\alpha-\epsilon_l]\left(\fD_f\fD_l(\fiy{\beta-\epsilon_f})\right)
+\sum_k c_{lfk}\fD_k(\fiy{\beta-\epsilon_f}).
 \]
We have $\beta-\epsilon_f \prec_f \beta$ and thus, by induction hypothesis,
$\fD_k(\fiy{\beta-\epsilon_f})=\fiy{\beta-\epsilon_f+\epsilon_k} + F$
where $F\in \FA[|\beta|]$, for all $k$, and in particular for $k=l$.
We apply then the induction hypothesis
on $\fD_f(\fiy{\beta-\epsilon_f+\epsilon_l})$ and 
on $\fD[\alpha-\epsilon_l] (\fiy{\beta+\epsilon_l})$, observing that
$\beta-\epsilon_f+\epsilon_l \prec_f \beta$
while $\alpha-\epsilon_l \prec_l \alpha$.
\end{proof} 

\begin{propy} \label{carmen} The map 
$\phi : \FA \rightarrow \FG$ defined in \propr{casimir}
is a  morphism of differential algebras
i.e.
$\phi \circ \fD_i = \iD_i \circ \phi$, for all $1\leq i \leq \dimx$.
\end{propy}

\begin{proof} 
We need to prove that
$$
H(i,\alpha): \quad \phi ( \fD_i (\fiy{\alpha}) )= \iD_i(\phi(\fiy{\alpha}) )
$$ 
for all $\alpha \in \Ni^m$.
If this is true for all $|\alpha|\leq k$ then 
$\phi ( \fD_i (F) )= \iD_i(\phi(F) )$  for all $F\in \FA[k]$.
The proof is an induction along the well founded pre-order:
\[
(j,\beta)\prec(i,\alpha) 
\equi 
\left\{\begin{array}{l}
|\beta|<|\alpha| \hb{or}\\
|\beta|=|\alpha| \hb{and} j<i .
\end{array} \right.
\]

$H(i,\alpha)$ is trivially true when $\alpha$ is zero 
or when $i\leq \fs[\alpha]$. It is therefore true whenever $i=1$.

Assume $H(j,\beta)$ holds for any $(j,\beta)\prec (i,\alpha)$.
Only the case $i>\fs[\alpha]=f$ needs scrutiny.
We have
$\fD_i(\fiy{\alpha})=\fD_f\left(\fD_i(\fiy{\alpha-\epsilon_f})\right)
 +\sum_k c_{ifk}\fD_k(\fiy{\alpha-\epsilon_f}) $.
Since $\fD_i(\fiy{\alpha-\epsilon_f}) \in \FA[|\alpha|]$ while $f<i$, 
the induction hypothesis implies that 
$
\phi\left(\fD_f\left(\fD_i(\fiy{\alpha-\epsilon_f})\right)\right)
=
\iD_f \left( \phi \left(\fD_i(\fiy{\alpha-\epsilon_f})\right) \right)
$.
And since $|\alpha-\epsilon_f|<|\alpha|$,
$ \phi \left(\fD_k(\fiy{\alpha-\epsilon_f})\right)
=\iD_k(\phi \left(\fiy{\alpha-\epsilon_f})\right)$, 
for any $k$ and in particular for $k=i$.
Therefore
$$
\phi\left( \fD_i(\fiy{\alpha})\right)
=\iD_f\iD_i \left(\phi(\fiy{\alpha})\right) 
 +\sum_k\cminv[ifk]\iD_k \left(\phi(\fiy{\alpha-\epsilon_f})\right) .
$$
This is equal to $\iD_i\left(\phi(\fiy{\alpha})\right)$
by \propr{commuterules}.
\end{proof}

\begin{exy} \label{sl2:algmon}
We carry on with \exr{sl2:invariantization},  
\ref{sl2:K} and \ref{sl2:minorder}.

The stabilization order was $s=1$ and we took a cross-section of that
order.

According to \theor{dolce}, or \propr{rosemary}, 
the set $\ninv[2]$ below forms a generating set of differential invariants:
$$\ninv[2]=
\{\inv x_1, \inv x_2, \inv u_{00},\inv u_{10},\inv u_{01},
  \inv u_{20},\inv u_{11},\inv u_{02}\}.$$ 
We accordingly introduce $\A[0]$ with coordinates
$$\A[0] :\quad (\fix[1]{00},\fix[2]{00}, \fiu[00]{00},
\fiu[10]{00},\fiu[01]{00},
\fiu[20]{00},\fiu[11]{00},\fiu[02]{00}).$$
The coordinates on $\A[k]$ are the
$\fiy{ij}$ where $i+j\leq k$ and $\fiy{}$ 
ranges over the differential indeterminates
$\{\,\fix[1]{},\fix[2]{}, \fiu[00]{},
\fiu[10]{},\fiu[01]{},\fiu[20]{},\fiu[11]{},\fiu[02]{}\,\}$:
$$\A[k] :\quad (\fix[1]{ij},\fix[2]{ij}, \fiu[00]{ij},
\fiu[10]{ij},\fiu[01]{ij},
\fiu[20]{ij},\fiu[11]{ij},\fiu[02]{ij}), \; i+j\leq k .$$
Their images through $\phi:\FA\rightarrow \FG$ are the monotone derivatives of $\ninv[2]$:
\[
\begin{array}{c}
\phi(\fix[1]{ij})=\iD[i]_1\iD[j]_2 (\inv x_1),\;
\phi(\fix[2]{ij})=\iD[i]_1\iD[j]_2 (\inv x_2),\;
\phi(\fiu[00]{ij})=\iD[i]_1\iD[j]_2 (\inv u_{00}),\\
\phi(\fiu[10]{ij})=\iD[i]_1\iD[j]_2 (\inv u_{10}),\;\ldots, \;
\phi(\fiu[02]{ij})=\iD[i]_1\iD[j]_2 (\inv u_{02}).
\end{array}
\]

Given that $[\iD_2,\iD_1]= 
(\inv u_{10}\inv u_{20}+\inv u_{01}\inv u_{11}) \, \iD_2
-(\inv u_{10}\inv u_{11}+\inv u_{01}\inv u_{02})\,\iD_1
$
we define on $\FA$ the derivations $\fD_1$ and $\fD_2$ recursively 
as follows. 
\[
\begin{array}{lcl} 
\fD_1 (\fiy{i,j}) & = & \fiy{i+1,j}, \\
\fD_2(\fiy{0,j}) &=&\fiy{0,j+1},\\
\fD_2(\fiy{i+1,j}) & =& \fD_1 \fD_2(\fiy{i,j}) 
+ (\fiu[10]{00}\fiu[20]{00}+\fiu[01]{00}\fiu[11]{00}) 
    \,  \fD_2(\fiy{i,j})
-(\fiu[10]{00}\fiu[11]{00}+\fiu[01]{00}\fiu[02]{00})
\,\fD_1(\fiy{i,j}).
\end{array}\]
According to \propr{carmen}, $\phi \circ \fD_i=\iD_i\circ \phi$.
We have for instance, with $a+b\leq 2$:
$$\phi(\fD_2(\fiu[ab]{kl}))= \iD_2 \iD[k]_1\iD[l]_2 (\inv u_{ab})$$ 
while
$$\phi(\fD_1(\fiu[ab]{kl}))= \iD_1 \iD[k]_1\iD[l]_2 (\inv u_{ab}) 
=\iD[k+1]_1\iD[l]_2 (\inv u_{ab})=
\phi(\fiu[ab]{k+1,l}).$$ 
\end{exy}

\subsection{Complete set of syzygies}

As a rather immediate  consequence of \theor{recurrences},
the following differential 
relationships hold among the first order derivatives of $\ninv[s+1]$:
\[ \begin{array}{rcll}
\iD_i(\inv x_j) & = & 
\ds  \delta_{ij}-\sum_{a=1}^\dimg K_{ia}\,\inv\left(\vf_a( x_j)\right),
  & 1\leq i,j,\leq\dimx
\\
\iD_i(\inv u_\alpha) &  =  &
   \inv u_{\alpha+\epsilon_i} 
\ds   - \sum_{a=1}^\dimg K_{ia} \, \inv\left( \vf_a( u_\alpha)\right),
 & |\alpha|\leq s
\\
 \iD_i (\inv u_\alpha) 
  -  \iD_j ( \inv u_{\beta}) 
 &  =  & \ds \sum_{a=1}^\dimg K_{ja} \inv \left( \vf_a( u_\beta) \right)
  -  K_{ia} \inv \left( \vf_a( u_\alpha) \right),
  & 
\begin{array}[t]{c}
 \alpha+\epsilon_i=\beta+\epsilon_j,\\ |\alpha|=|\beta|=s+1,
\end{array}
\end{array}
\]
where $\delta_{ij}=1$ or $0$ according to whether $i=j$ or not 
while  $\epsilon_i$ was defined in  \notr{tupleware}.

The first two sets of equations describe how the invariant 
derivations act on  the elements of $\ninv[s]$ in terms of $\ninv[s+1]$.
The last set of equations describes the cross-derivatives
 of the elements of $\ninv[s+1]\setminus \ninv[s]$.
The indices $\alpha$ and $\beta$ and the derivations $\iD_i$ and $\iD_j$
are chosen so that $u_\alpha$ and $u_\beta$ have a 
common derivative $u_\gamma=u_{\alpha+\epsilon_i}=u_{\beta+\epsilon_j}$.
The idea here is that 
there are more than one way to rewrite $\inv u_\gamma$ in terms
of the monotone derivatives of $\ninv[s+1]$:  on one hand 
$\inv u_\gamma = \iD_i (\inv u_\alpha) +K_{ia}\inv\left( \vf_a( u_\alpha)\right)$
and on the other hand
$\inv u_\gamma = \iD_j ( \inv u_{\beta}) 
+ \sum_{a=1}^\dimg K_{ja} \inv \left( \vf_a( u_\beta) \right)$;
both should be equivalent.

Using the setting introduced in the previous subsection
we  formalize and prove  that  those relationships form
a complete set of differential syzygies for $\ninv[s+1]$. 
We actually prove the result for a subset obtained 
by restricting the range of $(i,j)$ for the third type of
 relationships which bears  on $\ninv[s+1]\setminus\ninv[s]$. 
Indeed, some of those relationships 
can be deduced from the others. More specifically,
 if we write $T^{\alpha,i}_{\beta,j}$ for this latter relationship 
and if  $\gamma+\epsilon_k=\alpha+\epsilon_i=\beta+\epsilon_j$ then
$T^{\alpha,i}_{\beta,j}=T^{\alpha,i}_{\gamma,k} -  T^{\beta,j}_{\gamma,k}$.

\begin{defy}
Let $\phi: \FA[k] \rightarrow \FG[s+k+1]$ be as in  \propr{casimir}.
An element of $\FA[k]$  is a \emph{(differential) syzygy}
on the monotone derivatives of  $\ninv[s+1]$  if its 
image by $\phi$ is zero on the cross-section in  $\jva[k]$.
\end{defy}

Since differential invariants are locally determined by their
restriction to the cross-section, this is the same as requesting that
the image is zero on an open set that contains the cross-section.
Furthermore, by \propr{carmen}, the set of syzygies is a differential ideal: 
if $f$ is a syzygy then so is $\fD_i(f)$, for all $1\leq i \leq \dimx$.

\begin{theoy}\label{cnisyz}
Let $s$ be greater or equal to the stabilization 
order\footnote{Under our assumption of a locally effective action 
on $\jva[0]$, the generic orbits in \jva[s] are of the same 
dimension \dimg as the group.}
and assume a cross-section is defined as the zero set of
 $P=(p_1,\ldots,p_\dimg):\jva[s]\rightarrow \Ri^r$.
Let $\FA=\cup_{k\geq 0} \FA[k]$ be the differential algebra 
of monotone derivatives of $\ninv[s+1]$, 
the normalized invariants of order $s+1$.

Consider the map 
$\phi :  \FA  \rightarrow  \FG$
defined by 
$\phi\left(   \fix[]{\alpha} \right)= \iD[\alpha]( \inv x)$, 
and
$\phi \left(\fiu[\beta]{\alpha} \right) =  \iD[\alpha] (\inv u_\beta)$, 
$ \forall \alpha, \beta\in \Ni^\dimx, |\beta|\leq s+1$.
It is surjective and  its kernel is a differential ideal
for the formal invariant derivations,
 $\fD_1,\ldots, \fD_\dimx$ (\defr{hepatique}).
Let $\psi:\F[s+1]\rightarrow \FA[0]$ be the morphism define by 
\( \psi (x) = \fix{0}, \, \psi(u_\beta)=\fiu[\beta]{0} .\)
A generating set for the kernel of $\phi$ is given by the
 union of the three following finite subsets of $\FA[1]$  

\begin{itemize}
\item $\ds \Syzr=\{\,\ds p_1(\fix{0}, \fiu{0}), \ldots, p_\dimg(\fix{0}, \fiu{0})\,\} \, \subset \,\FA[0]$
\item $\ds \Syzs = \{\, \ds \syzs{i}{x_j} \,|\, 1\leq i,j\leq m\, \} 
\cup  \{\, \ds \syzs{i}{u_\alpha}\,|\, |\alpha|\leq s,\,1\leq i\leq m \,\}
 \, \subset \,\FA[1] $
where 
\[ \syzs{i}{x_j} =
\fix[j]{\epsilon_i} - \delta_{ij}- 
 \sum_{a=1}^\dimg \psi\left( {\mcm}_{ia} \vf_a( x_j) \right)  \]
and 
\[ \syzs{i}{u_\alpha} =\fiu[\alpha]{\epsilon_i} - 
      \fiu[\alpha+\epsilon_i]{0} 
     - \sum_{a=1}^\dimg \psi\left({\mcm}_{ia} \vf_a( u_\alpha) \right)
\]
\item $\Syzt= \{\,\syzt{i}{u_\beta} \,|\,
   |\beta|=s+1$ and $\fs[\beta]<i\leq m\, \}  \, \subset \,\FA[1]$
where, with  $f=\fs[\beta]$,
$$\syzt{i}{u_\beta}=  \fiu[\beta]{\epsilon_i}
  -  \fiu[\beta+\epsilon_i-\epsilon_f]{\epsilon_f} 
  - \sum_{a=1}^\dimg \psi\left({\mcm}_{ia} \vf_a( u_{\beta+\epsilon_i-\epsilon_f}) 
     -{\mcm}_{fa}\vf_a( u_\beta)\right).
$$

\end{itemize}

\end{theoy}

The result is deduced from the following lemma. It shows
that  any monotone derivative of $\ninv[s+1]$ can be rewritten in terms
of the normal derivatives modulo $\Syzs\cup\Syzt$. 

\begin{lemy} \label{caracas}
For any  $\alpha \in \Ni^m$ and $|\gamma| \leq s+1$
there exists a linear operator  $L^\alpha_{u_\gamma}$
of order $|\alpha|-1$ in $\seq{\fD}{\dimx}$ 
such that, for $\beta=\alpha+\gamma$, 
\[ \fiu[\gamma]{\alpha} - \fiu[\bas]{\cre}
 - L^\alpha_{u_\gamma}(\Syzs,\Syzt)
\in \FA[|\alpha|-1] .\]
\end{lemy}

\begin{proof}
We consider first the case where $|\gamma|= s+1$
and prove that there  exists a homogeneous linear operator  $H^\alpha_{u_\gamma}$
of order  $|\alpha|-1$ in $\seq{\fD}{\dimx}$ 
such that
$\fiu[\gamma]{\alpha}- \fiu[\bas]{\cre} 
     - H^\alpha_{u_\beta}(\Syzt)   \in \FA[|\beta|-1]$.
The proof is by induction along the following well 
founded pre-order on $\Ni^m$:
\[\gamma \prec \gamma'
\equi \left\{ \begin{array}{l}
 |\gamma|<|\gamma'|  \\
 \hb{or}  | \gamma|=|\gamma'| \hb{and} \ls[\gamma]<\ls[\gamma'] \\
 \hb{or}  |\gamma|=|\gamma'| \hb{and} l=\ls[\gamma]=\ls[\gamma']
  \hb{and} \gamma_l<\gamma_l'
\end{array}\right.
\]

Let
$E_{\beta}=\{\gamma' \,|\, \; |\gamma'|=s+1, \;\exists \alpha' \hb{such that} 
      \alpha'+\gamma'=\beta \}$. 
Note that $\gamma\in E_\beta$ and 
that  $\cre$ is the minimal element of $E_\beta$  according to
$\prec$. 

If $\ls[\alpha]\leq \fs[\gamma]$ 
then $\cre=\alpha$ and $\bas=\gamma$ 
and the result needs no further argument. 

Otherwise assume the result is true for all 
$\gamma'\in E_\beta$ with $\gamma'\prec \gamma$.
Let  $l=\ls[\alpha]>\fs[\gamma]=f$.
We have:

\[ 
\begin{array}{rl}
\fiu[\gamma]{\alpha}
 = & \fD[\alpha-\epsilon_l] (\fiu[\gamma]{\epsilon_l})
\\
= &\fD[\alpha-\epsilon_l]
  \left(
   \fiu[\gamma-\epsilon_f+\epsilon_l]{\epsilon_f} 
   + \syzt{l}{u_\gamma}
   + \sum_{a=1}^\dimg \psi \left({\mcm}_{la} \vf_a( u_{\gamma-\epsilon_f+\epsilon_l}) 
                  -{\mcm}_{fa}\vf_a( u_{\gamma})\right)\right)
.
\end{array}\]

On one hand, the argument of $\psi$ belongs to $\F[s+1]$ 
so that its image belongs to $\FA[0]$.
On the other hand 
$\fD[\alpha-\epsilon_l]
  \left(
   \fiu[\gamma-\epsilon_f+\epsilon_l]{\epsilon_f} \right)
-  \fiu[\gamma-\epsilon_f+\epsilon_l]{\alpha+\epsilon_f -\epsilon_l}
\in \FA[|\alpha|-1]$
according to \lemr{cortina}.
Thus
\[ 
\fiu[\gamma]{\alpha}
-  \fiu[\gamma-\epsilon_f+\epsilon_l]{\alpha+\epsilon_f -\epsilon_l}
- \fD[\alpha-\epsilon_l]\left(\syzt{l}{u_\gamma} \right)
 \in \FA[|\alpha|-1] .
\]
Since $\gamma-\epsilon_f+\epsilon_l \prec \gamma$
we can conclude our induction argument. 

We are left to prove that, 
for all $|\gamma|\leq s$ and $\alpha\in \Ni^m$,
there is a $\mu\in\Ni^\dimx$ with  $|\mu|=s+1-|\gamma|$ 
and a differential operator $L_{u_\gamma}^{\alpha}$
such that 
$$
\fiu[\gamma]{\alpha}-\fiu[\gamma+\mu]{\alpha-\mu}   
-L_{u_\gamma}^{\alpha}(\Syzs)   \in \FA[|\alpha|-1].
$$
For that it is sufficient to lead an inductive argument on the fact that
\[ \fiu[\gamma]{\alpha}= 
   \fD[\alpha-\epsilon_l]\left(\fiu[\gamma]{\epsilon_l}\right)
   =
      \fiu[\gamma+\epsilon_l]{\alpha-\epsilon_l}+
    \fD[\alpha-\epsilon_l]\left( \syzs{l}{u_\gamma}
     + \sum_{a=1}^r \psi \left( \mcm[{la}] \vf_a(u_\gamma)\right) \right),
\]
where $l=\ls[\alpha]$.
\end{proof}

\begin{proof} (of the theorem).
Taylor's formula with integral remainder shows the following 
\cite[Paragraph 2.5]{bourbaki67}. For a smooth function $f$ on an open set 
$U\times I_1\times \ldots \times  I_l \subset \Ri^k\times\Ri^l$, 
where the $I_i$ are intervals of $\Ri$ that contain zero,
there are smooth functions $f_0$ on $U$, and 
$f_i$ on $U\times I_1\times \ldots \times  I_i$, $1\leq i\leq l$
such that
$f(x,t_1, \ldots, t_l)=f_0(x)+ \sum_{j=1}^l t_j\, f_j(x,t_1,\ldots, t_j)$.

Let us restrict the $\A[k]$ to appropriate neighborhoods 
of the zero set of $\Syzs$, $\Syzt$ and their derivatives.
 Take $f\in \FA[k+1]$. 
By first applying \lemr{caracas} for
 $|\alpha+\gamma|=k+1$, we can first write it as:
\[ 
f( \fiu[\gamma]{\alpha}, \fiu[\gamma']{\alpha'}) 
  = 
f_1( \fiu[{\bas}]{\cre},  \fiu[\gamma']{\alpha'})
+ \sum_{|\alpha+\gamma|=k+1} L_{u_\gamma}^{\alpha}(\Syzs,\Syzt)\, F_{u_\gamma}^{\alpha}
\]
where $(\gamma,\alpha)$ range over $|\alpha+\gamma|=k+1$ while
 $(\gamma',\alpha')$ range over $|\alpha'+\gamma'|\leq k$
so that $\beta$ ranges over $|\beta|= k+1$
and $ F_{u_\gamma}^{\alpha}\in\FA[k+1]$. 
We can iterate this process on the 
 $\fiu[\gamma']{\alpha'}$, with $|\alpha'+\gamma'|=k$, in  $f_1$.
Induction then shows that
 \[ 
f( \fiu[\gamma]{\alpha}) 
  = 
F( \fiu[{\bas}]{\cre})
+ \sum_{|\alpha+\gamma|\leq k+1} L_{u_\gamma}^{\alpha} F_{u_\gamma}^{\alpha}
\]
where now $(\alpha,\gamma)$ range over $|\alpha+\gamma|\leq k+1$
and $\beta$ over $|\beta|\leq k+1$.

Thus  $\phi(f)=\phi(F)$. By \lemr{aioli},
if  $f$ belongs to the kernel of $\phi$ then 
$F$ is a linear combination of elements of $\Syzr$.
\sil{
Assume that a function $F$ of $|\ndinv[k]|$ variables  
that vanishes on $\ndinv[k]$:
$F\left(\iD[\alpha]( \inv z) | 
\inv z\in \ninv[s+1],|\alpha|\leq k\right)\equiv0$. 
It follows that
$F(f^\alpha_z | \inv z\in \ninv[s+1], |\alpha|\leq k)\equiv 0$
where $f^\alpha_z$ are the functions of $\ndinv[|\alpha|]$ defined in
the previous lemma.
By \lemr{aioli} $F(f^\alpha_z)$ is not independent of $\Syzr$.
}
\end{proof}

\begin{exy} \label{sl2:syz}
We carry on with 
\exr{sl2:invariantization}, \ref{sl2:K}, \ref{sl2:minorder}
 and \ref{sl2:algmon}.

Recall that
\[ 
 \vf_1= \pd{x_1}, \quad \vf_2=\pd{x_2}, \quad 
 \vf_3=x_1\,\pd{x_1}+x_2\,\pd{x_2}-\sum_{i,j\geq 0}(i+j)\,u_{ij}\,\pd{u_{ij}}
\]
while 
$P=\left(x_1,x_2,\frac{1}{2}-\frac{1}{2}(u_{10}^2+u_{01}^2)\right)$,
so that
\[ K = \left(\begin{array}{ccc} 1&0&\inv v \\ 0&1&\inv w
   \end{array}\right)
\hbox{ where }
v=-(u_{10}u_{20}+u_{01}u_{11}) \hb{and} w=-( u_{10}u_{11}+u_{01}u_{02}).
\]

According to \theor{cnisyz} 
a complete set of syzygies on \ninv[2],
i.e. a basis for the kernel of $\phi:\FA\rightarrow\FG$,
consists of the following elements.
\begin{description}
\item[{$\Syzr$},] 
the functional relationships implied by the choice of the cross-section.
$$
\fix[1]{00} , \;\fix[2]{00},\;
\frac{1}{2}-\frac{1}{2} \left( (\fiu[10]{00})^2+(\fiu[01]{00})^2\right)$$
\item[{$\Syzs$},]
 the  relationships describing 
the derivations of the elements of $\ninv[s]$:
\[ \begin{array}{lclclcl}
\syzs{1}{x_1}&:& \fix[1]{10} 
&\quad &
\syzs{2}{x_1}&:& \fix[1]{01}
\\
\syzs{1}{x_2}&:& \fix[2]{10} 
&\quad &
\syzs{2}{x_2}&:& \fix[2]{01}
\\
\syzs{1}{u_{00}} &:& 
\fiu[00]{10} -\fiu[10]{00},
&\quad &
\syzs{2}{u_{00}} &:& 
\fiu[00]{01} -\fiu[01]{00}, 
\\
\syzs{1}{u_{10}} &:& 
  \fiu[10]{10}-\fiu[20]{00} +\fiu[10]{00}\,\mathfrak{v}, 
&\quad &
\syzs{2}{u_{10}} &:& 
   \fiu[10]{01}-\fiu[11]{00} +\fiu[10]{00}\,\mathfrak{w},
\\
\syzs{1}{u_{10}} &:& 
    \fiu[01]{10}-\fiu[11]{00} +\fiu[01]{00}\,\mathfrak{v} ,
&\quad &
\syzs{2}{u_{10}} &:& 
\fiu[01]{01}-\fiu[02]{00} +\fiu[01]{00}\,\mathfrak{w} ,
\end{array}
\]
where 
\[
\mathfrak v 
   =-\fiu[10]{00}\,\fiu[20]{00}-\fiu[01]{00}\,\fiu[11]{00}
   =\psi(v),
\quad 
 \mathfrak w 
  =-\fiu[10]{00}\,\fiu[11]{00}-\fiu[01]{00}\,\fiu[02]{00},
=\psi(w) .
\]

\item[{$\Syzt$},]
 the  relationships obtained by cross-differentiating the elements
 $\ninv[s+1]\setminus\ninv[s]$:
\[ \begin{array}{lcl}
 \syzt{2}{u_{20}} & : & 
   \fiu[20]{01} -\fiu[11]{10}
    - 2\, \fiu[20]{00} \,\mathfrak{w} + 2\, \fiu[11]\, \mathfrak{v}
\\
 \syzt{2}{u_{11}} & : & 
   \fiu[11]{01} -\fiu[02]{10} 
  - 2\, \fiu[11]{00} \,\mathfrak{w} + 2\, \fiu[02]{00} \,\mathfrak{v}.
\end{array}
\]
\end{description}


Yet from \theor{minorder} we know that 
$\{\inv u, \, \inv v,\, \inv w\}$ form a generating set. 
As $\inv v$ and $\inv w$ are the coefficients of the commutation rules, 
we can perform a differential elimination to obtain a complete set of 
syzygies bearing on  $\{\inv u, \, \inv v,\, \inv w\}$ 
\citep{hubert03d,hubert05}. We obtain:
\[ 
 \fder[1](\mathfrak w)-\fder[2](\mathfrak v) =0,\qquad
 \fder[1](\mathfrak u)^2+\fder[2](\mathfrak u)^2=1 .
\]
\end{exy}

\section{Classical examples}
     \label{examples} 

We treat two very classical  geometries, 
curves and surfaces in Euclidean 3-space, 
in order to illustrate the general theory of this paper on well-known cases.

For surfaces we shall use the classical cross-section, 
show how the mean and Gauss curvature relate to the 
exhibited generating set of differential invariants 
and how the Gauss-Codazzi equation on the principal
 curvatures arises as the syzygy. 

For curves we shall choose some non classical cross-sections
 that can come of use.
We first illustrate \theor{minorder} 
for a cross-section of minimal order that is not a 
coordinate cross-section and therefore not covered by \citet{olver07}.
The edge invariants are explicitly shown to form a generating set 
of differential invariants and endowed with a rewriting procedure.
The syzygies there are trivial.

We then consider the  cross-section introduced by
 \citet{olver07} to show that  the minimal order condition on the cross-section 
is necessary for  \theor{minorder} to hold, 
i.e. for the edge invariants to be a generating set of differential
 invariants.
There are then non trivial differential syzygies on the generating set 
of normalized invariants. 
Elimination on those allows to diminish the number of generators.

As should come clear from those examples, the only data we start
 with are the infinitesimal generators of the action 
and a choice of cross-section.
Of course, the art of choosing the appropriate cross-section 
for a given application should not be underestimated.

For the benefit of a lighter notation system,
we skip the Gothic notation of the formalism introduced in \secr{syzygies}
when formalizing the notion of syzygies. Therefore $\inv u_\alpha$ 
will in turn represent a local invariant, i.e. an element of $\FG$, 
or the coordinate function  $\fiu[\alpha]{0}$ of $\A$. 

\subsection{Surfaces in Euclidean geometry}
 \label{euclidean:cn}

We shall show how to retrieve the Codazzi equation as the syzygy between 
the two  generators for the differential invariants.

We choose coordinate functions $(x_1,x_2,u)$ for $\Ri^2\times\Ri$.
We consider $x_1,x_2$ as the independent variables and $u$ 
as the dependent variable.

The infinitesimal generators of the classical action 
of the Euclidean group $SE(3)$  on $\Ri^3$
 are:
\[ \begin{array}{c} 
\ds \vf[0]_1= \pdif{ }{x_1},\;  \vf[0]_2= \pdif{ }{x_2}, \; \vf[0]_3=\pdif{ }{u}, \\  
\ds \vf[0]_4= x_1\,\pdif{ }{u} -u\,\pdif{ }{x_1},  
\ds \vf[0]_5= x_2\,\pdif{ }{u} -u\, \pdif{ }{x_2}, 
\ds \vf[0]_6= x_1\,\pdif{ }{x_2}-x_2\, \pdif{ }{x_1},
 \end{array} 
\]
so that their prolongations are
\[ \begin{array}{c} 
\ds \vf_1= \pdif{ }{x_1},\;  \vf_2= \pdif{ }{x_2}, \; \vf_3=\pdif{ }{u}, \\  
\ds \vf_4=\sum_{\alpha}\D[\alpha](x_1+u_{00}u_{10})\,\pdif{ }{u_\alpha}-u_{00}\,\der[1],  
\ds \vf_5=\sum_{\alpha}\D[\alpha](x_2+u_{00}u_{01})\,\pdif{ }{u_\alpha} -u_{00}\, \der[2], \\
\ds \vf_6= x_1\,\der[2]-x_2\,\der[1]+\sum_{\alpha}\D[\alpha](x_2u_{10}-x_1u_{01})\,\pdif{ }{u_\alpha}.
 \end{array} 
\]

Let us choose the classical cross-section defined by
 $P=(x_1,x_2,u_{00},u_{10},u_{01},u_{11})$.
The Maurer-Cartan matrix of \theor{recurrences} is
\[ \mcm = \left( \begin{array}{cccccc}
 1 & 0 & 0 & \inv u_{20} & 0 & \ds \frac{\inv u_{21}}{\inv u_{20}-\inv u_{02}} 
\\ \\
0 & 1 & 0 & 0 & \inv u_{02} & \ds \frac{\inv u_{12}}{\inv u_{20}-\inv u_{02}} 
 \end{array}\right). \]
Applying \propr{commuterules} we have
\begin{equation} \label{malbec}
[\iD_2, \iD_1 ] = 
\frac{\inv u_{21}}{\inv u_{20}-\inv u_{02}} \,\iD_1 
 + \frac{\inv u_{12}}{\inv u_{20}-\inv u_{02}} \, \iD_2 .
\end{equation}
Given that $\inv x_1,\inv x_2,\inv u_{00}, \inv u_{10},\inv u_{01},\inv u_{11}=0$
the non zero elements of $\Syzs$ in \theor{cnisyz} are
\[\begin{array}{lclllcl}
\syzs{1}{u_{20}}& = &\ider[1](\inv u_{20})-\inv u_{30},
& \quad &
\syzs{2}{u_{20}}& = & \ider[2] (\inv u_{20})-\inv u_{21},
\\
\syzs{1}{u_{02}}& =  &\ider[1](\inv u_{02})-\inv u_{12}, 
& & 
\syzs{2}{u_{02}}& = &\ider[2] (\inv u_{02})-\inv u_{03},
\end{array}\]
while the elements of $\Syzt$ are
\[\begin{array}{lcl}
\ds
\syzt{2}{u_{12}} &= & \ds
\ider[2](\inv u_{12})-\ider[1](\inv u_{03}) -
\frac{\inv u_{12}}{\inv u_{20}-\inv u_{02}}\,(\inv u_{21}+\inv u_{03}),
\\ \ds
\syzt{2}{u_{30}}& = & \ds
\ider[2]({\inv u_{30}})-\ider[1]({\inv u_{21}})
-{\frac {{\inv u_{21}}}{{\inv u_{20}}-{\inv u_{02}}}}
\,({\inv u_{12}} +{\inv u_{30}}),
\\ \ds
\syzt{2}{u_{21}} & = &  \ds
\iD_2({\inv u_{21}})-\iD_1({\inv u_{12}}) 
-\frac {{\inv u_{21}}\,{\inv u_{03}}+{\inv u_{12}}\,{\inv u_{30}}
        -2\,{{\inv u_{21}}}^{2}-2\,{{\inv u_{12}}}^{2}}
       {{\inv u_{20}}-{\inv u_{02}}}
+ \left( {\inv u_{20}}-{\inv u_{02}} \right) {\inv u_{02}}\,{\inv u_{20}} .
\end{array}\]

\theor{minorder}  predicts that 
$\{\inv u_{20},\inv u_{02},\inv u_{21},\inv u_{12}\}$ form a generating set.
From $\Syzs$ we see furthermore that all the third order \cnis
 can be expressed as derivatives of $\{\inv u_{20}, \inv u_{02}\}$. 
This latter set therefore already forms a generating set of
invariants. 
Indeed, with \theor{thomas}, we can write  
the Gauss and mean curvatures in terms of
 $\{\inv u_{20}, \inv u_{02}\}$
 \citep[(10.6.5)]{berger88},  \citep[(1.3)]{ivey03}
\[\begin{array}{c}
\ds \sigma
 = \frac{u_{20}u_{02}-u_{11}^2}{(1+u_{10}^2+u_{01}^2)^2} 
 = \inv u_{20}\, \inv u_{02}, 
\\ \\
\ds \pi
= \frac{1}{2}\frac{(1+u_{01}^2)u_{20}-2_{10}u_{01}u_{11}+(1+u_{10}^2)u_{02}}
                  {(1+u_{10}^2+u_{01}^2)^{\frac{3}{2}}}
=\frac{1}{2}(\inv u_{20}+\inv u_{02}).
\end{array}\] 
Our generators $\{\inv u_{20}, \inv u_{02}\}$
 are thus the principal curvatures.
Let us write $\kappa=\inv u_{20}$ and $\tau=\inv u_{02}$.
From $\Syzs$ we have
$$ \inv u_{30}=\ider[1](\kappa), \; \inv u_{21}=\ider[2](\kappa),\; 
  \inv u_{12}= \ider[1](\tau), \hb{and} \inv u_{03}=\ider[2](\tau).$$
Making the substitution in  $\Syzt$ we obtain
\[ 
\begin{array}{l} \ds
\ds 
\ider[2]\ider[1]( \tau) 
 -\ider[1]\ider[2](\tau)  -
\frac{\ider[1]( \tau)}{\kappa -\tau}\,(\ider[2](\kappa)+\ider[2](\tau) )
\\\ds
\ds
\ider[2]\ider[1](\kappa) -\ider[1]\ider[2](\kappa) 
-\frac { \ider[2] (\kappa)}{\kappa -\tau}
\,(\ider[1](\kappa)+\ider[1](\tau) )
\\ 
\ds
\iD[2]_2( \kappa) -\iD[2]_1 (\tau) 
-\frac {\ider[1](\kappa)\,\ider[1](\tau)+\ider[2](\kappa) \,\ider[2](\tau) 
        -2\,{\ider[2](\kappa)}^{2}-2\,{\ider[1](\tau)}^{2}}
       {\kappa -\tau}
+ \left( \kappa -\tau \right)\kappa\, \tau .
\end{array}\]

The two first functions vanish when one rewrites 
$\ider[2]\ider[1](\tau)$ and $\ider[2]\ider[1]( \kappa)$ 
in terms of monotone derivatives using \Ref{malbec}.
The last function provides the Gauss-Codazzi equation 
 \citep[Exercise 2.3.1]{ivey03}.

\subsection{Curves in Euclidean geometry} \label{curves}

For this example we will first work with a
cross-section of minimal order.
The edge invariants are then generating and submitted 
to no non trivial syzygies.
When we then use a cross-section that is not of minimal order, 
a non trivial syzygie appears on the predicted generating sets.

We consider the classical action of $\hbox{SE}(3)$ on space curves.
We have $\jva[0]=\xva^1\times\uva^2$ with coordinate $(x,u,v)$.
The infinitesimal generators of the action are: 
\[ 
\begin{array}{lll}
\ds \vf[0]_1= \pdif{}{x}, 
 & \ds \vf[0]_2=\pdif{}{u}, 
 & \ds\vf[0]_3=\pdif{}{v} 
\\
  \ds \vf[0]_4= v\,\pdif{}{u}-u\,\pdif{}{v} , 
 & \ds\vf[0]_5= x\pdif{}{u}-u\pdif{}{x}, 
 &\ds \vf[0]_6= x\pdif{}{v} -v\pdif{}{x}
\end{array} \]
so that their prolongations are given by 
\[ 
\begin{array}{l}
 \ds \vf_1= \pdif{}{x}, 
 \quad \vf_2=\pdif{}{u}, 
 \quad \vf_3=\pdif{}{v},
 \quad 
  \ds \vf_4= \sum_{k} v_k\,\pdif{}{u_k}-u_k\,\pdif{}{v_k} , 
 \\\ds \vf_5= -u_0 \, \D + \sum_{k}\D[k](x-u_0\,u_1)\pdif{}{u_k}
              -\sum_k \D[k](u_0\,v_1)\pdif{}{v_k}, 
 \\ \ds \vf_6= -v_0\,\D  -\sum_k \D[k](v_0\,u_1)\pdif{}{u_k}
             + \sum_{k}\D[k](x-v_0\,v_1)\pdif{}{v_k} .
\end{array} \]

The action is transitive on $\jva[1]$ 
and becomes locally free on $\jva[2]$ with 
 generic orbits of codimension 1.

\subsubsection*{Minimal order cross-section}

We choose a non classical cross-section of minimal order:
$P=(x,u_0,v_0,u_1,v_1,v_2-u_2)$.
Then: 
\[ \inv (\D(P)) 
= \left( \begin{array}{cccccc}
  1& 0& 0& \inv u_2& \inv u_2& \inv (v_3-u_3)
\end{array} \right).\]

On one hand we know from \theor{dolce}  that 
$\ninv[3]=\{\inv x, \inv u_0, \inv u_1, \inv v_1, \inv u_3, \inv v_3\}$
is a generating set of differential invariants and 
rewriting any differential invariants in terms of them 
is a recursive process described  in \secr{geni}, or more specifically 
by \propr{rosemary}. One can check that
the complete set of  syzygies  on $\ninv[3]$ given in \theor{cnisyz}
boils down to 
$\Syzr=\{\inv x, \inv u_0,\inv v_0,\inv u_1,\inv v_1,\inv v_2-\inv u_2\}$
since $\Syzs=\{0\}$ and $\Syzt=\emptyset$.

On the other hand \theor{minorder} implies that 
 $\moinv =\{ \inv u_2, \, \inv w \}$, 
where $w=v_3-u_3$, is a generating set of differential invariants.
For the purpose of rewriting any other differential invariants
 in terms of them
we write every element of $\ninv[3]$ in terms of $\moinv$. 

From  \theor{recurrences} we have 
\( \iD (\inv u_2)= \inv u_3 - \frac{1}{2} \,\inv w \)
since 
\[\ds K = \left( \begin{array}{cccccc}
   1 & 0 & 0 & \ds \frac{\inv w}{2\,\inv u_2} & \inv u_2 & \inv u_2 
   \end{array}\right) \]
while 
$\inv\left( \vf(u_2) \right) 
= \left( \begin{array}{cccccc}
        0&0&0&\inv u_2&0 & 0
 \end{array}\right)^T .$
Thus 
\[ \inv v_2 = \inv u_2,\; 
   \inv u_3 = \iD(\inv u_2)+ \frac{\inv w}{2},\;
\hb{and}
\inv v_3= \iD(\inv u_2)-\frac{\inv w}{2}. \]

Note that $\inv u_2$ is a differential invariant of order 2 
and is therefore a function of the curvature,
while $\inv (u_3-v_3)$, as a differential invariant of order $3$ 
is a function of the curvature $\kappa$ and the torsion $\tau$. 
There are several ways to compute the algebraic expression for 
$\inv u_2, \inv u_3$ and $\inv v_3$ \citep{fels99,hubert07,hubert07b}.
But conversely, given the analytic expression for
 the curvature and the torsion \citep[(8.4.13.1) and (8.6.10.2)]{berger88} 
it is easy to write them in terms of 
 $\inv u_2, \inv u_3$ and $\inv v_3$ thanks to \theor{thomas}.
\[ \kappa 
   = \sqrt{2\, \inv u_2^2}, \quad
\tau = 
\frac{ \inv u_3 -\inv v_3}{2\, \inv u_2} .\]

\subsubsection*{Non minimal cross-section}

We consider now the third order cross-section 
$ P= (x,u_0,v_0,v_1,v_2, v_3-1)$.
 \cite{olver07} introduced it
to show that the minimal order condition
 is necessary for \theor{minorder}.

As a consequence of \theor{dolce}, 
$\{\inv u_1, \, \inv u_2, \, \inv u_3, \, \inv u_4, \inv v_4\}$ 
is a generating set of differential invariants.
According to \theor{cnisyz}
the following functions form a complete set of
 differential syzygies.

\[ \begin{array}{lcl}
\ider(\inv { u_1}) 
&=& 
\ds \inv { u_2}
+
 {\frac {1+\inv { u_1}^2}{3\,\inv { u_1}}}
\,    \left( \frac{ \inv { u_3}}{\inv { u_2}}-\, \inv {v_4} \right)
\\
\ider(\inv { u_2})
&=& \ds
2\,\inv { u_3}
-\inv { u_2}\,\inv { v_4}
\\
\ider(\inv { u_3} ) 
&=& \ds
\inv  u_4
- \left( \frac{4}{3}\,\inv u_3+{\frac {{\inv u_2}^{2}}{\inv { u_1}}} \right) \inv { v_4}
+{\frac {{\inv u_1}^{2}+1}{\inv u_2}}
+\frac{4}{3}\,{\frac {{\inv u_3}^{2}}{\inv { u_2}}}
+{\frac {\inv { u_2}\,\inv { u_3}}{\inv { u_1}}}
\end{array}
\]
From the two first equations we can deduce 
$\inv u_3$ and $\inv v_4$ in terms of 
$\{\inv u_1,\,\inv u_2\}$ and their derivatives. 
Substituting in the last equation we can do the same for  $\inv u_4$
so that $\{\inv u_1, \inv u_2\}$ is a generating set. 
Concomitantly, 
given their explicit expressions, we can write the curvature 
and the torsion in terms of those through \theor{thomas}:
\[ \kappa = \sqrt{\frac{\inv u_2^2}{(1+\inv u_1^2)^3}}, 
\quad
\tau = \frac{1}{\inv u_2 (1+\inv u_1^2)} .\]

\section{Three independent variables} 
  \label{d3} 

The indefinite orthogonal group 
$O(m_1,m_2)$ is defined as the subgroup of $GL(m_1+m_2)$ that 
leaves the bilinear form 
$x_1^2+\ldots+x_{m_1}^2-x_{m_1+1}^2-\ldots-x_{m_1+m_2}^2$ invariant.
The groups   $O(m_1,m_2)$  and $O(m_1,m_2) \ltimes \Ri^{m_1+m_2}$ 
arise as symmetries of physical differential systems.
For instance,  $O(m,0) \ltimes \Ri^m$ 
is a group of symmetry for the Laplacian, 
$u_{x_1x_1}+\ldots+u_{x_mx_m}$, 
while $O(m-1,1) \ltimes \Ri^m$ 
is a symmetry group for the D'Alembert equation
 $u_{x_1x_1}-u_{x_2x_2}-\ldots-u_{x_mx_m}$.
Their differential invariants of all orders were determined by \citet{xu98}.

In the case $m=m_1+m_2=3$ we offer here a
classification of the generating sets
 in the differential sense.
As far as we know very few examples dealing with 
three independent variables 
have been studied by a moving frame approach.
To provide those examples we have substantially applied 
our symbolic computation software \textsc{aida} \citep{aida}. 
The corresponding worksheet 
is  available at 
\url{http://www-sop.inria.fr/cafe/Evelyne.Hubert/aida/syzygies/E3l.html}

   \subsection{Linear action of $O(3-l,l)$ on the independent variables}
        \label{SO3l}      

With the help of a parameter $\epsilon$ we shall treat both 
the orthogonal group $O(3,0)$ and $O(2,1)$ at once. 
When we specialize $\epsilon=1$  we shall retrieve the result 
for $O(3,0)$ and when $\epsilon=-1$  we shall retrieve the result for $O(2,1)$.
The moving frame approach was applied to this action of $O(3,0)$ in
\citep[Example 15.3]{fels99} which provides us with a double check 
while providing the analysis for $O(2,1)$.

From the knowledge of the infinitesimal generators for the action
of $O(3-l,l)$ and a choice of a (minimal order) 
cross-section we exhibit a complete set of 
syzygies for the second order normalized invariants.  
This is a direct application of \theor{cnisyz}. 
The set of second order normalized invariants is generating 
but so is the much smaller set of edge invariants. 
This allows us to prove that  an alternative set of equal 
cardinality is also generating. 
This latter makes computations easier. 
By differential elimination on the syzygies of the
second order normalized invariants
 we can retrieve a complete 
set of syzygies for this new and smaller generating set. 

Those above computations are performed without the knowledge 
of the moving frame nor the explicit expression to the invariants.
We nonetheless provide the expressions for those latter elements 
for illustration.

\subsubsection{Action}

We accordingly define $O(3-l,l)$, for $l=0$ or $1$, 
as the subgroup of  $GL(3)$ 
that preserves the bilinear form $x_1^2+x_2^2+\epsilon\,x_3^2$, 
where $\epsilon=(-1)^l$.
We consider its linear action on the independent variables $\{x_1,x_2,x_3\}$.
The dependent variable is left unchanged by those transformations.

The infinitesimal generators are then:
\[
\begin{array}{c} 
\ds \vf[0]_1= x_2\,\pdif{ }{x_1} -x_1\,\pdif{ }{x_2},  
\ds \vf[0]_2= x_3\,\pdif{ }{x_1} -\epsilon\,x_1\, \pdif{ }{x_3}, 
\ds \vf[0]_3= x_3\,\pdif{ }{x_2}- \epsilon\,x_2\, \pdif{ }{x_3}
.
\end{array}
\]

\subsubsection{Generators}

We choose the minimal order cross-section 
$$ \sva:\; x_1=0,\,x_2=0,\,u_{100}=0 .$$
The edge invariants:
$\moinv =\{\inv x_3, \inv u_{000},\inv u_{200},\inv u_{110},\inv u_{101}\}$
form a generating set (\theor{minorder}).
Furthermore, since  $u_{000}$ is invariant, and therefore $\vf(u_{000})=0$,
we deduce from \theor{recurrences}  that
$\inv u_{010}=\ider[2] (\inv u_{000})$
and
$\inv u_{001}=\ider[3] (\inv u_{000})$.
It will be convenient to use the following set of generators:

\[
\rho={\frac{1}{\inv x_3}},\;
\sigma=\inv u_{000},\;
\phi=\frac{\inv u_{200}}{\inv u_{010}}- \epsilon\,\frac{\inv u_{001}}{{\inv u_{010}}\, {\inv x_3}},\;
\zeta={\frac {\inv u_{110}}{\inv u_{010}}},\;
\psi={\frac {\inv u_{101}}{\inv u_{010}}}
\]
as we can then write the matrix $K$ of \theor{recurrences} as
\[ K=\left( \begin {array}{ccc} 
\phi &\rho &0
\\\noalign{\medskip}
\zeta &0&\rho 
\\\noalign{\medskip}
\psi &0&0
\end {array} \right).\]
\cite{hubert07m} actually shows that the entries of this matrix,
 the Maurer-Cartan matrix, together with $\ninv[0]$,
 always form a generating set of invariants.
It is arguably a more appropriate generating set for practical purposes. 
Here, for instance, it saves us dealing with denominators.
We then deduce from \propr{commuterules} that:
\[
[{ \ider[1]},{ \ider[2]}]=\phi\,\ider[1]+\zeta\,\ider[2],\;
[{ \ider[1]},{ \ider[3]}]=\rho\,\ider[1]+\psi\,\ider[2],\;
[{ \ider[3]},{ \ider[2]}]=\psi\,\ider[1]-\rho\,\ider[2]
.
\]

\subsubsection{Syzygies}

According to \theor{cnisyz} a complete set of syzygies for $\ninv[2]$
 is given by:

$$\Syzr\left\{\; \inv x_1=0,\quad \inv x_2 = 0,\quad \inv u_{100}=0, \right.$$

\[\Syzs \left\{\begin{array}{l}
\ider[1]( \inv x_3)=0,\quad \ider[2]( \inv x_3 )=0,\quad \ider[3]( \inv x_3)=1,
\\
\ider[1](\inv u_{000})=0,\quad
\ider[2](\inv u_{000})={\inv u_{010}},\quad
\ider[3](\inv u_{000})={\inv u_{001}},
\\
\ider[1](\inv u_{010})={\inv u_{110}},\;
\ider[2](\inv u_{010})={\inv u_{020}}-\epsilon\,\rho\,{\inv u_{001}},\;
\ider[3](\inv u_{010})={\inv u_{011}},
\\
\ider[1](\inv u_{001})={\inv u_{101}},\;
\ider[2](\inv u_{001})={\inv u_{011}}+\rho\,{\inv u_{010}},\;
\ider[3](\inv u_{001})={\inv u_{002}},
\end{array} \right.\]
and
\[ \Syzt \left\{ \begin{array}{l}
\ider[3](\inv u_{011})-\ider[2](\inv u_{002})=
\psi\,{\inv u_{101}}-2\,\rho\,{\inv u_{011}},
\\
\ider[3](\inv u_{020})-\ider[2](\inv u_{011})=
2\,\psi\,{\inv u_{110}}-\zeta\,{\inv u_{101}}
-\rho\,{\inv u_{020}}+\epsilon\,\rho\,{\inv u_{002}},
\\
\ider[2](\inv u_{101})-\ider[1](\inv u_{011})=
-\zeta\,{\inv u_{011}}-\phi\,{\inv u_{101}},
\\
\ider[3](\inv u_{101})-\ider[1](\inv u_{002})=
-\psi\,{\inv u_{011}}-2\,\rho\,{\inv u_{101}},
\\
\ider[2](\inv u_{110})-\ider[1](\inv u_{020})=
\zeta\,{\inv u_{200}}-\zeta\,
{\inv u_{020}}-2\,\phi\,{\inv u_{110}}-\epsilon\,\rho\,{\inv u_{101}},
\\
\ider[3](\inv u_{110})-\ider[1](\inv u_{011})=
\psi\,{\inv u_{200}}-\psi\,{\inv u_{020}}
-\phi\,{\inv u_{101}}-\rho\,{\inv u_{110}},
\\
\ider[2](\inv u_{200})-\ider[1](\inv u_{110})=
-2\,\zeta\,{\inv u_{110}}-\phi\,{\inv u_{200}}
+\phi\,{\inv u_{020}}+\epsilon\,\rho\,{\inv u_{011}},
\\
\ider[3](\inv u_{200})-\ider[1](\inv u_{101})=
-2\,\psi\,{\inv u_{110}}+\phi\,{\inv u_{011}}
-\rho\,{\inv u_{200}}+\epsilon\,\rho\,{\inv u_{002}}
.
\end{array} \right.
\]
By differential elimination 
we can rewrite any normalized invariants of order 2 and less in terms of 
$\{\rho,\sigma,\psi,\phi,\zeta\}$ and find a complete set of syzygies for those.
The first part
\[ \begin{array}{c}
\ds {\inv x_1}=0,\;\inv x_2=0,\;{ \inv x_3} =\frac{1}{\rho },\;
{ \inv u_{000}} =\sigma ,
\\
\inv u_{100} =0,\;
{ \inv u_{010}} =\ider[2](\sigma),\;
{ \inv u_{001}} =\ider[3](\sigma),\;
\\
{ \inv u_{200}} =   \phi \,\ider[2](\sigma) +\epsilon\,\rho\,\ider[3](\sigma) ,\;
{ \inv u_{110}} = \zeta\, \ider[2](\sigma),\;
{ \inv u_{101}} = \psi\,\ider[2](\sigma) ,
\\
{ \inv u_{020}} =\ider[2]^2(\sigma)+\epsilon\,\rho\,\ider[3](\sigma) ,\;
{ \inv u_{011}} =\ider[2]\ider[3](\sigma)-\rho\,\ider[2](\sigma) ,\;
{ \inv u_{002}} =\ider[3]^2(\sigma)
\end{array}\]
allows us to rewrite any other differential invariants in terms of 
$\{\rho,\sigma,\psi,\phi,\zeta\}$.
 The second part provides a complete set of syzygies for 
$\{\rho,\sigma,\psi,\phi,\zeta\}$:
\[\begin{array}{c}
\ider[1](\zeta)-\ider[2](\phi)={\zeta }^{2}+{\phi }^{2}+\epsilon\,{\rho }^{2},\\
\ider[1](\psi)-\ider[3](\phi)=\phi\, \rho +\psi\, \zeta ,\\
\ider[1](\sigma)=0,
\;\ider[1](\rho)=0,\;\ider[2](\rho)=0,\;\ider[3](\rho)=-{\rho }^{2}
\end{array}\]
We observe that  $\zeta$ can actually be written in terms of 
$\{\phi,\psi,\rho,\sigma\}$ so that this latter is already a generating set.
The first two syzygies then become:
\[\begin{array}{r}
{\psi }\left(\ider[1]^2(\psi)- \ider[1]\ider[3](\phi)\right)= 
{\ider[3](\phi)}^{2}+2\,{\ider[1](\psi)}^{2}-3\,\ider[1](\psi)\ider[3](\phi)
+\phi \rho\,( 2\,\ider[3](\phi)-3\, \ider[1](\psi))
\\+{\psi }^{2}\ider[2](\phi) +\psi \rho  \ider[1](\phi)
+ {\phi }^{2}\,{\rho }^{2}+\epsilon\,{\psi }^{2}\,{\rho }^{2}
 +{\psi }^{2}{\phi }^{2}
\end{array}\]
since
\[\zeta ={\frac {\ider[1](\psi)-\ider[3](\phi)-\phi \rho }{\psi }}.\]

\subsubsection{Generating differential invariants}

For completion on this example, let us give the explicit 
expressions for a set of generating differential invariants. 
We split that into giving the expressions for  
$\{\inv x_3,\inv u_{001},\inv u_{010}\}$ as we can 
 write $\{\rho,\sigma,\phi,\psi\}$ in terms of them.
To determine $\{\inv x_3,\inv u_{001},\inv u_{010}\}$ we follow  
\cite{hubert07,hubert07b}  so as to compute global invariants
 through algebraic elimination. 
We accordingly avoid introducing radicals by
giving the algebraic combinations of 
 $\{\inv x_3,\inv u_{001},\inv u_{010}\}$ 
that are global invariants.
The actual expression for $\{\inv x_3,\inv u_{001},\inv u_{010}\}$, 
with sign determination, should be deduced from those according 
to the point of the cross-section in the neighborhood of which 
we wish to work. In this neighbourhood $\inv f$ and $f$ must 
agree on the cross-section.

{ \small
\[\begin{array}{l}
{{ \inv x_3}}^{2}  =
{{ x_3}}^{2}+\epsilon\,{{ x_1}}^{2}+\epsilon\,{{ x_2}}^{2},
\\
{ \inv x_3}\,{ \inv u_{001}}
={ x_3}u_{001}+{ x_1}u_{100}+{ x_2}u_{010},
\\
{{ \inv x_3}}^{2}{{ \inv u_{010}}}^{2}
 =
  {{ x_1}}^{2}\left( {u_{001}}^{2}+\epsilon\,{u_{010}}^{2} \right) 
+ {{ x_2}}^{2}\left( {u_{001}}^{2}+\epsilon\,{u_{100}}^{2} \right) 
+  {{ x_3}}^{2} \left( {u_{100}}^{2}+{u_{010}}^{2} \right)
\\ \hspace{5mm}
-2\,\epsilon\,{ x_3}{ x_1}u_{100}u_{001}
-2\,\epsilon\,{ x_3}{ x_2}u_{001}u_{010}
-2\,\epsilon\,{ x_1}{ x_2}\,u_{100}\,u_{010}
,
\\
{{ \inv x_3}}^{2}{{ \inv u_{010}}}^{2}{ \inv u_{200}}
 =
 {{ x_1}}^{2}\left( {u_{010}}^{2}u_{002}
        +{u_{001}}^{2}u_{020}-2\,u_{001}u_{011}u_{010} \right) 
\\ \hspace{5mm}
+{{ x_2}}^{2}\left( u_{002}{u_{100}}^{2}
      -2\,u_{001}u_{101}u_{100}+u_{200}{u_{001}}^{2} \right) 
+  {{ x_3}}^{2} \left( u_{200}{u_{010}}^{2}+u_{020}{u_{100}}^{2}
        -2\,u_{100}u_{110}u_{010} \right) 
\\ \hspace{5mm}
-2\,\epsilon\, { x_3}{ x_1} \left( u_{100}u_{001}u_{020}
    +u_{101}{u_{010}}^{2}-u_{110}u_{001}u_{010}
     -u_{100}u_{010}u_{011} \right) 
\\ \hspace{5mm}
-2\,\epsilon\, { x_3}{ x_2} \left( u_{011}{u_{100}}^{2}-u_{010}u_{101}u_{100}
   +u_{200}u_{001}u_{010}-u_{100}u_{110}u_{001} \right)
\\ \hspace{5mm}
+  { x_2}{ x_1} \left( 2\,u_{100}u_{011}u_{001}
  +2\,u_{010}u_{101}u_{001}-2\,u_{002}u_{100}u_{010} 
   -2\,{u_{001}}^{2}u_{110} \right) 
,
\\
{{ \inv x_3}}^{2}{ \inv u_{010}}\,{ \inv u_{101}}
 =
\epsilon\,{{ x_1}}^{2} \left(- u_{101}u_{010}+u_{110}u_{001} \right) 
-\epsilon\, {{ x_2}}^{2} \left(u_{110}u_{001} -u_{011}u_{100} \right) 
+ {{ x_3}}^{2}\, \left( u_{101}u_{010}-u_{011}u_{100} \right) 
\\ \hspace{5mm} 
+{ x_3}\,{ x_1} \left( u_{200}u_{010}-\epsilon\,u_{002}u_{010}
    +\epsilon\,u_{011}u_{001}-u_{110}u_{100} \right) 
\\ \hspace{5mm}
+ { x_3}\,{ x_2} \left(u_{110}u_{010} -u_{020}u_{100}
    -\epsilon\,u_{101}u_{001}+\epsilon\,u_{002}u_{100} \right)
\\ \hspace{5mm}
-\epsilon\,  { x_2}\,{ x_1} \left( u_{200}u_{001}+u_{011}u_{010}-u_{020}u_{001}-u_{101}u_{100} \right) 
\end{array}\]
} 

\subsubsection{Invariant derivations}

We can obtain an expression, depending on $\epsilon$ 
for the moving frame of both the action of $O(3,0)$ and $O(2,1)$. 
It is given by the matrix $A_\epsilon$ below
 
\[\left( \begin {array}{ccc}
\ds {\frac {{ x_3}u_{010}-\epsilon\,{ x_2}u_{001}}
         {{ \inv u_{010}}\,{ \inv x_3}}}
&\ds 
{\frac {\epsilon { x_1}u_{001}-{ x_3}u_{100}}
         {{ \inv u_{010}}\,{ \inv x_3}}}
&\ds 
{\frac {{ x_2}\,u_{100}-{ x_1}u_{010}}
         {{ \inv u_{010}}\,{ \inv x_3}}}
\\\noalign{\medskip}
a_\epsilon & b_\epsilon & c_\epsilon 
\\\noalign{\medskip}\ds 
{\frac {{ \epsilon x_1}}{{ \inv x_3}}}
&\ds 
{\frac {{ \epsilon x_2}}{{ \inv x_3}}}
&\ds
{\frac {{ x_3}}{{ \inv x_3}}}
\end {array} \right) ,
\]
where
\[
\begin{array}{l}
\ds a_\epsilon = 
{\frac {
   ({{ x_3}}^{2}+\epsilon{{ x_2}}^{2})u_{100}
  -\epsilon{ x_1}\,{ x_2}\,u_{010}-\epsilon{ x_1}\,{ x_3}u_{001}}
 {{{ \inv x_3}}^{2}{ \inv u_{010}}}}  
\\
\ds b_\epsilon = {\frac { ({{ x_3}}^{2}+\epsilon {{ x_1}}^{2})u_{010}
 - \epsilon { x_2}{ x_3}\,u_{001} -\epsilon\,{ x_1}{ x_2}u_{100}
}{{{ \inv x_3}}^{2}{ \inv u_{010}}}}
\\
\ds c_\epsilon = {
\frac {  
   \epsilon\,({{ x_2}}^{2}+{{ x_1}}^{2})u_{001}
  -{ x_3}\,{ x_1}u_{100} -{ x_3}\,{ x_2}\,u_{010}
}
{{{ \inv x_3}}^{2}{ \inv u_{010}}}}.
\end{array}
\]
When $\epsilon=1$ the matrix belongs to $O(3,0)$
 so that $A_{1}^{-1}=A_1^T$.
The invariant derivations are then given by $\iD = A_1 \D$.
When $\epsilon=-1$ the matrix belongs to $O(2,1)$ and the invariant derivations are then given by $\iD = A_{-1}^{-T} \D$.

    \subsection{Affine action of $O(3-l,l)\ltimes \Ri^3$ 
              on the independent variables}
         \label{E3l}       

The action of $E(3)=O(3,0)\ltimes \Ri^3$ 
was considered by \citet{mansfield01}
in the context of the  symmetry reduction for a
 differential elimination problem.

We show here that the differential invariants of 
second order are generating.
We provide a complete set of syzygies for a generating set of three
second order differential invariants. 

\subsubsection{Action}

Compared with the action of $O(3-l,l)$ treated above, we have
additionally translation. The infinitesimal generators are then:

\[
\begin{array}{c} 
\ds \vf[0]_1= x_2\,\pdif{ }{x_1} -x_1\,\pdif{ }{x_2},  \;
\ds \vf[0]_2= x_3\,\pdif{ }{x_1} -\epsilon\,x_1\, \pdif{ }{x_3},\; 
\ds \vf[0]_3= x_3\,\pdif{ }{x_2}- \epsilon\,x_2\, \pdif{ }{x_3},\;
\\
\ds \vf[0]_4= \pdif{ }{x_1}, \;  
 \vf[0]_5= \pdif{ }{x_2},\; 
 \vf[0]_6= \pdif{ }{x_3}.
\end{array}
\]

\subsubsection{Generators}

We choose the minimal order cross-section 
$$ \sva:\; x_1=0,\,x_2=0,x_3=0,\,u_{100}=0,\,u_{010}=0,\,u_{110}=0. $$
The edge invariants
$\moinv =\{ \inv u_{000},\inv u_{200},\inv u_{020},\inv u_{101},\inv u_{011},
 \,\inv u_{210},\,\inv u_{120},\,\inv u_{111}\}$ thus form a generating set 
(\theor{minorder}). 
Furthermore, since  $u_{000}$ is invariant, and therefore $\vf(u_{000})=0$,
we know from \theor{recurrences}  that
$\inv u_{001}=\ider[3] (\inv u_{000})$. 
It will be convenient to use the following set of generators:
\[ \begin{array}{c}
\sigma = \inv u_{000},\;
\ds \phi=\epsilon\,{\frac {\inv u_{200}}{\inv u_{001}}},\;
\psi=\epsilon\,{\frac {\inv u_{020}}{\inv u_{001}}},\;
\kappa={\frac {\inv u_{101}}{\inv u_{001}}},\;
\tau={\frac {\inv u_{011}}{\inv u_{001}}},
\\ \\
\ds \Gamma={\frac {\inv u_{200}\inv u_{011}-\inv u_{210}\inv u_{001}}
        {\inv u_{001} \left( \inv u_{200}-\inv u_{020} \right) }},\;
\Lambda={\frac {\inv u_{020}\inv u_{101}-\inv u_{120}\inv u_{001}}
        {\inv u_{001} \left( \inv u_{200}-\inv u_{020} \right) }},\;
\ds \Omega={\frac{2\,\inv u_{011}\inv u_{101}-\inv u_{111}\inv u_{001}}
        {\inv u_{001} \left( \inv u_{200}-\inv u_{020}\right) }},
\end{array}\]
as we can then write the Maurer-Cartan matrix $K$ of \theor{recurrences} as
\[ K=
\left( \begin {array}{cccccc} \Gamma&\phi&0&1&0&0
\\
\Lambda&0&\psi&0&1&0
\\
\Omega&\epsilon\,\kappa&
\epsilon\,\tau&0&0&1\end {array} \right).
\]
We then deduce from \propr{commuterules} that:
\[ \begin{array}{c}
[\ider[1],\ider[2]]=\Gamma\,\ider[1]+\Lambda\,\ider[2],
\quad
[\ider[1],\ider[3]]=\phi\,\ider[1]+\Omega\,\ider[2]-\kappa\,\ider[3],
\\ \hbox{}
[\ider[3],\ider[2]]=\Omega\,\ider[1]-\psi\,\ider[2]-\tau\,\ider[3]
.
\end{array}
\]

\subsubsection{Syzygies}

According to \theor{cnisyz} a complete set of syzygies for $\ninv[2]$
 is given by:

$$\Syzr\, \left\{\quad \inv x_1=0,\quad\inv x_2=0,\quad
\inv x_3=0,\quad\inv u_{100}=0,\quad\inv \inv u_{010}=0,\quad\inv u_{110}=0,
\right.
$$

{\small 
\[\Syzs \left\{\begin{array}{l}
\ider[1](\inv u_{000})=0
,
\ider[2](\inv u_{000})=0
,
\ider[3](\inv u_{000})=\inv u_{00}
,
\\
\ider[1](\inv u_{001})=\inv u_{101}
,
\ider[2](\inv u_{001})=\inv u_{011}
,
\ider[3](\inv u_{001})=\inv u_{002}
,
\\
\ider[1](\inv u_{200})=\inv u_{300}-2\,\phi\,\epsilon\,\inv u_{101}
,
\ider[2](\inv u_{200})=\inv u_{210}
,
\ider[3](\inv u_{200})=\inv u_{201}-2\,\kappa\,{\epsilon}^{2}\inv u_{101}
,
\\
\ider[1](\inv u_{101})= \inv u_{201}-\Gamma\,\inv u_{011}
        +\phi\,\inv u_{200}-\phi\,\epsilon\,\inv u_{002}
,
\ider[2](\inv u_{101})=\inv u_{111}-\Lambda\,\inv u_{011}
,
\\ \hspace{15mm}
\ider[3](\inv u_{101})=\inv u_{102}-\Omega\,\inv u_{011}+\kappa\,\epsilon
      \,\inv u_{200}-\kappa\,{\epsilon}^{2}\inv u_{002}
,
\\
\ider[1](\inv u_{020})=\inv u_{120}
,
\ider[2](\inv u_{020})=\inv u_{030}-2\,\psi\,\epsilon\,\inv u_{011}
,
\ider[3](\inv u_{020})=+\inv u_{021}-2\,\tau\,{\epsilon}^{2}\inv u_{011}
,
\\
\ider[1](\inv u_{011})=\inv u_{111}+\Gamma\,\inv u_{101}
,
\ider[2](\inv u_{011})=\inv u_{021}+\Lambda\,\inv u_{101}
       +\psi\,\inv u_{020}-\psi\,\epsilon\,\inv u_{002}
,
\\ \hspace{15mm}
\ider[3](\inv u_{011})=\inv u_{012}+\Omega\,\inv u_{101}
    +\tau\,\epsilon\,\inv u_{020}-\tau\,{\epsilon}^{2}\inv u_{002}
,
\\
\ider[1](\inv u_{002})=+\inv u_{102}+2\,\phi\,\inv u_{101}
,
\ider[2](\inv u_{002})=+\inv u_{012}+2\,\psi\,\inv u_{011}
,
\\ \hspace{15mm}
\ider[3](\inv u_{002})=+\inv u_{003}+2\,\kappa\,\epsilon\,\inv u_{101}
    +2\,\tau\,\epsilon\,\inv u_{011}
,
\end{array} \right.\]
}
and 
{\small 
\[ \Syzt \left\{ \begin{array}{l}
\ider[3](\inv u_{012})-\ider[2](\inv u_{003})=
\Omega\,\inv u_{102}+2\,\kappa\,\epsilon\,\inv u_{111}
+2\,\tau\,\epsilon\,\inv u_{021}
-\tau\,{\epsilon}^{2}\inv u_{003}-3\,\psi\,\inv u_{012},
\\
\ider[3](\inv u_{021})-\ider[2](\inv u_{012})=
2\,\Omega\,\inv u_{111}+\kappa\,\epsilon\,\inv u_{120}
+\tau\,\epsilon\,\inv u_{030}-2\,\tau\,{\epsilon}^{2}\inv u_{012}
-\Lambda\,\inv u_{102}
\\ \hspace{35mm}
-2\,\psi\,\inv u_{021} +\psi\,\epsilon\,\inv u_{003},
\\
\ider[3](\inv u_{030})-\ider[2](\inv u_{021})=
3\,\Omega\,\inv u_{120}-3\,\tau\,{\epsilon}^{2}\inv u_{021}
-2\,\Lambda\,\inv u_{111}-\psi\,\inv u_{030}
+2\,\psi\,\epsilon\,\inv u_{012},
\\
\ider[2](\inv u_{102})-\ider[1](\inv u_{012})=
+2\,\psi\,\inv u_{111}-\Lambda\,\inv u_{012}
-\Gamma\,\inv u_{102}-2\,\phi\,\inv u_{111},
\\
\ider[3](\inv u_{102})-\ider[1](\inv u_{003})=
-\Omega\,\inv u_{012}+2\,\kappa\,\epsilon\,\inv u_{201}
-\kappa\,{\epsilon}^{2}\inv u_{003}
+2\,\tau\,\epsilon\,\inv u_{111}-3\,\phi\,\inv u_{102},
\\
\ider[2](\inv u_{111})-\ider[1](\inv u_{021})=
\Lambda\,\inv u_{201}-\Lambda\,\inv u_{021}
+\psi\,\inv u_{120}-\psi\,\epsilon\,\inv u_{102}
-2\,\Gamma\,\inv u_{111}-\phi\,\inv u_{120},
\\
\ider[3](\inv u_{111})-\ider[1](\inv u_{012})=
\Omega\,\inv u_{201}-\Omega\,\inv u_{021}
+\kappa\,\epsilon\,\inv u_{210}
-\kappa\,{\epsilon}^{2}\inv u_{012}
+\tau\,\epsilon\,\inv u_{120}
\\ \hspace{35mm}
  -\tau\,{\epsilon}^{2}\inv u_{102}-\Gamma\,\inv u_{102}-2\,\phi\,\inv u_{111},
\\
\ider[2](\inv u_{120})-\ider[1](\inv u_{030})=
2\,\Lambda\,\inv u_{210}-\Lambda\,\inv u_{030}
-2\,\psi\,\epsilon\,\inv u_{111}-3\,\Gamma\,\inv u_{120},
\\
\ider[3](\inv u_{120})-\ider[1](\inv u_{021})=
2\,\Omega\,\inv u_{210}-\Omega\,\inv u_{030}
-\kappa\,{\epsilon}^{2}\inv u_{021}
-2\,\tau\,{\epsilon}^{2}\inv u_{111}-2\,\Gamma\,\inv u_{111}
-\phi\,\inv u_{120},
\\
\ider[2](\inv u_{201})-\ider[1](\inv u_{111})=
\psi\,\inv u_{210}-2\,\Lambda\,\inv u_{111}
-\Gamma\,\inv u_{201}+\Gamma\,\inv u_{021}
-\phi\,\inv u_{210}+\phi\,
\epsilon\,\inv u_{012},
\\
\ider[3](\inv u_{201})-\ider[1](\inv u_{102})=
\kappa\,\epsilon\,\inv u_{300}-2\,\Omega\,\inv u_{111}
-2\,\kappa\,{\epsilon}^{2}\inv u_{102}
+\tau\,\epsilon\,\inv u_{210}+\Gamma\,\inv u_{012}
\\ \hspace{35mm}
-2\,\phi\,\inv u_{201}+\phi\,\epsilon\,\inv u_{003},
\\
\ider[2](\inv u_{210})-\ider[1](\inv u_{120})=
\Lambda\,\inv u_{300}-2\,\Lambda\,\inv u_{120}
-\psi\,\epsilon\,\inv u_{201}
-2\,\Gamma\,\inv u_{210}+\Gamma\,\inv u_{030}+\phi\,\epsilon\,\inv u_{021},
\\
\ider[3](\inv u_{210})-\ider[1](\inv u_{111})=
\Omega\,\inv u_{300}-2\,\Omega\,\inv u_{120}
-2\,\kappa\,{\epsilon}^{2}\inv u_{111}
-\tau\,{\epsilon}^{2}\inv u_{201}
\\ \hspace{35mm}
-\Gamma\,\inv u_{201}
+\Gamma\,\inv u_{021}-\phi\,\inv u_{210}
+\phi\,\epsilon\,\inv u_{012},
\\
\ider[2](\inv u_{300})-\ider[1](\inv u_{210})=
2\,\Gamma\,\inv u_{120}+2\,\phi\,\epsilon\,\inv u_{111}
-3\,\Lambda\,\inv u_{210}-\Gamma\,\inv u_{300},
\\
\ider[3](\inv u_{300})-\ider[1](\inv u_{201})=
2\,\phi\,\epsilon\,\inv u_{102}
-3\,\Omega\,\inv u_{210}-3\,\kappa\,{\epsilon}^{2}\inv u_{201}
+2\,\Gamma\,\inv u_{111}-\phi\,\inv u_{300}.
\end{array} \right.
\]
}
By differential elimination 
we can rewrite any normalized invariants of order 2 and less in terms of 
$\{\Omega,\Lambda,\Gamma,\kappa,\phi,\tau,\psi,\sigma\}$ and find a
complete set of syzygies for those. 
It is nonetheless easier to obtain 
the syzygies for the Maurer-Cartan invariants
$\{\Omega,\Lambda,\Gamma,\kappa,\phi,\tau,\psi\}$ 
 from the structure equations  \citep{mansfield06,hubert07m}.
The first part
\[ \begin{array}{c}
\inv u_{001} =\ider[3](\sigma)
\\
\inv u_{200} =
\phi \epsilon \ider[3](\sigma),
\;
\inv u_{101} =\kappa \ider[3](\sigma),
\;
\inv u_{020} =\psi \epsilon \ider[3](\sigma),
\;
\inv u_{011} =\tau \ider[3](\sigma),
\;
\inv u_{002} =\ider[3]^2(\sigma),
\\
\inv u_{300} =\epsilon \ider[3](\sigma) 
   \left( \ider[1](\phi)+3\,\phi \kappa  \right) ,
\;
\inv u_{210} =\epsilon \ider[3](\sigma) 
   \left( \Gamma \psi -\Gamma \phi +  \tau \phi \right) ,
\\
\inv u_{120} =\epsilon \ider[3](\sigma)
   \left(\kappa \psi +\Lambda \psi -\Lambda \phi  \right) ,
\;
\inv u_{111} =\ider[3](\sigma) 
   \left(  2\,\kappa \tau-\epsilon \Omega \phi+\epsilon \Omega \psi 
     \right) ,
\\
\inv u_{102} =2\,\kappa \ider[3]^2(\sigma)
  +\ider[3](\kappa)\ider[3](\sigma)
  +\Omega \tau \ider[3](\sigma)
  -\phi \kappa \ider[3](\sigma),
\\
\inv u_{201} =
  \phi \epsilon \ider[3]^2(\sigma)
  +\ider[3](\sigma)\ider[1](\kappa)
  -\epsilon {\phi }^{2} \ider[3](\sigma)
  +\Gamma \tau \ider[3](\sigma) 
  +{\kappa }^{2}\ider[3](\sigma),
\\
\inv u_{030} =\epsilon \ider[3](\sigma) 
   \left( \ider[2](\psi)+3\,\psi \tau  \right) ,
\\
\inv u_{021} =  \epsilon \psi \ider[3]^2(\sigma)
   +\ider[3](\sigma)\ider[2](\tau)
   -\epsilon {\psi }^{2} \ider[3](\sigma)
   -\Lambda \kappa \ider[3](\sigma)
   +{\tau }^{2}\ider[3](\sigma),
\\
\inv u_{012} =2\,\tau \ider[3]^2(\sigma)
    +\ider[3](\tau)\ider[3](\sigma)
   -\Omega \kappa \ider[3](\sigma)
   -\psi \tau \ider[3](\sigma)
,
\\
\inv u_{003} =\ider[3]^3(\sigma)-2\,\epsilon 
   {\kappa }^{2}\ider[3](\sigma)-2\,\epsilon {\tau }^{2}
   \ider[3](\sigma),

\end{array}\]
allows us to rewrite any other differential invariants in terms of 
$\{\Omega,\Lambda,\Gamma,\kappa,\phi,\tau,\psi,\sigma\}$. 
The second part consist of a complete set of syzygies for 
$\{\Omega,\Lambda,\Gamma,\kappa,\phi,\tau,\psi,\sigma\}$:
\[\begin{array}{c}
\ider[2](\Gamma)-\ider[1](\Lambda)=
 -\epsilon\, \phi\, \psi -{\Gamma }^{2}-{\Lambda }^{2},
\\
\ider[1](\Omega)-\ider[3](\Gamma)=
 \Omega\, \Lambda+ \phi\,\Gamma   +\kappa \,\Omega +\phi\, \tau ,
\\
\ider[3](\Lambda)-\ider[2](\Omega)=
\Omega \,\Gamma - \psi\, \Lambda -\tau\, \Omega +\kappa\, \psi ,
\\
\ider[2](\phi)=
 \left( \psi -\phi  \right)\, \Gamma  ,
\qquad
\ider[1](\psi)=  \left( \psi -\phi  \right) \,\Lambda ,
\\
\ider[2](\kappa)=\epsilon  (\psi - \phi)\, \Omega
-\tau\, \Lambda +\kappa \,\tau ,
\quad
\ider[1](\tau)=\epsilon \, (\psi - \phi)\, \Omega 
+\kappa\, \Gamma +\kappa\, \tau  ,
\\
\ider[3](\phi)-\epsilon \,\ider[1](\kappa)=
\epsilon\,  \tau\, \Gamma
-\epsilon \,{\kappa }^{2}-{\phi }^{2},
\\
 \ider[2](\tau)-\epsilon\,\ider[3](\psi)=
\kappa\, \Lambda  + {\tau }^{2}+\epsilon\,{\psi }^{2},

\\
\ider[1](\sigma)=0,\; \ider[2](\sigma)=0.
\end{array}\]
We see that we can actually write the third order differential invariants
$\{\Gamma,\Omega,\Lambda\}$ in terms of the second order
differential invariants
$\{\phi,\psi,\kappa,\tau\}$, so that this latter is already a generating set.

\[\begin{array}{c}
\ds \Gamma =
{\frac {\ider[2](\phi)}{\psi-\phi }},
\qquad
\Lambda ={\frac {\ider[1](\psi)}{\psi-\phi }},
\\ \\
\ds  \Omega = 
\epsilon\,{\frac {\tau\,\ider[1](\psi) }
        { \left( \psi-\phi \right) ^{2}}}
+\epsilon\, {\frac { \ider[2](\kappa)}
         { \psi-\phi }}
-\epsilon{\frac {\tau\,\kappa}
        { \psi-\phi  }}
\end{array}\]

\sil{
\[ \begin{array}{c}
 \Omega =
{\frac {\epsilon \ider[1](\kappa)\ider[1](\psi)}{
 \left( \phi-\psi  \right)
 \ider[2](\phi)}}
-{\frac {\ider[1](\psi)\ider[3](\phi)}
 { \left( \phi-\psi  \right) 
 {\ider[2](\phi)}}
-{\frac { \left( {\phi }^{2}
+\epsilon {\kappa }^{2} \right) \ider[1](\psi)}
{ \left( \phi-\psi  \right) 
\ider[2](\phi)}}-{\frac {\kappa \ider[3](\phi)}
{\ider[2](\phi)}}
-{\frac {\epsilon  \ider[2](\kappa)}{
\left( \phi-\psi  \right) }}
+{\frac {\epsilon \kappa \ider[1](\kappa)}{\ider[2](\phi)}}
-{\frac { \left( {\phi }^{2}
+\epsilon {\kappa }^{2} \right) \kappa }
{\ider[2](\phi)}}}.
\\
\tau ={\frac {\epsilon \left(\psi-\phi \right) \ider[3](\phi)}{ \ider[2](\phi)}}
+{\frac { \left( \phi-\psi  \right) \ider[1](\kappa)}
{\ider[2](\phi)}}
+{\frac { \left( {\phi }^{2}+\epsilon {\kappa }^{2} \right) 
 \left(\psi - \phi \right) }{\epsilon \ider[2](\phi)}}.
\end{array}\]}

The coefficient of the commutation rules can now be expressed in terms
of the first order derivatives of $\{\phi,\psi,\kappa,\tau\}$. We can
therefore still apply the differential elimination of \citet{hubert05} to
obtain a complete set of syzygies on the generating set
$\{\phi,\psi,\kappa,\tau\}$. We obtain: 
\[\begin{array}{c}
\ds \ider[2](\tau)-\epsilon\,\ider[3](\psi)
=
\epsilon\,{\psi}^{2}+{\tau}^{2}
+ {\frac {\kappa}{\psi-\phi}} \,\ider[1](\psi)
,
\\
\ds \ider[1](\kappa)-\epsilon\,\ider[3](\phi)
=
\epsilon\,{\phi}^{2}+{\kappa}^{2}-{\frac {\tau}{\psi-\phi}}\, \ider[2](\phi)
,
\\
\ds \ider[1](\tau) -\ider[2](\kappa)=
\tau\,{\frac {\ider[1](\psi)}{\psi-\phi}}
+ \kappa\,{\frac {\ider[2](\phi)}{\psi-\phi}}
,
\end{array}
\]
\[\begin{array}{l}
\ds \ider[2]^2(\phi)-\ider[1]^2(\psi)=
{\frac {\ider[1](\psi)\,\ider[1](\phi)}{\psi-\phi}}
+{\frac {\ider[2](\phi)\,\ider[2](\psi)}{\psi-\phi}}
-2\,{\frac {{\ider[1](\psi)}^{2}}{\psi-\phi}}
-2\,{\frac {{\ider[2](\phi)}^{2}}{\psi-\phi}}
-  \epsilon\left( \psi-\phi \right)\phi\psi,
\\
\ds
\ider[1]^2(\tau)- \epsilon \ider[2]\ider[3](\phi)
=
\kappa\,{\frac { \ider[1]\ider[2](\phi)}{\psi-\phi}}
+2\,{\frac {\ider[1](\tau)\ider[1](\psi)}{\psi-\phi}}
+2\,\epsilon{\frac {\ider[3](\phi)\ider[2](\phi)}{\psi-\phi}}
-\epsilon{\frac {\ider[3](\psi)\ider[2](\phi)}{\psi-\phi}}
\\ \ds \hspace{1cm} 
+\kappa \,
   {\frac {\ider[1](\phi)\ider[2](\phi)}{\left( \psi-\phi \right) ^{2}}}
-3\,\kappa \, 
   \frac {\ider[1](\psi)\ider[2](\phi)}{ \left( \psi-\phi \right) ^{2}}
-\tau \, 
    \frac {{\ider[2](\phi)}^{2}}{ \left( \psi-\phi \right) ^{2}}
\\\ds \hspace{1cm} 
-\tau^2
   \frac{ \ider[2](\phi)}{\psi-\phi}
-2\,\tau\,\kappa\,
   {\frac {\ider[1](\psi)}{\psi-\phi}}
-\epsilon\, \psi\left({\psi}-2\,\phi \right)\, 
   {\frac { \ider[2](\phi)}{\psi-\phi}}
+2\,\kappa \, \ider[1](\tau)
+\epsilon\,\tau\,\phi\,\psi
.
\end{array}\]

From the first equation we see that  $\kappa$ can be written in terms of
$\{\phi,\psi, \tau\}$. Substituting the expression for $\kappa$ in the other
three equations we obtain a complete set of syzygies for those. As the
expression grow considerably we do not give them explicitly here.

We can actually compute the expressions for the normalized 
second order differential invariants by algebraic elimination
(\cite{hubert07,hubert07b}). 
Alternatively
\citet{fushchich92,xu98} provided a functionally independent set
of second order differential invariants for this action. They can be
easily rewritten in terms of $\ninv[2]$. With additional manipulation
we can then find the expression for our generating set.

\section*{Acknowledgment}

I am indebted to Elizabeth Mansfield for both introducing
 me to the subject of differential invariants 
and her continuous support.
I have interacted very openly on the specific problem 
of syzygies with Irina Kogan, at the occasion of mutual visits,
and  Peter Olver, in particular during my visit to
the Institute for Mathematics and its 
Applications at the University of Minnesota during the
 thematic year on \emph{Applications of Algebraic Geometry}.
Those interactions have been very influential and
 for this I am grateful.
I would also like to thank the referees, as well as G. Labahn,  P. Olver, and 
F. Valiquette for corrections and pertinent suggestions on the manuscript.

\bibliographystyle{elsart-harv}

\begin{thebibliography}{48}
\expandafter\ifx\csname natexlab\endcsname\relax\def\natexlab#1{#1}\fi
\expandafter\ifx\csname url\endcsname\relax
  \def\url#1{\texttt{#1}}\fi
\expandafter\ifx\csname urlprefix\endcsname\relax\def\urlprefix{URL }\fi

\bibitem[{Anderson and et~al.(2007)}]{vessiot}
Anderson, I., et~al., 2007. The Maple11 library DifferentialGeometry (formerly
  Vessiot). Utah State University.

\bibitem[{Berger and Gostiaux(1988)}]{berger88}
Berger, M., Gostiaux, B., 1988. Differential geometry: manifolds, curves, and
  surfaces. Vol. 115 of Graduate Texts in Mathematics. Springer-Verlag, New
  York, translated from the French by Silvio Levy.

\bibitem[{Boulier and Hubert(1998)}]{diffalg}
Boulier, F., Hubert, E., 1998. {\textsc{diffalg}: description, help pages and
  examples of use}. Symbolic Computation Group, University of Waterloo,
  Ontario, Canada, \url{www.inria.fr/cafe/Evelyne.Hubert/diffalg}.

\bibitem[{Bourbaki(1967)}]{bourbaki67}
Bourbaki, N., 1967. \'{E}l\'ements de math\'ematique. {F}asc. {XXXIII}.
  {V}ari\'et\'es diff\'erentielles et analytiques. {F}ascicule de r\'esultats
  ({P}aragraphes 1 \`a 7). Actualit\'es Scientifiques et Industrielles, No.
  1333. Hermann, Paris.

\bibitem[{Cartan(1935)}]{cartan35}
Cartan, E., 1935. La m\'{e}thode du rep\`{e}re mobile, la th\'{e}orie des
  groupes continus, et les espaces g\'{e}n\'{e}ralis\'{e}s. Vol.~5 of
  Expos\'{e}s de G\'{e}om\'{e}trie. Hermann, Paris.

\bibitem[{Cartan(1937)}]{cartan37}
Cartan, E., 1937. La th\'eorie des groupes finis et continus et la
  g\'eom\'etrie diff\'erentielle trait\'ees par la m\'ethode du rep\`ere
  mobile. No.~18 in Cahiers scientifiques. Gauthier-Villars.

\bibitem[{Cartan(1953)}]{cartan53}
Cartan, E., 1953. {\OE}uvres compl\`etes. {P}artie {II}. {V}ol. 1. {A}lg\`ebre,
  formes diff\'erentielles, syst\`emes diff\'erentiels. {V}ol. 2. {G}roupes
  infinis, syst\`emes diff\'erentiels, th\'eories d'\'equivalence.
  Gauthier-Villars, Paris.

\bibitem[{Dridi and Neut(2006{\natexlab{a}})}]{dridi06}
Dridi, R., Neut, S., 2006{\natexlab{a}}. The equivalence problem for fourth
  order differential equations under fiber preserving diffeomorphisms. J. Math.
  Phys. 47~(1), 013501, 6.

\bibitem[{Dridi and Neut(2006{\natexlab{b}})}]{dridi06a}
Dridi, R., Neut, S., 2006{\natexlab{b}}. On the geometry of {$y\sp
  {(4)}=f(x,y,y',y'',y''')$}. J. Math. Anal. Appl. 323~(2), 1311--1317.

\bibitem[{Fels and Olver(1999)}]{fels99}
Fels, M., Olver, P.~J., 1999. Moving coframes. {I}{I}. {R}egularization and
  theoretical foundations. Acta Appl. Math. 55~(2), 127--208.

\bibitem[{Fushchich and Yegorchenko(1992)}]{fushchich92}
Fushchich, W.~I., Yegorchenko, I.~A., 1992. Second-order differential
  invariants of the rotation group {${\rm O}(n)$} and of its extensions:
  {$E(n),\;P(1,n),\;G(1,n)$}. Acta Appl. Math. 28~(1), 69--92.

\bibitem[{Gardner(1989)}]{gardner89}
Gardner, R.~B., 1989. The method of equivalence and its applications. SIAM,
  Philadelphia.

\bibitem[{Green(1978)}]{green78}
Green, M.~L., 1978. The moving frame, differential invariants and rigidity
  theorems for curves in homogeneous spaces. Duke Math. Journal 45, 735--779.

\bibitem[{Griffiths(1974)}]{griffiths74}
Griffiths, P.~A., 1974. On {C}artan's method of {L}ie groups as applied to
  uniqueness and existence questions in differential geometry. Duke Math.
  Journal 41, 775--814.

\bibitem[{Hubert(2003)}]{hubert03d}
Hubert, E., 2003. Notes on triangular sets and triangulation-decomposition
  algorithms {II}: {D}ifferential systems. In: Winkler, F., Langer, U. (Eds.),
  Symbolic and Numerical Scientific Computing. No. 2630 in Lecture Notes in
  Computer Science. Springer Verlag Heidelberg, pp. 40--87.

\bibitem[{Hubert(2005{\natexlab{a}})}]{ncdiffalg}
Hubert, E., 2005{\natexlab{a}}. {diffalg}: extension to non commuting
  derivations. INRIA, Sophia Antipolis.

\bibitem[{Hubert(2005{\natexlab{b}})}]{hubert05}
Hubert, E., 2005{\natexlab{b}}. Differential algebra for derivations with
  nontrivial commutation rules. Journal of Pure and Applied Algebra 200~(1-2),
  163--190.

\bibitem[{Hubert(2007{\natexlab{a}})}]{hubert07m}
Hubert, E., 2007{\natexlab{a}}. Generation properties of {Maurer-Cartan}
  invariants, \url{http://hal.inria.fr/inria-00194528}.

\bibitem[{Hubert(2007{\natexlab{b}})}]{aida}
Hubert, E., 2007{\natexlab{b}}. The \textsc{maple} package \textsc{aida} -
  Algebraic Invariants and their Differential Algebra. INRIA.

\bibitem[{Hubert(2008)}]{hubert08a}
Hubert, E., 2008. Algebra of differential invariants, in preparation.

\bibitem[{Hubert and Kogan(2007{\natexlab{a}})}]{hubert07}
Hubert, E., Kogan, I.~A., 2007{\natexlab{a}}. Rational invariants of a group
  action. {C}onstruction and rewriting. Journal of Symbolic Computation
  42~(1-2), 203--217.

\bibitem[{Hubert and Kogan(2007{\natexlab{b}})}]{hubert07b}
Hubert, E., Kogan, I.~A., 2007{\natexlab{b}}. Smooth and algebraic invariants
  of a group action. {L}ocal and global constructions. Foundations of
  Computational Mathematics 7~(4).

\bibitem[{Hubert and Olver(2007)}]{hubert07c}
Hubert, E., Olver, P.~J., 2007. Differential invariants of conformal and
  projective surfaces. Symmetry Integrability and Geometry: Methods and
  Applications 3~(097).

\bibitem[{Ivey and Landsberg(2003)}]{ivey03}
Ivey, T.~A., Landsberg, J.~M., 2003. Cartan for beginners: differential
  geometry via moving frames and exterior differential systems. Vol.~61 of
  Graduate Studies in Mathematics. American Mathematical Society, Providence,
  RI.

\bibitem[{Janet(1929)}]{janet29}
Janet, M., 1929. Sur les syst\`emes d'\'equations aux d\'eriv\'ees paritelles.
  Gauthier-Villars.

\bibitem[{Jensen(1977)}]{jensen77}
Jensen, G.~R., 1977. Higher order contact of submanifolds of homogeneous
  spaces. Springer-Verlag, Berlin, lecture Notes in Mathematics, Vol. 610.

\bibitem[{Kolchin(1973)}]{kolchin}
Kolchin, E.~R., 1973. Differential Algebra and Algebraic Groups. Vol.~54 of
  Pure and Applied Mathematics. Academic Press.

\bibitem[{Kumpera(1974)}]{kumpera74}
Kumpera, A., 1974. Invariants diff\'erentiels d'un pseudogroupe de {L}ie. In:
  G\'eom\'etrie diff\'erentielle (Colloq., Univ. Santiago de Compostela,
  Santiago de Compostela, 1972). Springer, Berlin, pp. 121--162. Lecture Notes
  in Math., Vol. 392.

\bibitem[{Kumpera(1975{\natexlab{a}})}]{kumpera75a}
Kumpera, A., 1975{\natexlab{a}}. Invariants diff\'erentiels d'un pseudogroupe
  de {L}ie. {I}. J. Differential Geometry 10~(2), 289--345.

\bibitem[{Kumpera(1975{\natexlab{b}})}]{kumpera75b}
Kumpera, A., 1975{\natexlab{b}}. Invariants diff\'erentiels d'un pseudogroupe
  de {L}ie. {II}. J. Differential Geometry 10~(3), 347--416.

\bibitem[{Mansfield(2001)}]{mansfield01}
Mansfield, E.~L., 2001. Algorithms for symmetric differential systems.
  Foundations of Computational Mathematics 1~(4), 335--383.

\bibitem[{Mansfield(2008)}]{mansfield:bk}
Mansfield, E.~L., 2008. Invariant Calculus for Differential and Discrete
  Problems. Cambridge University Press.

\bibitem[{Mansfield and van~der Kamp(2006)}]{mansfield06}
Mansfield, E.~L., van~der Kamp, P.~H., 2006. Evolution of curvature invariants
  and lifting integrability. J. Geom. Phys. 56~(8), 1294--1325.

\bibitem[{Mu{\~n}oz et~al.(2003)Mu{\~n}oz, Muriel, and
  Rodr{\'{\i}}guez}]{munoz03}
Mu{\~n}oz, J., Muriel, F.~J., Rodr{\'{\i}}guez, J., 2003. On the finiteness of
  differential invariants. J. Math. Anal. Appl. 284~(1), 266--282.

\bibitem[{Neut(2003)}]{neut03}
Neut, S., 2003. Implantation et nouvelles applications de la m\'ethode
  d'\'equivalence de Cartan. Ph.D. thesis, Universit\'e des Sciences et
  Technologie de Lille, \url{http://www.lifl.fr/~neut}.

\bibitem[{Neut and Petitot(2002)}]{neut02}
Neut, S., Petitot, M., 2002. La g\'eom\'etrie de l'\'equation
  {$y'''=f(x,y,y',y'')$}. C. R. Math. Acad. Sci. Paris 335~(6), 515--518.

\bibitem[{Olver(1986)}]{olver:yellow}
Olver, P.~J., 1986. Applications of Lie Groups to Differential Equations. No.
  107 in Graduate texts in Mathematics. Springer-Verlag, New York.

\bibitem[{Olver(1995)}]{olver:purple}
Olver, P.~J., 1995. Equivalence, Invariants and Symmetry. Cambridge University
  Press.

\bibitem[{Olver(2005)}]{olver05s}
Olver, P.~J., 2005. A survey of moving frames. In: Li, H., Olver, P.~J.,
  Sommer, G. (Eds.), Computer Algebra and Geometric Algebra with Applications.
  Vol. 3519 of Lecture Notes in Computer Science. Springer-Verlag, New York,
  pp. 105--138.

\bibitem[{Olver(2007{\natexlab{a}})}]{olver07a}
Olver, P.~J., 2007{\natexlab{a}}. Differential invariants of surfaces.
  Preprint.

\bibitem[{Olver(2007{\natexlab{b}})}]{olver07}
Olver, P.~J., 2007{\natexlab{b}}. Generating differential invariants. Journal
  of Mathematical Analysis and Applications 333, 450--471.

\bibitem[{Olver and Pohjanpelto(2007)}]{olver07b}
Olver, P.~J., Pohjanpelto, J., 2007. Differential invariant algebras of lie
  pseudo-groups. Preprint.

\bibitem[{Ovsiannikov(1982)}]{ovsiannikov78}
Ovsiannikov, L.~V., 1982. Group analysis of differential equations. Academic
  Press Inc. [Harcourt Brace Jovanovich Publishers], New York, translated from
  the Russian by Y. Chapovsky, Translation edited by William F. Ames.

\bibitem[{Riquier(1910)}]{riquier10}
Riquier, C., 1910. Les syst\`emes d'\'equations aux d\'eriv\'ees partielles.
  Gauthier-Villars, Paris.

\bibitem[{Ritt(1950)}]{ritt}
Ritt, J.~F., 1950. Differential Algebra. Vol. XXXIII of Colloquium
  publications. American Mathematical Society, {\tt
  http://www.ams.org/online\_bks}.

\bibitem[{Tresse(1894)}]{tresse94}
Tresse, A., 1894. Sur les invariants des groupes continus de transformations.
  Acta Mathematica 18, 1--88.

\bibitem[{Xu(1998)}]{xu98}
Xu, X., 1998. Differential invariants of classical groups. Duke Math. J.
  94~(3), 543--572.

\bibitem[{Yaffe(2001)}]{yaffe01}
Yaffe, Y., 2001. Model completion of {L}ie differential fields. Annals of Pure
  and Applied Logic 107~(1-3), 49--86.

\end{thebibliography}

\end{document}